\documentclass[naustrian,english,a4paper,11pt,twoside,openright,ansinew]{report}
\usepackage[T1]{fontenc}
\usepackage[latin9]{inputenc}
\usepackage{amsmath}
\usepackage{amssymb}
\usepackage{graphicx}
\usepackage{esint}

\makeatletter

\providecommand{\tabularnewline}{\\}
\newcommand{\lyxdot}{.}

\usepackage[german,english]{babel}
\usepackage{geometry}
\usepackage{graphicx}
\usepackage[bf]{caption2}
\usepackage{amsmath}
\usepackage{amsthm}
\usepackage{mathrsfs}
\usepackage{bm}
\usepackage{bbm}
\usepackage{layout}
\usepackage{verbatim}

\usepackage{times}

\usepackage{float}
\usepackage{fancyhdr}
\usepackage{amsfonts}
\usepackage{amssymb}

\usepackage{comment}

\usepackage{color}

\normalsize

\newcommand\openone{\leavevmode\hbox{\small1\kern-3.3pt\normalsize1}}




\newcommand{\ket}[1]{|#1\rangle}
\newcommand{\bra}[1]{\langle#1|}

\makeatother

\usepackage{babel}
\begin{document}
\selectlanguage{english}
\begin{titlepage}
\begin{figure}[t]
\vspace{-1.75cm}
\begin{flushright}
\includegraphics{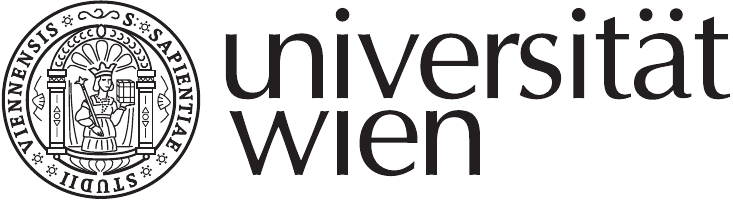}
\end{flushright}
\vspace{1.75cm}
\end{figure}
\vspace{4cm}
\centering{\Huge{\bfseries{\textsf{DISSERTATION}}}}
\vspace{1.75cm}
\\
\centering{\large{\textsf{Titel der Dissertation}}}
\\
\vspace{0.5cm} 
\centering{\LARGE{\textsf{Mutual information in interacting spin systems\\} 
\vspace{0.35cm} 
\textsf{}}} 
\vspace{2.65cm} \\
\centering{\large{\textsf{Verfasser}}} 
\vspace{0.5cm} \\ 
\centering{\LARGE{\textsf{Johannes Wilms}}} 
\vspace{1.25cm} \\
\centering{\large{\textsf{angestrebter akademischer Grad}}} 
\vspace{0.5cm}\\
\centering{\LARGE{\textsf{Doktor der Naturwissenschaften (Dr. rer. nat.) }}} 
\vspace{2cm}\\
\flushleft{ \begin{tabular}{ll} \textsf{Studienkennzahl laut Studienblatt:}      & \textsf{A 091 411} \vspace{0.3cm}\\ 
\textsf{Dissertationsgebiet laut Studienblatt:}  & \textsf{Physik} \vspace{0.3cm}\\ 
\textsf{Betreuer:} 					      & \textsf{Univ.-Prof.\  Dr.\ Frank Verstraete}  \end{tabular}} \vspace{1cm} \\
\flushleft{\large{\textsf{Wien, im Juni 2012}}}
\vspace{0.75cm}
\end{titlepage}
\newpage
\thispagestyle{empty}

\tableofcontents{}

\chapter*{Abstract}

\addcontentsline{toc}{chapter}{Abstract}

\selectlanguage{english}

This thesis combines ideas from the fields of quantum information
and condensed matter. This combination has already proven extraordinarily
successful: While one would naively assume that in condensed matter
theory one would always have to work with an exponentially large Hilbert
space, we are now beginning to understand that only a small part of
these states is physically relevant. This understanding is primarily
driven by insights about entanglement properties of the states actually
occurring in nature; those insights have been delivered exactly by
the application of quantum information theory.

In the spirit of these ideas, this thesis applies a quantity that
is defined and justified by information theory -- mutual information
-- to models of condensed matter systems. More precisely, we will
study models which are made up out of ferromagnetically interacting
spins. Quantum information theory often focuses on the ground state
of such systems; we will however be interested in what happens at
finite temperature.

Using mutual information, which can be seen as a generalization of
entanglement entropy to the finite-temperature case, we can study
the different phases occurring in these models, and in particular
the phase transitions between those. We examine broadly two different
classes of models: classical spins on two-dimensional lattices \cite{Wilms11},
and fully-connected models of quantum-mechanical spin-1/2 particles
\cite{wilms2012finite}. We find that in both cases the mutual information
is able to characterize the different phases, ordered and unordered,
and shows clear features at the phase transition between those.

In particular, in the case of classical spins on a lattice, we numerically
find a divergence of the first derivative of the mutual information
at the critical point, accompanied by a maximum within the high-temperature
phase. We justify the location of the maximum by studying the behaviour
of Fortuin-Kasteleyn clusters of aligned spins.

For fully-connected spins, we find a rather different behaviour: a
mutual information that logarithmically diverges at the phase transition,
as long as it is of second order; for a first-order phase transition
there is no divergence at all. Analytical calculations in the classical
limit support the numerical evidence. The behaviour is consistent
with what one would have predicted from existing studies of entanglement
entropy at the zero-temperature phase transition in these models.

\chapter*{Zusammenfassung}

\addcontentsline{toc}{chapter}{Zusammenfassung}

\selectlanguage{german}

\enlargethispage{1cm}

\selectlanguage{naustrian}%
Diese Dissertation kombiniert Ideen aus den Bereichen Quanteninformation
und Festkörperphysik. Diese Verbindung hat sich bereits als außerordentlich
erfolgreich erwiesen: Es hat sich gezeigt, dass anstelle des gesamten,
exponentiell großen, Hilbertraums nur ein kleiner Teil der Zustände
tatsächlich physikalisch relevant ist. Durch Einsichten in die Verschränkungseigenschaften
von Zuständen hat Quanteninformation hier wesentlich zu unserem Verständnis
beigetragen.

In Anlehnung an solche Ideen wird in dieser Arbeit die Verwendung
von mutual information (auch: Transinformation, gegenseitige Information,
oder Synentropie) zur Beschreibung von Korrelationen in Festkörpersystemen
motiviert und beschrieben. Dazu werden ferromagnetisch wechselwirkende
Spinsysteme betrachtet. Im Bereich der Quanteninformation werden häufig
die Grundzustände solcher Systeme diskutiert; in dieser Arbeit werden
dagegen thermische Zustände bei endlicher Temperatur erörtert werden.

Mittels mutual information, die dabei als Verallgemeinerung von entanglement
entropy (Verschränkungsentropie) für den Fall gemischter Zustände
gesehen werden kann, können wir dann die verschiedenen Phasen dieser
Modelle untersuchen, und insbesondere die Phasenüber-gänge zwischen
ihnen. Wir werden im Wesentlichen zwei Klassen von Modellen betrachten:
zum einen klassische Spins auf zweidimensionalen Gittern \cite{Wilms11},
und zum anderen vollständig verbundene Modelle quantenmechanischer
Spin-1/2-Teilchen \cite{wilms2012finite}. In beiden Fällen zeigt
die mutual information charakteristisches Verhalten am Phasenübergang
und innerhalb der geordneten beziehungsweise ungeordneten Phasen.

Insbesondere finden wir für klassische Spins auf einem Gitter mittels
numerischer Me-thoden eine Divergenz der ersten Ableitung der mutual
information am kritischen Punkt, verbunden mit einem Maximum innerhalb
der paramagnetischen Phase. Wir motivieren die Existenz dieses Maximums
durch Betrachtungen von Fortuin-Kasteleyn-Clustern von parallel ausgerichteten
Spins.

Für die vollständig verbundenen Modelle zeigt sich ein anderes Verhalten:
Die mutual information divergiert logarithmisch am Phasenübergang,
sofern dieser ein Übergang zweiter Ordnung ist; für einen Übergang
erster Ordnung gibt es keine Divergenz. Analytische Rechnungen im
klassischen Grenzfall unterstützen hierbei die numerischen Ergebnisse.
Das gefundene Verhalten ist konsistent mit bestehenden Resultaten,
die bereits in anderen Arbeiten für die entanglement entropy des Grundzustandes
gefunden wurden.

\selectlanguage{english}%

\chapter{Introduction and overview}

The purpose of this thesis is to explore mutual information as a correlation
measure, in the setting of a few selected spin models, at finite temperature.

In order to do this, I will try to generally motivate the study of
mutual information, as a correlation measure, in chapter \ref{chap:mutual-information}.
This will mostly be based on an information-theoretic point of view.
In particular, we will of course see the connection to entanglement
entropy, to which mutual information sometimes reduces, and which
was certainly a motivation for its study. I would however like to
point out already at this point that this thesis is not concerned
with a measure for {}``quantum entanglement'', which would allow
to distinguish entanglement from {}``classical correlations''. It
is well known that mutual information cannot provide that \cite{vedral1997quantifying}
-- and while such a distinction is without any doubt very interesting,
we will see that it might not actually be so relevant if one wants
to use correlations to study phase transitions.

The bulk of the thesis will then be dealing with actually calculating
mutual information in interesting model systems. They can all be understood
as {}``spin systems''. There are essentially two very different
classes that we will discuss: On the one hand, classical spins on
fixed lattice positions, with local interactions, discussed in chapters
\ref{chap:classical-spins-lattice} through \ref{chap:More-lattice-spin}.
Some key results out of this group have been published as \cite{Wilms11}.
On the other hand, there are fully-connected spin models, where the
spins all interact with each other rather than only with a small set
of neighbours; this will be the subject of chapters \ref{chap:lmg}
and \ref{chap:m-n-models}. Again, some of the results have already
been published, as \cite{wilms2012finite}. The thesis finishes with
some concluding remarks and an outlook.

\chapter{\label{chap:mutual-information}Mutual information}

In this chapter I will try to generally motivate the study of mutual
information in physical systems. The following chapters will then
deal with its calculation for some specific systems, and also contain
sometimes more specific motivation.

\section{\label{sec:mi-motivation}Motivation}

\begin{figure}
\centering{}\includegraphics[width=0.8\textwidth]{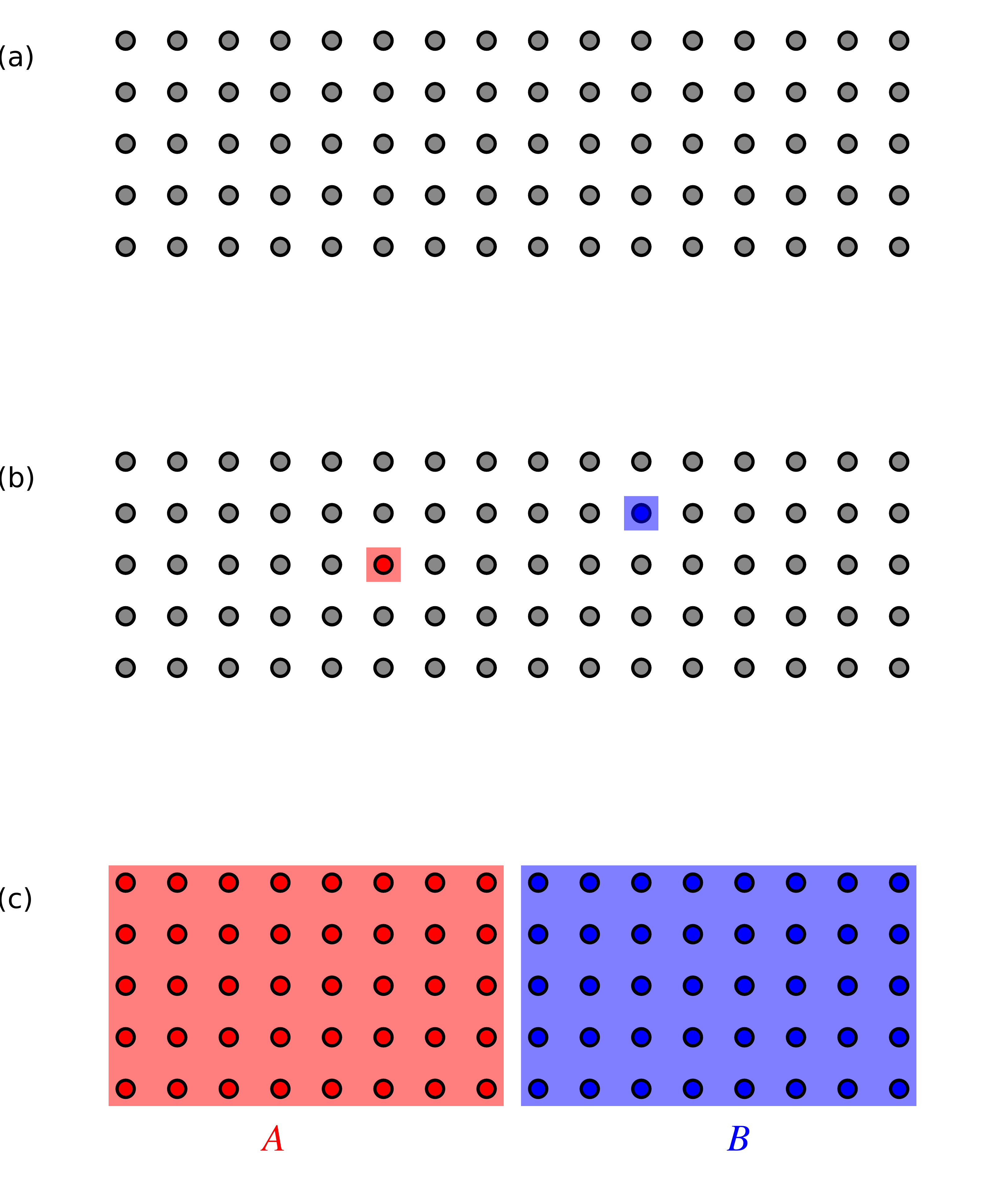}\caption{\label{fig:corrfun-vs-mi}Correlation functions versus mutual information}
\end{figure}

I believe the most intuitive way to motivate the study of mutual information
is the following: Take a look at figure \ref{fig:corrfun-vs-mi}(a).
This tries to represent a rather generic physical system: some bigger
system that is made up out of many smaller constituents. These constituents
could for example be individual nucleons, or atoms, or molecules.
Let us maybe call them just {}``particles'' for now. They could
be arranged in a regular lattice (such as in the figure, and also
as discussed in the next few sections), or they might be more randomly
distributed.

For the system to be interesting, the particles will have to have
some interaction. This interaction makes them behave differently than
if they were just isolated individual particles. In fact, there will
now be \emph{correlations} between the particles. For example, in
a solid the individual particles are arranged in a regular lattice,
which is obviously a very strong correlation between the particles.

As another example, closer to the systems that will be considered
in this thesis, a \emph{ferromagnetic} interaction means that particles
(which you may in this case imagine as some sort of elementary magnets)
will tend to all align in the same direction, which is clearly also
some sort of correlation. But they will not always be perfectly aligned
-- there might be other factors that tend to counteract the perfect
alignment, such as external magnetic fields, or temperature and entropy.
In fact, at high enough temperature or fields, the ferromagnetic order
typically vanishes, in a phase transition to an unordered (paramagnetic)
phase.

It is therefore certainly important to try to actually quantify such
correlations. Let's stay with the example of a ferromagnetic interaction,
and call the individual particles \emph{spins. }A very well-established
way to describe the correlations is the following: just consider two
of these spins, as indicated in part (b) of figure \ref{fig:corrfun-vs-mi}.
Maybe they could be described by classical values $s_{i}$ and $s_{j}$.
These could for example be allowed to be only $+1$ and $-1$, representing
orientation along some given axis. Maybe they would instead have to
be described quantum-mechanically, in terms of spin operators, for
which we could however just use the same symbols. In both cases, the
concept of a correlation function is applicable: It is an expression
like
\[
\langle s_{i}s_{j}\rangle-\langle s_{i}\rangle\langle s_{j}\rangle
\]
that tells us how the expectation value of the product of the spin
{}``operators'' is different from just looking at the two spins
individually. This is clearly an expression that tells us something
about correlations between these two spins. It is called a correlation
{}``function'' because we can now study how this expression depends,
for example, on interaction strength, temperature, external fields,
or the distance of the two spins.

But don't you find it a bit lacking? In particular, don't you think
it is rather arbitrary to pick out just two spins out of all of them?
Certainly, in general there will be correlations in the system that
we miss in this way, for which we would have to look at more than
two spins at a time. Of course, this is well understood, and there
are corresponding {}``higher order'' correlation functions that
involve more than two spins. But now, it is no longer so obvious which
of these correlation functions we should refer to if we want to quantify
the correlations, so is there maybe something else we could do?

Indeed there is: Imagine that you (virtually, not physically) divide
the system into two parts, as indicated in figure \ref{fig:corrfun-vs-mi}(c).
Then we would like to ask the following question: how much information
can we gain about one part of it by looking only at the other one?
It turns out that this is exactly what is known as \emph{mutual information}!

So what is this mutual information? For a much more general (but also
more abstract) discussion you can consult e.g. \cite{coverthomas},
but let me try to describe it in the setting I have started to lay
out. The first thing we need to understand is that the state of a
system at finite temperature is described by a probability distribution.
This is also needed for the averaging in the correlation functions
of course, but now we will make even more explicit use of the probabilities.
In classical physics, in thermal equilibrium, the probability distribution
is the well known Boltzmann-Gibbs distribution. In quantum physics,
we will need a corresponding Boltzmann-Gibbs density operator instead.

The Shannon entropy \cite{Shannon48} of a probability distribution
$\{p_{i}\}$
\[
S=-\sum_{i}p_{i}\log p_{i}
\]
is a quantity that describes the uncertainty we have about the state
of the system. If we now do our virtual bipartitioning, we write the
index of a state instead of $i$ as $(a,b)$ with the understanding
that $a$ and $b$ describe the state of the parts respectively.

This means we can now write the total entropy as
\[
S_{AB}=-\sum_{ab}p_{ab}\log p_{ab}\,,
\]
or
\[
S_{AB}=-\mathrm{Tr\,\rho_{AB}\log\rho_{AB}}
\]
for the corresponding von Neumann entropy in the quantum case (see
e.g. chapter 11 of \cite{nielsenchuang2000} for more about the quantum
case). It is of course clear that the classical case is just the special
case of a diagonal density matrix; I still find it useful to write
out the classical sums as well, as they make explicit the operations
that are implicit in the quantum-mechanical trace.

Let us realize now that we can also find the probability distributions
for the parts as the marginal probability distributions
\[
p_{a}=\sum_{b}p_{ab}\,,\; p_{b}=\sum_{a}p_{ab}\,.
\]
In the quantum mechanical setting, we would have to take partial traces
of the density matrix, {}``tracing out'' the respective other part
(\cite{nielsenchuang2000}, chapter 2),
\[
\rho_{A}=\mathrm{Tr_{B}\,\rho_{AB}\,,\;\rho_{B}=\mathrm{Tr_{A}\,\rho_{AB}\,.}}
\]
We can now also calculate entropies of the marginal distributions,
or reduced density matrices:
\[
S_{A}=-\sum_{a}p_{a}\log p_{a}\,,\; S_{B}=-\sum_{b}p_{b}\log p_{b}
\]
or
\[
S_{A}=-\mathrm{Tr\,\rho_{A}\log\rho_{A}}\,,\; S_{B}=-\mathrm{Tr\,\rho_{B}\log\rho_{B}}\,.
\]
Now we are ready to define mutual information as
\begin{equation}
I(A:B)=S_{A}+S_{B}-S_{AB}.\label{eq:mi-definition-shannon}
\end{equation}
How can we understand that this is what we want? I very much like
the following point of view: it is the uncertainty ({}``what we do
not know'') about $A$ and {}``what we do not know about $B$''
minus {}``what we do not know about the whole system''. Quite intuitively,
this is the information that $A$ and $B$ have in common. Or, as
described before, the information that we can extract from $A$ about
$B$ and vice versa (note that the definition is symmetric).

\begin{figure}
\begin{centering}
\includegraphics[width=0.38\textwidth]{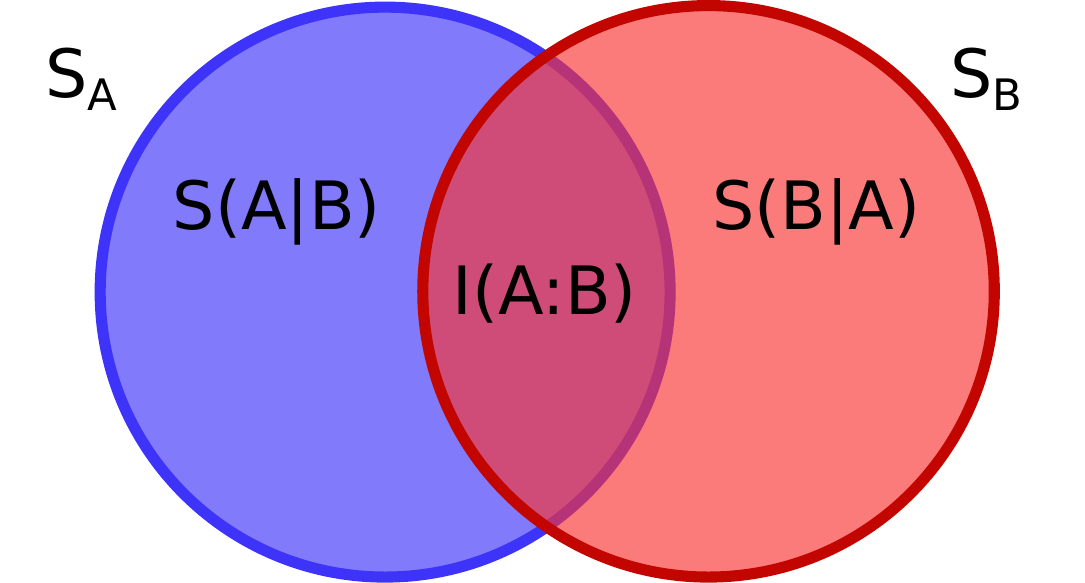}
\par\end{centering}

\caption{\label{fig:Information-diagram}Information diagram for (Shannon)
mutual information}
\end{figure}

We can try to illustrate this with an {}``information diagram''
as shown in figure \ref{fig:Information-diagram} \cite{coverthomas,cerf1997negative}.
The red circle represents $S_{A}$, the blue one $S_{B}$. The uncertainty
about the whole system, $S_{AB}$, is the whole shaded area in the
figure. If you subtract that from $S_{A}+S_{B}$, you see that what
remains is the {}``overlap'' $I(A:B)$.

Also indicated in the figure is the relation to conditional entropies.
As you can see there, we can also write the mutual information as
\begin{equation}
I(A:B)=S_{A}-S(A|B)=S_{B}-S(B|A)\label{eq:mi-cond-ent}
\end{equation}
where $S(A|B)$ is the entropy of $A$ given that we already know
$B$. You can easily see that this leads to the same interpretation
of mutual information.

There is another important way to view mutual information: Let us
write it out in terms of classical probabilities:
\begin{equation}
I(A:B)=\sum_{ab}p_{ab}\log\frac{p_{ab}}{p_{a}p_{b}}\label{eq:kullback-leibler}
\end{equation}
and you can see that this in fact a sort of {}``distance measure''
between the actual joint probability distribution $p_{ab}$ of the
total system, and the product of the marginal distributions $p_{a}p_{b}$.
It does not actually fulfill the axioms of a distance (you can immediately
notice that it is not symmetric between $p_{ab}$ and $p_{a}p_{b}$).
It is therefore labeled a {}``divergence'' instead. More specifically:
the Kullback-Leibler divergence of these two distributions (this is
also called their relative entropy).

It seems intuitive to measure correlations by looking at how far the
actual distribution is from a product distribution, so this is another
nice interpretation of mutual information. It is also relevant for
a different reason: The arguments I presented before rely on adding,
or subtracting, entropies (or conditional entropies). This makes sense
for Shannon/von Neumann type entropies, as explained for example in
\cite{coverthomas}. But these are not the only entropies out there,
and such a property is not generally true for other definitions of
entropy. If we wanted to work with Rényi entropies \cite{renyi1960,renyi1965}
instead, the way towards a definition of mutual information should
therefore certainly start from such a distance measure instead. More
about this in section \ref{sec:mi-renyi}.

\section{Mutual information and entanglement entropy}

There is another strong motivation for mutual information that has
already been mentioned in the introduction: It has been applied very
successfully in a special case where it is usually known as {}``entanglement
entropy''. You could also say that mutual information is a generalization
of this by now pretty well-established concept. The special case is
that the system is in a pure quantum state, which has therefore zero
total entropy. A classical state with zero entropy is usually uninteresting,
because the entropies of the parts will be zero as well. But pure
quantum states can be strongly entangled, which reflects in nonzero
entropies of the reduced density matrices. In that case, $S_{A}=S_{B}$
is called entanglement entropy. You will read more about mutual information
as a generalization of entanglement entropy in particular in chapters
\ref{chap:lmg} and \ref{chap:m-n-models}.

\section{The success story of mutual information}

I would now like to argue that mutual information has proven a very
useful concept in a huge number of applications, in many different
branches of physics and other sciences. If you are already convinced
of that, or do not care either way, feel free to skip this section.

In fact, I think the story of mutual information is such a success
that it is impossible to even give an appropriate selection of references,
not least because I am far from an expert in most of the applications.
That is why I would like to point out only a very small number of
references that I personally found particularly interesting: Mutual
information seems to have turned out to be extremely useful for the
alignment of (in particular medical) images: Imagine you have two
different pictures of the same thing -- say, one obtained by x-rays
(computer tomography) and one obtained by magnetic resonance imaging,
and you want to overlay them to get a better total picture (since
they show different features differently well). The most effective
way to get them to {}``match'' seems indeed to be minimization of
their mutual information, in a suitably defined sense, as originally
described in \cite{viola1997alignment,viola1995alignment,wells1996multi,collignon1995automated,collignon1998multi}.

Or, for an example that is maybe closer to your idea of physics, see
\cite{fraser1986independent} for an application in the context of
dynamical systems.

All of these references are widely cited, but I found that many physicists
are still pretty unfamiliar with the concept of mutual information
and hence sometimes doubtful if it really can be that relevant. Therefore,
I would like to go one step farther in trying to convince you that
it has in fact proven useful. Indeed, let us try to quantify how much
use it has found: First, let us look at articles submitted to arXiv.org,
which is a preprint server with a strong focus on physics (and {}``neighbouring''
sciences, such as mathematics or quantitative biology). We could even
restrict to just papers that are explicitly categorized as physics,
but the results turn out to be rather similar, so let us work with
the largest possible data set.

\begin{figure}
\begin{centering}
\includegraphics[width=1\textwidth]{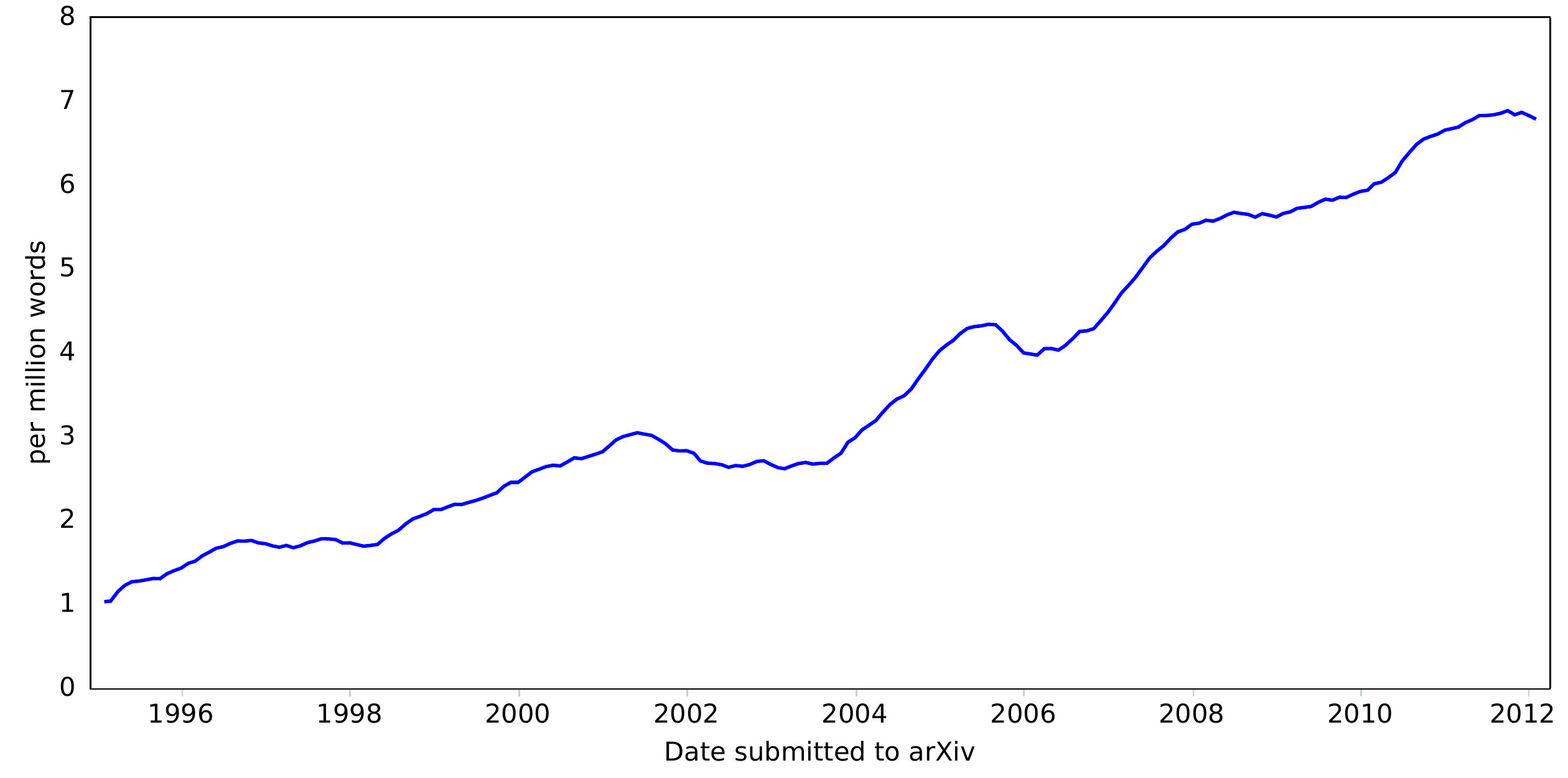}
\par\end{centering}

\caption{\label{fig:arxiv}Relative frequency of the (case-insensitive) phrase
{}``mutual information'' in articles submitted to arXiv.org. 12-month
rolling average, spanning the period January 17, 1995 to February
22, 2012. Graph produced by the web application at arxiv.culturomics.org.}
\end{figure}

Figure \ref{fig:arxiv} shows the frequency of the phrase {}``mutual
information'' in the articles submitted to arXiv over a period from
about 1995 (which is about the earliest date from which there is a
reasonably good data basis) to the present. This graph is produced
by arxiv.cultoromics.org; since we are not interested in any particular
short-term trends or fashions, the data are smoothed by carrying out
a rolling average over 12 months. It seems to be very clear to me
that mutual information must have proven a useful tool -- at least
that is the only reasonable explanation I can offer for why it is
now used more than five times as often that it was about 15 years
ago. Of course that by itself would not be sufficient to argue that
it is also relevant in the current context, but I certainly find this
encouraging! I would like to point out that I only think this is a
good indicator for relevance because the data spans many years --
if it were just a short-term trend, you could easily argue that it
is just currently {}``fashionable'' and may well prove to have very
limited use after all.

\begin{figure}
\begin{centering}
\vspace*{\bigskipamount}
\includegraphics[width=1\textwidth]{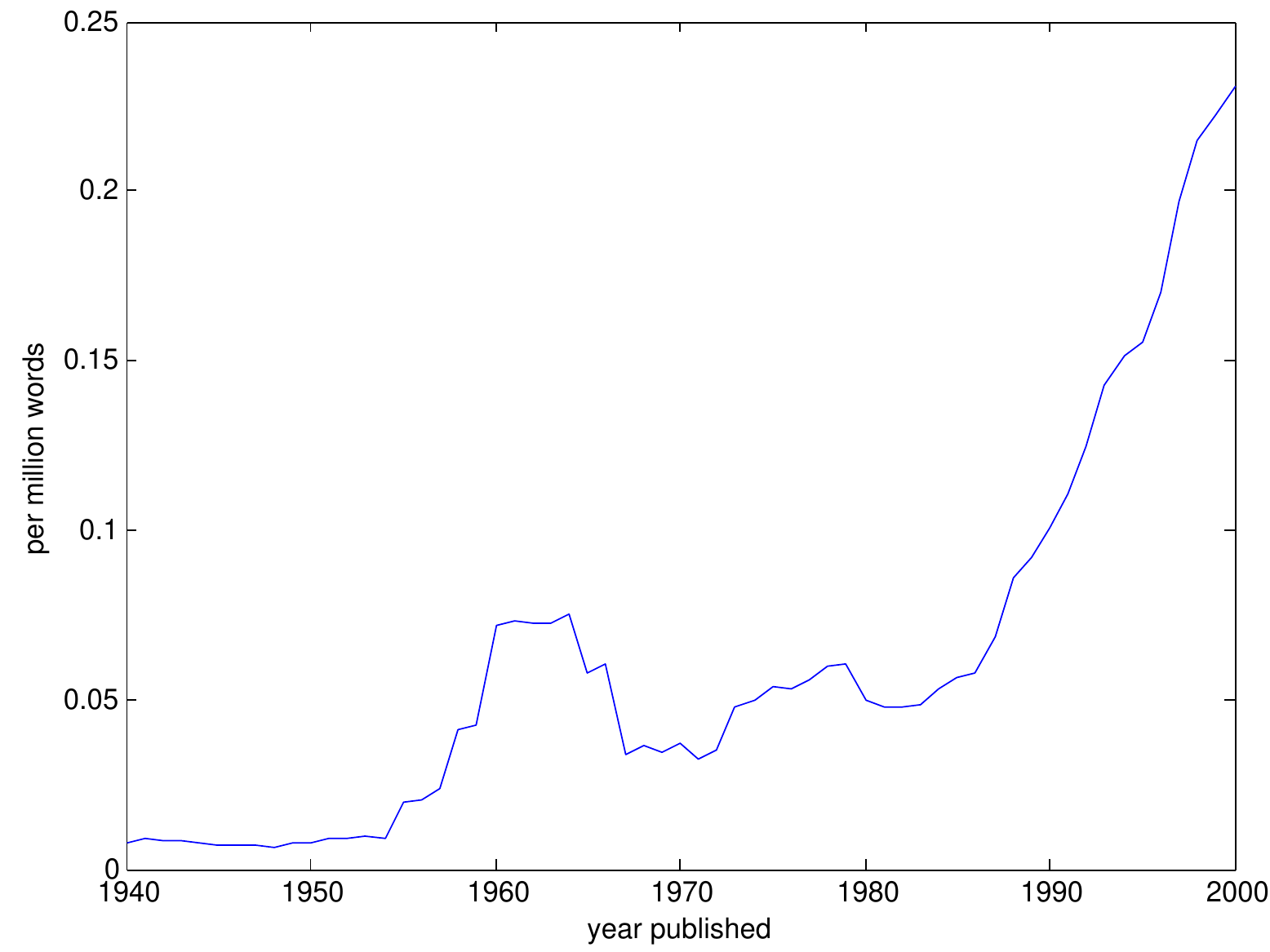}
\par\end{centering}

\caption{\label{fig:books}Relative frequency of the (case-insensitive) phrase
{}``mutual information'' in books published between 1940 and 2000,
see reference \cite{michel11}. 7-year rolling average.}
\end{figure}

Since arXiv data only go back a relatively short period of time, I
hope figure \ref{fig:books} can convince you that the success of
mutual information is in fact a long-term one. The figure is based
on \cite{michel11} and the data available at books.google.com/ngrams
($n$-gram is just a fancy term for a sequence of $n$ words; in this
language, {}``mutual information'' is called a 2-gram, or bigram).
The figure is produced from the raw data {}``English, Version 20090715''
which is considered the highest quality data set currently available.
Of course, the frequency of the term {}``mutual information'' in
a general corpus of books is much lower than on the arXiv, but I believe
we can again call it a pretty consistent success story. As an aside,
you may try the phrase {}``information theory'' at books.google.com/ngrams
to convince yourself that it was indeed started by Shannon's famous
paper in 1948 \cite{Shannon48}.

\section{\label{sec:what-done}What has been done?}

Maybe you are now convinced that it should be a pretty obvious thing
to try and look at mutual information. So, the next question is: has
anyone else already done it? Judging by the title {}``Mutual Information
in Ising Systems'' of the paper \cite{matsuda1996} you would think
it has been done at least in a setting similar to the one which will
be discussed in chapter \ref{cha:Ising-lattice}. However, that paper
studies the mutual information of just two spins that are picked out,
quite similarly to the study of correlation functions in fact. This
is certainly also interesting (and turns out to be much simpler than
what we are going to do), but cannot yield the new kind of information
about many-body correlations we are looking for. You can also find
a similar kind of study among the results of \cite{sole1996}. Yet
another different definition of mutual information, as occurring between
the steps of a Monte Carlo simulation of an Ising system, is presented
in \cite{arnold1996}. While very different from the approach presented
here, it is also very interesting, and turns out to work well to determine
the phase transition, something that we will also repeatedly be discussing
(with our definition) in the following chapters.

Much closer to the spirit of this thesis are the works \cite{Hastings10,Melko10,Singh11}.
They present the same physically motivated idea of mutual information
as a generalization of entanglement entropy. They study quantum spins
on 2-dimensional lattices, which is not among the cases considered
in this thesis, but it turns out that their results are not too dissimilar
from what we will find for classical systems on these lattices, as
will be shown in the following chapters. However, their (numerical)
results are from my point of view severely limited by the fact that,
in their Quantum Monte Carlo simulations, it is only possible to calculate
Rényi entropies of reduced density matrices, with integer Rényi parameters
larger than 1. They go on to use these to define mutual information
according to \eqref{eq:mi-definition-shannon}. It does seem to produce
useful results, but in principle the usage of this formula is not
justified for Rényi entropies; as mentioned before, adding and subtracting
entropies only really makes sense for Shannon/von Neumann entropies.
Even just generally working with {}``raw'' (non-smoothed) Rényi
entropies can sometimes be very misleading. If you're wondering now
why all that is, or what Rényi entropies are at all, read on to the
next section!

\section{\label{sec:mi-renyi}Generalizations}

Of course there are some obvious generalizations to our definition
of mutual information, the most obvious one being maybe dividing the
system into more than just two parts. The aforementioned studies \cite{matsuda1996,sole1996}
could indeed be considered an extreme limiting case, where the subsystems
picked would just be single spins, with the rest of the system understood
as a sort of background. If we wanted to study this more generally
though, there will be a lot of freedom in our choice of the parts,
and that is simply beyond the scope of this investigation. Also, in
many cases the calculations will become infeasibly hard.

Another very interesting route would however be to study not just
Shannon entropies (or their quantum generalization, von Neumann entropies).
Possibly the most interesting other class of entropies are Rényi entropies
\cite{renyi1960,renyi1965}
\[
S^{(\kappa)}=\frac{1}{1-\kappa}\log\sum_{i}p_{i}^{\kappa}
\]
for a probability distribution $\{p_{i}\}$, or in the quantum case
\[
S^{(\kappa)}=\frac{1}{1-\kappa}\log\mathrm{Tr}\,\rho^{\kappa}
\]
where $\kappa$ is a {}``Rényi parameter'', $\kappa\geq0$ and $\kappa\neq1$.

If you carefully take the limit for $\kappa\to1$, you will see that
you reproduce exactly the Shannon/von Neumann entropy (you will probably
need l'Hôpital's rule). This shows that this is obviously a possible
generalization of the Shannon definition (because it reduces to that
in this limit).

One advantage of Rényi entropies is that (integer) powers of density
matrices may sometimes be calculated even where logarithms may not,
such as was the case in \cite{Hastings10,Melko10,Singh11}.

Even more interesting is the fact that in some settings Rényi-type
entropies may have a more concrete operational meaning. As outlined
in \cite{smoothrenyi,Koenig09}, Shannon entropy generally has an
{}``asymptotic'' interpretation: In the limit of a large number
of realizations of a system, the uncertainty it describes can be made
more tangible as both the amount of randomness that can be extracted
from it and and its optimum encoding length (think of a compressed
description of the system). If you have however just one realization
of a system, these two things can be very different. If you want a
measure of these quantities that is appropriate for a single realization
of a system, you should look at Rényi entropies $S^{(\kappa)}$ --
maybe in particular the limiting cases $S^{(0)}$ and $S^{(\infty)}$,
also called max- and min-entropy respectively. Written out explicitly
for a probability distribution $\{p_{i}\}$, these are
\begin{eqnarray*}
S^{(0)} & = & \log\,|\mathrm{supp\,}\{p_{i}\}|\\
S^{(\infty)} & = & -\log\max_{i}p_{i}
\end{eqnarray*}
and from this you may already see that they are related to compressibility
on the one hand and amount of extractable randomness on the other
hand. Max-entropy, which is simply the logarithm of the number of
nonzero probabilities, is also known as Hartley entropy \cite{hartley1928}
and as such even precedes Shannon entropy.

In fact, it turns out that you might not even have to use these extreme
cases, but that there are essentially two classes of Rényi entropies,
namely the ones for $\kappa<1$ and the ones for $\kappa>1$. However,
all of this is only true if you apply {}``smoothing'', which means
in calculating the entropy you optimize over all probability distributions
within a certain distance of the one you are looking at. If you do
not do this, Rényi entropies can become misleading and have very undesirable
behaviour, because small changes in the probability distributions
can lead to large changes in entropy. Maybe this is most easily seen
in the definition of $S^{(0)}$, which would be clearly changed a
lot by adding very many very small probabilities (while slightly reducing
a large one, for conservation of probability). Smoothing is the procedure
to deal with such undesired behaviour.

This extra optimization step however (which is not required for Shannon/von
Neumann entropies) makes those smoothed Rényi entropies much more
complicated to deal with, at least numerically. Add to this the fact
that for Rényi mutual information, we cannot start from \eqref{eq:mi-definition-shannon}
or \eqref{eq:mi-cond-ent}, but need to find a good generalization
of \eqref{eq:kullback-leibler}, and you see why this actually becomes
significantly more involved than Shannon/von Neumann mutual information.
And after all, the usual thermodynamic entropy is just Shannon/von
Neumann entropy! Therefore, we will focus on the Shannon/von Neumann
definitions from now on.

\section{\label{sec:pt-area}Phase transitions; area laws}

As has already been hinted in the previous sections, we will generally
be interested to study mutual information in the vicinity of phase
transitions -- we have argued that mutual information measures correlations,
and phase transitions are where we expect the correlations to change
strongly.

We will see that mutual information may even be used to {}``detect''
phase transitions, meaning that by looking at the behaviour of mutual
information you can conclude where (i.e., at what system parameters)
the phase transition occurs. In the systems studied here, this is
usually well known already, and can typically be found with less effort
by looking at more traditional, {}``classical thermodynamical''
quantities. In the systems studied here, you do not even need to bother
with things like correlation functions, but straightforward {}``order
parameters'' will suffice. However, understanding the possible behaviours
of mutual information is important anyway, because there are systems
where traditional order parameters are not applicable and even correlation
functions might fail. These are sometimes called novel or topological
phases.

There is another aspect of mutual information that I should already
mention in connection with phase transitions, which will be discussed
in more detail within the following chapters: Remember the {}``classical
thermodynamic'' quantities such as correlation functions, or simply
order parameters, with which we contrasted mutual information. In
certain types of phase transitions, these show a very regular behaviour,
independent of microscopic details of a system, but rather {}``universal''
for a whole class of systems, which is then suitably called a universality
class. The typical manifestation of such behaviour is in universal
{}``critical exponents'' that describe the functional behaviour
of these quantities.

So is there anything similar for mutual information? There is indeed.
Probably the most well known manifestation of such universal behaviour
comes in the form of \emph{area laws} \cite{arealaws}. They describe
how mutual information scales with the size of the system, which is
in many interesting cases not with the volume of the system, but rather
its surface area -- therefore the name area law. Area laws are something
that has also been studied extensively in the context of entanglement
entropy, see e.g. \cite{vidal2003entanglement,calabrese2004entanglement,plenio2005entropy,its2005entanglement,verstraete2006matrix,verstraete2006criticality,hastings2007area,eisert2010colloquium}.
We will see a clear instance of area laws in the following chapters!

While the correlations in classical systems will always strictly bounded
by an exact area law \cite{arealaws}, there may in general be corrections
to the area law behaviour; and {}``universal exponents'' might actually
manifest in corrections, see also e.g. \cite{stephanshannon,renyi2dising}.

In the systems studied in chapters \ref{chap:lmg} and \ref{chap:m-n-models},
there is no notion of dimension (or you could say they are infinite-dimensional
models). Hence, there is no such thing as area laws. Still, there
are typical scaling behaviours that have been identified in the ground
states of such models, e.g. in the entanglement entropy \cite{Latorre05_2,Barthel06_2,Vidal07,Orus08_2}.
In section \ref{sec:mi-scaling-vs-qpt} we will discuss similar behaviour
that we found in the mutual information.

\chapter{\label{chap:classical-spins-lattice}Classical spin models on a lattice}

In this chapter, we will consider how to calculate mutual information
in a classical spin model on a lattice. The initial considerations
will be independent of the lattice, but we will then specialize to
two-dimensional lattices and in particular consider the case of a
square lattice.

Key results of the calculations in this chapter and the next two have
been published in \cite{Wilms11}, but a big part of the methods sections
is previously unpublished, in particular the connection to matchgates
as laid out in section \ref{sec:ising-matchgates}.

Let us start from the definition of mutual information in the form
of equation \eqref{eq:kullback-leibler}. This involves a sum over
all the states of the system. Even for classical spins, these are
exponentially many states: if we have $N$ spins that can take just
two possible values, say $\{-1,+1\}$, there are $2^{N}$ states.
Clearly, carrying out such a sum quickly becomes intractable -- even
a pretty modestly sized $10\times10$ lattice is out of reach.

One idea you might immediately have is to replace the exact sum by
a Monte Carlo sampling \cite{metropolis1953,mcbook1}. However, take
another look at equation \eqref{eq:kullback-leibler}: what you then
need to evaluate for each sample is a still pretty complicated expression
involving marginal probabilities, which are not easily accessible
-- tracing out one subsystem involves an exponentially large sum again.
Of course, you could argue that this could again be approximated using
Monte Carlo methods. But if we had to do that in each step of the
{}``outer'' Monte Carlo run, it would be very inefficient -- with
Monte Carlo methods, you need to keep the updates simple, so that
you can have the large number of samples you generally require for
a good approximation.

However, such initial considerations are what finally lead to the
following approach, which can in the end be used even without any
involvement of Monte Carlo methods: You may notice that our formula
bears some resemblance to a partition function, which is also a sum
over all possible states of a system, although a simpler one:
\begin{equation}
Z=\sum_{ab}p_{ab}.\label{eq:def-part-func}
\end{equation}
In many cases, partition functions can actually be calculated pretty
efficiently. However, our problem currently does not have the form
of a partition function. But let us now also remember the result mentioned
in section \ref{sec:pt-area}, that a classical spin system always
has to fulfill an area law \cite{arealaws}. In a way, this implies
that instead of summing over all the spins in the system, it should
be sufficient to sum over just those that make up the {}``boundary''
between the systems.

\begin{figure}
\begin{centering}
\includegraphics[width=0.8\textwidth]{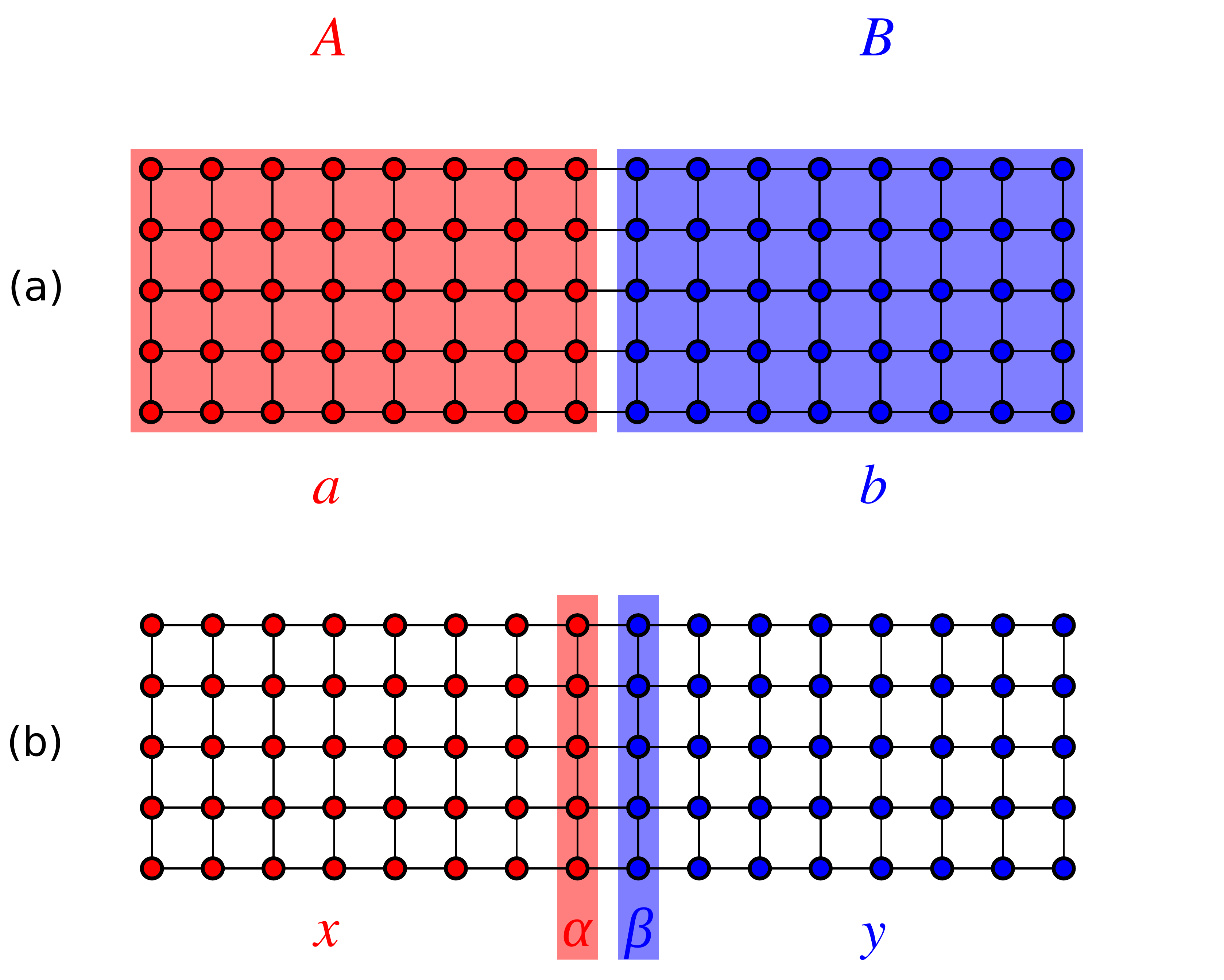}
\par\end{centering}

\caption{\label{fig:systems-borders}Spins on a strip/cylinder: (a) subsystems;
(b) their interiors and borders.}
\end{figure}

Let us try to combine these ideas into a way to rewrite the definition
\eqref{eq:kullback-leibler}. We divide each subsystem into its {}``interior''
and its {}``border'' with the other system. The border is defined
as the set of all those sites that share a bond with a site that belongs
to the other subsystem; the rest makes up the interior. This is illustrated
in figure \ref{fig:systems-borders}: part (a) shows you the most
straightforward way to divide a rectangular system into two parts
$A$ and $B$ with states $a$ and $b$. Part (b) shows how we identify
the borders, for the states of which we will now use labels $\alpha$
and $\beta$, and the interior (or {}``bulk'') parts which we will
label by $x$ and $y$ respectively. Therefore, we now divide the
labels $a$ and $b$ for the states of the subsystems, as we used
them in chapter \ref{chap:mutual-information}, into $a=(\alpha,x)$
and $b=(\beta,y)$.

Now, we can rewrite the joint probability $p_{ab}$ as $p_{ab}=p_{\alpha x\beta y}=p_{x}p_{\alpha x}p_{\alpha}p_{\alpha\beta}p_{\beta}p_{\beta y}p_{y}$
where $p_{x}$ describes the contribution from bonds between spins
all in the interior of $A$, $p_{\alpha x}$ the contribution from
the bonds between the interior and the border of $A$, $p_{\alpha}$
the one from bonds within the border, and so on. Plugging in this
{}``factorization'' of the probabilities, we find that we can rewrite
the Shannon mutual information \eqref{eq:kullback-leibler} as
\begin{align}
I(A:B) & =\sum_{\alpha x\beta y}p_{\alpha x\beta y}\log\frac{p_{\alpha x\beta y}}{\sum_{\beta'y'}p_{\alpha x\beta'y'}\sum_{\alpha'x'}p_{\alpha'x'\beta y}}\nonumber \\
 & =\sum_{\alpha\beta}p_{\alpha}p_{\alpha\beta}p_{\beta}\underbrace{\sum_{x}p_{x}p_{\alpha x}}_{Z_{A}(\alpha)}\underbrace{\sum_{y}p_{y}p_{\beta y}}_{Z_{B}(\beta)}\nonumber \\
 & \cdot\log\frac{p_{\alpha\beta}}{\underbrace{\sum_{\beta'y'}p_{\alpha\beta'}p_{\beta'}p_{\beta'y'}p_{y'}}_{Z_{\tilde{B}}(\alpha)}\underbrace{\sum_{\alpha'x'}p_{\alpha'\beta}p_{\alpha'}p_{\alpha'x'}p_{x'}}_{Z_{\tilde{A}}(\beta)}}.\label{eq:I}
\end{align}
Here the curly underbrackets identify {}``partial'' partition functions,
e.g. $Z_{A}(\alpha)=\sum_{x}p_{x}p_{\alpha x}$ is the partition function
of the system $A$ with the border $\alpha$ fixed -- we are summing
over all sites $x$ in the interior of $A$, but the border $\alpha$
of $A$ has some fixed value.

We also have definitions of partition functions of {}``extended''
systems. For example, $Z_{\tilde{A}}(\beta)$ means the partition
function of the enlarged system $\tilde{A}$, which has all of $A$
(both $x$ and $\alpha)$ in its interior, and is bounded by $\beta$,
which is the fixed border configuration for this partition function.

As you can see, the final sum is only over the spins in the boundary
region, $\alpha$ and $\beta$. So, if we have an efficient way to
calculate the partition functions in the formula -- with given fixed
boundary conditions -- then we have reduced the sum over all the possible
states to a sum over just the possible states of the borders.

In section \ref{sec:calc-part-func}, we will indeed see how partition
functions can be efficiently approximated for arbitrary classical
spin models in two dimensions. Chapter \ref{cha:Ising-lattice} will
then deal with the classical Ising model, where there are even ways
to calculate the partition function exactly, which we will also discuss
in some detail there. Most of the results even for the Ising model
will however be based on the approach of section \ref{sec:calc-part-func},
since it has some additional advantages. It is also the method used
in chapter \ref{chap:More-lattice-spin}, which deals with some models
that are not exactly solvable.

\section{\label{sec:MC-sampling}Monte Carlo sampling}

What remains to decide -- given a method to evaluate the partial partition
functions -- is how we carry out the remaining sum over the border
states. For small enough systems we can certainly just carry out the
sum exactly. For larger systems we will indeed use Monte Carlo sampling
for this problem, which is now much more tractable than the original
one. In particular, we no longer need to sample states of the whole
system, but only of the border region consisting of $\alpha$ and
$\beta$.

Let us spend a few words on why equation \eqref{eq:I} is in fact
in a form directly suitable for Monte Carlo: Monte Carlo sampling
generally works for an expression of the form

\begin{equation}
\frac{\sum_{i}p_{i}f_{i}}{\sum_{i}p_{i}}\label{eq:mc-sum-form}
\end{equation}
where the $p_{i}$ form a probability distribution in the sense that
they are real and nonnegative, and that we only get the value of $\sum_{i}p_{i}f_{i}$
up to the normalization of the $p_{i}$, which we therefore need to
have access to. Both of these conditions are fulfilled by the expression
\eqref{eq:I}, where the obvious choice for the probability part is
$p_{i}=p_{\alpha\beta xy}=p_{\alpha}p_{\alpha\beta}p_{\beta}Z_{A}(\alpha)Z_{B}(\beta$).

What about the error we make by using Monte Carlo sampling rather
than the exact sum? It is of statistical nature, and therefore proportional
to $1/\sqrt{n}$ where $n$ is the number of samples. It is also proportional
to the standard deviation of the $f_{i}$. But it also makes a big
difference if we manage to have truly independent samples or if the
samples are correlated. In fact, our samples will almost certainly
be correlated; we will be working with what probably most people mean
when they say Monte Carlo sampling, and that is Markov Chain Monte
Carlo. Even more specifically, we will use the Metropolis algorithm,
where, in essence, we start with a random sample, and new samples
$i'$ are proposed as updates of the current one, and accepted with
a probability $p_{i'}/p_{i}$ (or with probability 1 if $p_{i'}$
is greater than $p_{i}$). In the next section we will see a way to
calculate partition functions that allows for a particularly simple
form of updates, i.e., updates that can be done very efficiently,
with little computational effort.

There are a few more subtleties about the Monte Carlo algorithm, in
particular the fact that you do not want to start with a completely
randomly picked initial state, but only start the actual averaging
after a certain amount of equilibration. Also, in order to estimate
the error accurately for the case of correlated samples, it is not
enough to just calculate the standard deviation of the samples, but
one needs to take into account their correlation as well. We will
follow the common approach and do it by a coarse-graining procedure:
one combines neighbouring samples into bins and examines how the standard
deviation increases as a function of the bin size. Once it is converged,
one has found the effective standard deviation $s$ of uncorrelated
samples, which one can use to estimate the error made in the Monte
Carlo approximation.

\section{\label{sec:calc-part-func}Calculating partition functions}

\begin{figure}
\begin{centering}
\includegraphics[width=0.45\textwidth]{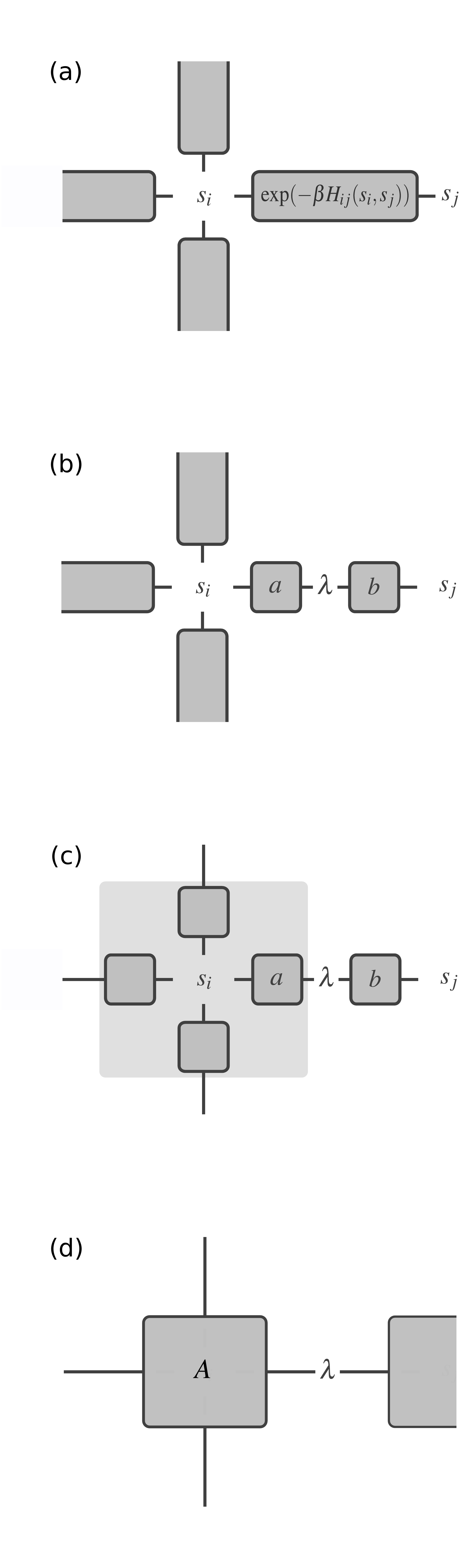}
\par\end{centering}

\caption{\label{fig:tensorconstruction}From partition function (a) to tensor
network contraction (d).}
\end{figure}

Now let us think about how to calculate partition functions for classical
spin systems on a two-dimensional lattice. We follow ideas presented
in \cite{verstraete2006criticality,arealaws}. Let us assume the Hamiltonian
has the form
\[
H=\sum_{\langle ij\rangle}H_{ij}(s_{i},s_{j})
\]
where $s_{i}$ and $s_{j}$ are the (classical) spins and $\langle ij\rangle$
defines interacting sites, e.g. nearest neighbours. The partition
function then is

\begin{equation}
Z=\sum_{\{s_{k}\}}\exp(-\beta_{\mathrm{th}}\sum_{\left\langle ij\right\rangle }H_{ij}(s_{i},s_{j}))=\sum_{\{s_{k}\}}\prod_{\left\langle ij\right\rangle }\exp(-\beta_{\mathrm{th}}H_{ij}(s_{i},s_{j})).\label{eq:spin-partitionfunction}
\end{equation}
with $\beta_{th}$ the {}``thermal'' $\beta=1/k_{B}T$ as opposed
to the $\beta$ used to label boundary configurations. Let us rewrite
the contribution of each bond $\left\langle ij\right\rangle $ as

\[
\exp(-\beta_{\mathrm{th}}H_{ij}(s_{i},s_{j}))=\sum_{\lambda}a(\lambda,s_{i})b(\lambda,s_{j})
\]
where we introduced a new variable $\lambda$. In order for the above
identity to be satisfied, $\lambda$ needs to take at most as many
different values as any of the classical spins involved. You can see
that by taking e.g. a singular value decomposition $USV^{\dagger}$
of the left hand side (understood as a matrix with row and column
indices $s_{i}$ and $s_{j}$). Then simply absorb the singular values
into either $U$ or $V^{\dagger}$ and you have exactly the desired
form. This rewriting of a single bond is graphically illustrated in
figure \ref{fig:tensorconstruction} (a) and (b).

We can now carry out the sums over the $s_{k}$ in \eqref{eq:spin-partitionfunction},
and let sums over the $\lambda$ remain. Doing this, we create at
each site a tensor which is constructed as the sum (over $s_{k})$
of direct products of $a$s and $b$s. The number of terms in the
direct product (and therefore the order of the resulting tensor) is
given by the number of bonds meeting at this site. This is illustrated
in figure \ref{fig:tensorconstruction} (c) and (d).

\begin{figure}
\begin{centering}
\includegraphics[width=0.5\textwidth]{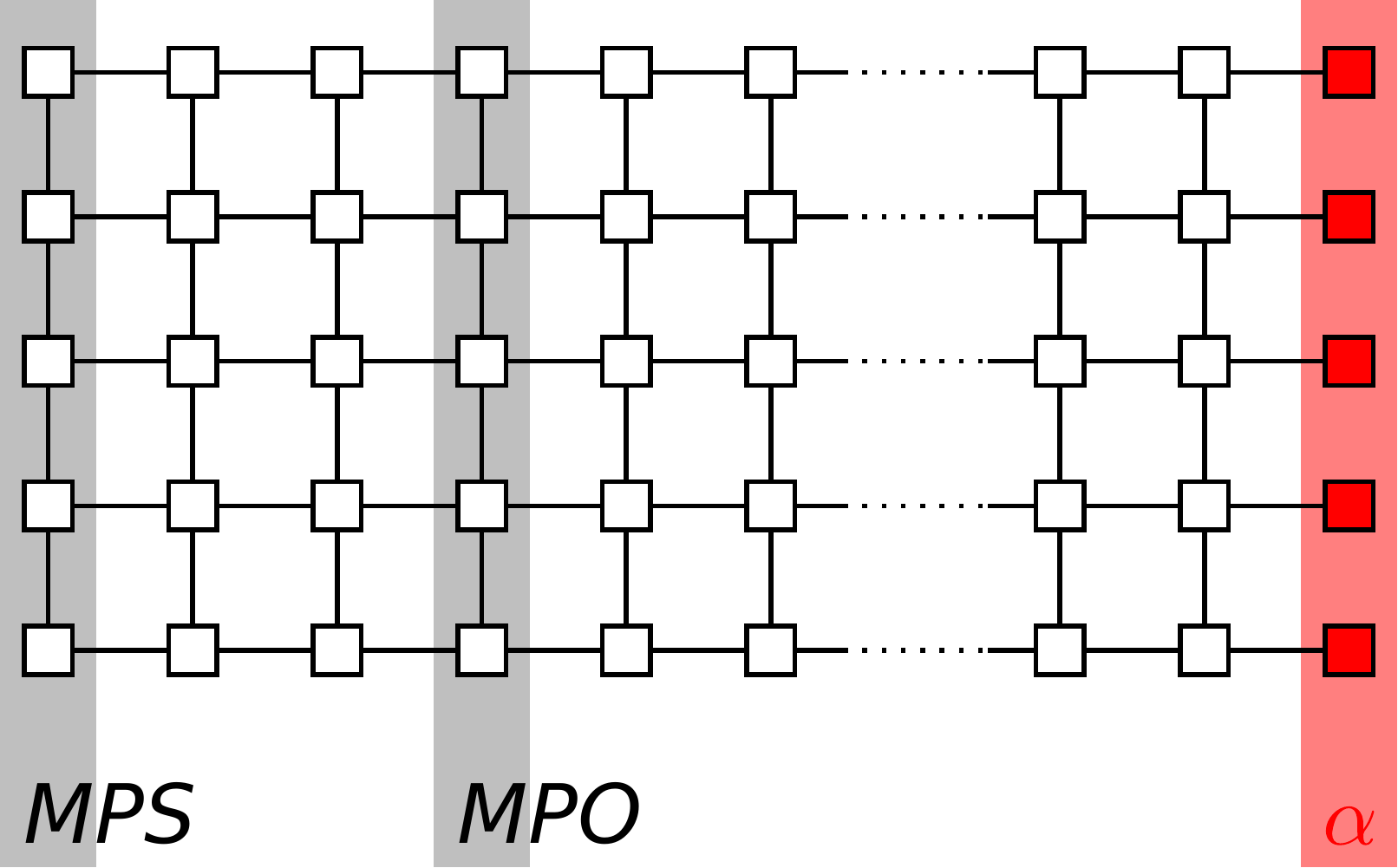}
\par\end{centering}

\caption{\label{fig:tensornetwork}Identifying MPSs and MPOs in the resulting
tensor network. Each square in this picture represents a tensor $A$
as constructed in figure \ref{fig:tensorconstruction}; each line
represents a tensor contraction (generalized matrix multiplication).}
\end{figure}

What remains to be done for the calculation of the partition function
is the summation over the $\lambda$-indices, which corresponds to
the contraction of the resulting tensor network. An example of such
a network is shown in figure \ref{fig:tensornetwork}, corresponding
to the evaluation of the partial partition function $Z_{A}(\alpha)$
in the geometry discussed before. Fixing the state of the border $\alpha$
can be understood as a specific choice of (unconnected) tensors on
the boundary.

In this network, we can now understand each interior column as a \emph{transfer
matrix} which has a special form -- it is a Matrix Product Operator
(MPO) \cite{mpsreview,pirvu2010matrix}. For non-square lattices,
you would need to be a bit more careful, but the principle remains
applicable. The boundary (in figure \ref{fig:tensornetwork}, the
leftmost column) can be understood as a Matrix Product State (MPS).
The network can now be contracted by repeatedly applying MPOs to a
MPS, resulting in another MPS to which we apply the next MPO, and
so on. In principle, this is exactly the same as contracting the tensor
network in any other way (although it is already much more efficient
than if you were to write each transfer matrix in full form rather
than as an MPO). The required dimension of the MPS does however grow
exponentially.

The approximation will come from reducing the dimension of the MPS
in this scheme. After we apply another MPO, we approximate the MPS
by another one of lower bond dimension (typically chosen fixed in
advance), thereby preventing the exponential growth. Of course, we
lose information this way, and we will in fact examine the quality
of the approximation in section \ref{sub:mps-error-analysis}.

One case is of particular importance, namely the case of a translationally
invariant system (more precisely, it just needs to be translationally
invariant along the horizontal direction in figure \ref{fig:tensornetwork}).
In that case, all the MPOs are identical. We can now consider the
case of a system that is very large in the horizontal direction. Now,
what we do is repeatedly apply the same transfer matrix $T$ (given
as an MPO) to a vector. It is clear that after many applications this
vector will in fact be proportional to the eigenvector of the MPO
that corresponds to its eigenvalue with largest magnitude (assuming
that it is unique) \cite{kramerswannier1941-1,kramerswannier1941-2,onsager1944,exactlysolvedbaxter1982}.
Now, there exist specific algorithms \cite{nishino1,nishino2,mpsreview,pbcmps}
that can very efficiently determine an approximation for this eigenvector,
in the form of an MPS with a given bond dimension. The algorithms
can also make use of translational invariance in what is the vertical
direction in figure \ref{fig:tensornetwork}. 

Let us indeed assume the number of columns $N_{\mathrm{bulk}}$ in
the bulk is so large that we can make this approximation. We can label
the largest eigenvalue $\Lambda$, the corresponding eigenvector $|\Lambda\rangle$,
and now the whole partial partition function discussed can be approximated
as $Z_{A}(\alpha)=\Lambda^{N_{\mathrm{bulk}}}\langle\Lambda|\alpha\rangle$.
Note how each application of the transfer matrix MPO contributes a
factor of $\Lambda$. What is really great about this formula, and
remains true in an analogous way for the case of a network without
translational symmetry, is that for a different border configuration
$\alpha$ we need only calculate the new overlap with $|\Lambda\rangle,$
rather than do the whole (much more complicated) tensor network contraction
again!

It should be obvious that the other partial partition functions in
equation \eqref{eq:I} can be approximated in a similar way as $Z_{A}(\alpha)$.
Let us now work with a rectangular system of $N$ columns total, in
the limit $N\to\infty$, and divide it as in figure \ref{fig:systems-borders},
exactly in the middle. If we assume periodic boundary conditions in
vertical direction, this is actually a very long cylinder, cut in
the middle. The partition functions are then $Z_{A}(\alpha)=\Lambda^{N/2-1}\langle\Lambda|\alpha\rangle$,
$Z_{B}(\beta)=\Lambda^{N/2-1}\langle\Lambda|\beta\rangle$, $Z_{\tilde{A}}(\beta)=\Lambda^{N/2}\langle\Lambda|\beta\rangle$,
and $Z_{\tilde{B}}(\alpha)=\Lambda^{N/2}\langle\Lambda|\alpha\rangle$.
The partition function of the whole system, which is needed for normalization
purposes, is $Z_{AB}=\Lambda^{N}.$ Let us further introduce the shorthand
$L(\alpha,\beta)=\langle\Lambda|\alpha\rangle\langle\Lambda|\beta\rangle$.
Let us also introduce \emph{unnormalized} Boltzmann weights, $q_{ab}=\exp(-\beta_{\mathrm{th}}E_{ab})$
where $E_{ab}$ is the energy of configuration $(a,b)$. Let us now
work with these $q_{ab}$ rather than the normalized probabilities
$p_{ab}=q_{ab}/\sum_{ab}q_{ab}=q_{ab}/Z_{AB}$. The $q_{ab}$ can
be factored exactly like the $p_{ab}$ in the previous section, and
we then get the following formula for the mutual information 
\begin{equation}
I(A,B)=\frac{1}{\Lambda^{2}}\sum_{\alpha\beta}q_{\alpha}q_{\alpha\beta}q_{\beta}L(\alpha,\beta)\log\frac{q_{\alpha\beta}}{L(\alpha,\beta)}\label{eq:I_strip}
\end{equation}
which is independent of $N$, so that the $N\to\infty$ limit is unproblematic.

\subsection{\label{sub:strip-simplifications}Simplifications}

Can we simplify formula \eqref{eq:I_strip} even further? It would
be a lot simpler if the logarithmic term were not there, because then
we would just have

\begin{equation}
\sum_{\alpha\beta}\langle\Lambda|\alpha\rangle q_{\alpha}q_{\alpha\beta}q_{\beta}\langle\beta|\Lambda\rangle=\langle\Lambda|TT|\Lambda\rangle\label{eq:simplesumTT}
\end{equation}
with $T$ the Ising transfer matrix, and because that can be written
as a MPO, this expression would be easy to calculate as the contraction
of a tensor network (and it is actually just the total partition function).

Of course we cannot simply drop the logarithmic term; but we can separate
the logarithmic term into three parts, $\log\left(q_{\alpha\beta}/\left(\langle\Lambda|\alpha\rangle\langle\Lambda|\beta\rangle\right)\right)=\log q_{\alpha\beta}-\log\langle\Lambda|\alpha\rangle-\log\langle\Lambda|\beta\rangle$.

Now, $q_{ab}$ is actually an exponential, of the sum over all the
bonds between two columns: $q_{ab}=\exp(-\beta_{\mathrm{th}}\sum_{i}H(\alpha_{i},\beta_{i}))$,
where the sum goes over the rows $i$ and $\alpha_{i}$ and $\beta_{i}$
are the components of the configurations $\alpha$ and $\beta,$ respectively.

The logarithm of the exponential is of course just the exponent $-\beta_{\mathrm{th}}\sum_{i}H(\alpha_{i},\beta_{i})$,
so we have to calculate
\begin{equation}
-\beta_{\mathrm{th}}\sum_{\alpha\beta}\langle\Lambda|\alpha\rangle q_{\alpha}q_{\alpha\beta}q_{\beta}\langle\beta|\Lambda\rangle\sum_{i}H(\alpha_{i},\beta_{i}).\label{eq:simplesum1}
\end{equation}
We can now consider the terms separately for each $i$, and notice
that they are all local and can therefore efficiently be calculated
as a contraction with suitably modified MPOs.

What about the remaining parts, with $\log\langle\Lambda|\alpha\rangle$
and $\log\langle\Lambda|\beta\rangle$? Let us look only at the one
with $\log\langle\Lambda|\alpha\rangle$; the one with $\log\langle\Lambda|\beta\rangle$
is exactly symmetric. We have
\[
\begin{aligned} & -\beta_{\mathrm{th}}\sum_{\alpha\beta}q_{\alpha}q_{\alpha\beta}q_{\beta}\langle\Lambda|\alpha\rangle\langle\Lambda|\beta\rangle\log\langle\Lambda|\alpha\rangle\\
= & -\beta_{\mathrm{th}}\sum_{\alpha}q_{\alpha}\underbrace{\sum_{\beta}q_{\alpha\beta}q_{\beta}\langle\Lambda|\beta\rangle}_{\Lambda\langle\Lambda|\alpha\rangle}\langle\Lambda|\alpha\rangle\log\langle\Lambda|\alpha\rangle\\
= & -\beta_{\mathrm{th}}\Lambda\sum_{\alpha}q_{\alpha}\langle\Lambda|\alpha\rangle^{2}\log\langle\Lambda|\alpha\rangle\\
= & -\frac{1}{2}\beta_{\mathrm{th}}\Lambda\sum_{\alpha}q_{\alpha}\langle\Lambda|\alpha\rangle^{2}\log\langle\Lambda|\alpha\rangle^{2}\\
= & -\frac{1}{2}\beta_{\mathrm{th}}\Lambda\sum_{\alpha}q_{\alpha}\langle\Lambda|\alpha\rangle^{2}\log q_{\alpha}\langle\Lambda|\alpha\rangle^{2}\\
 & +\frac{1}{2}\beta_{\mathrm{th}}\Lambda\sum_{\alpha}q_{\alpha}\langle\Lambda|\alpha\rangle^{2}\log q_{\alpha}
\end{aligned}
\]
where we introduced a $q_{\alpha}/q_{\alpha}$ unit term in the logarithm
and used that to separate it into two parts, one that has the form
of an entropy, 

\begin{equation}
-\sum_{\alpha}\pi_{\alpha}\log\pi_{\alpha}\quad,\label{eq:ising-ent}
\end{equation}
and an additional term again containing the logarithm of an exponential,
which becomes a sum of local terms that can be handled easily.

So what remains is just the term \eqref{eq:ising-ent} that describes
the entropy $-\sum_{\alpha}\pi_{\alpha}\log\pi_{\alpha}$ of the marginal
distribution $\pi_{\alpha}=q_{\alpha}\langle\Lambda|\alpha\rangle^{2}$,
normalized by $\sum_{\alpha}\pi_{\alpha}=\Lambda$. If we have the
eigenvector $|\Lambda\rangle$, we can calculate this entropy: Just
as discussed in section \ref{sec:MC-sampling}, for small numbers
of rows the sum can be carried out exactly, and for large numbers
of rows it can be approximated by Monte Carlo sampling.

In particular, note that for sampling the coefficients $\langle\Lambda|\alpha\rangle$
it is sufficient to know the eigenvector as a MPS, without ever needing
it in exponentially large full form (the $q_{\alpha}$ are straightforward
to calculate anyway). As a consequence, we can do the sampling particularly
efficiently by using Monte Carlo updates that sweep back and forth
along the MPS and store intermediate contraction results. It might
be worth noting that similar Monte Carlo sampling procedures for matrix
product states have also been used in a rather different (variational)
context \cite{sandvik2007,schuch2008}.

\chapter{\label{cha:Ising-lattice}The classical Ising model}

In this chapter, we consider a particular instance of the classical
spin models discussed in the previous one. The model under consideration
is the famous Ising model \cite{ising1925} in two spatial dimensions.
It has been solved exactly by Onsager \cite{onsager1944}. In this
case, that implies that there exist algorithms that allow an efficient
evaluation of its partition function, in exact form rather than just
as an approximated tensor network contraction. However, we will start
by reviewing how the general tensor network approach specializes to
the Ising model in section \ref{sec:ising-tnc}. In sections \ref{sec:ising-matchgates}
and \ref{sec:ising-fkt} we will then see two algorithms that make
use of the special features of the Ising model. At the end of the
chapter, we will discuss the results of the calculations, and see
if they match our ideas about mutual information.

Let us start with the model definition: The classical Ising Hamiltonian
is
\begin{equation}
H=-\sum_{\langle ij\rangle}J_{ij}s_{i}s_{j}\label{eq:isinghamiltonian}
\end{equation}
where the spins have possible values $+1$ or $-1$, and interactions
are between nearest neighbours. We will generally consider the homogenous
case $J_{ij}=J$, and restrict to the ferromagnetic $J>0$. On bipartite
lattices such as we will study, the antiferromagnetic case can directly
be obtained by simply flipping all the spins of one of the sublattices.

\section{\label{sec:ising-tnc}Tensor network contraction}

Let us go through the steps outlined in section \ref{sec:calc-part-func}
again, and see what they look like for the particular example of the
Ising model.

\subsection{Tensors}

Introducing the shorthand notation $K_{ij}=\beta_{\mathrm{th}}J_{ij}=J_{ij}/(k_{B}T)$,
the partition function is

\begin{equation}
Z=\sum_{\{s_{k}\}}\exp(\sum_{\left\langle ij\right\rangle }K_{ij}s_{i}s_{j})=\sum_{\{s_{k}\}}\prod_{\left\langle ij\right\rangle }\exp(K_{ij}s_{i}s_{j})\label{eq:partitionfunction}
\end{equation}
and we can rewrite the contribution of each bond $\left\langle ij\right\rangle $
as

\begin{equation}
\exp(K_{ij}s_{i}s_{j})=\sum_{\lambda}a(\lambda,s_{i})b(\lambda,s_{j})\label{eq:id-ising}
\end{equation}
where the new index $\lambda$ only takes two values here, say $0$
and $1$. Equation \eqref{eq:id-ising} can be satisfied by 
\[
\begin{aligned}a(\lambda=0,s=+1) & =\sqrt{\cosh K_{ij}}\\
a(\lambda=0,s=-1) & =\sqrt{\cosh K_{ij}}\\
a(\lambda=1,s=+1) & =\sqrt{-\sinh K_{ij}}\\
a(\lambda=1,s=-1) & =-\sqrt{-\sinh K_{ij}}
\end{aligned}
\]
and $b(\lambda,s)=a(\lambda,s)$. Unsymmetric choices are also possible,
where for example the $K$-dependence is only in the $b$s; it is
even possible to completely take out the $K$-dependence, and write
\[
\exp(K_{ij}s_{i}s_{j})=\sum_{\lambda}g(\lambda,s_{i})g(\lambda,s_{j})h(\lambda)
\]
with

\[
\begin{aligned}g(\lambda=0,s=+1) & =1\\
g(\lambda=0,s=-1) & =1\\
g(\lambda=1,s=+1) & =1\\
g(\lambda=1,s=-1) & =-1
\end{aligned}
\]
and

\[
h(\lambda=0)=\cosh K_{ij}\,,\; h(\lambda=1)=-\sinh K_{ij}\;.
\]
We can now carry out the sums over the $s_{k}$ in \eqref{eq:partitionfunction},
and let the newly introduced sums over the $\lambda$ remain. Doing
this, we create at each site $k$ a tensor, which is constructed as
the sum (over $s_{k})$ of direct products of $a$s and $b$s (or
$g$s and $h$s). The number of terms in the direct product (and therefore
the order of the resulting tensor) is given by the number of bonds
meeting at this site. What remains to be done for the calculation
of the partition function is the summation over the $\lambda$-indices,
which corresponds to the contraction of the resulting tensor network.

Let us write down the tensors, using $g$ and $h$. They have $4$
indices, say $\lambda_{1}$, $\lambda_{2}$, $\lambda_{3}$, and $\lambda_{4}$.
We will later sometimes want to understand the tensor as a gate acting
on two incoming indices, let these be $\lambda_{1}$ and $\lambda_{2}$.
The outgoing indices are then $\lambda_{3}$ and $\lambda_{4}$. Putting
the $K$-dependencies in the outgoing indices only, we get
\begin{equation}
T=\sum_{s_{k}}g(\lambda_{1},s_{k})g(\lambda_{2},s_{k})g(\lambda_{3},s_{k})g(\lambda_{4},s_{k})h(\lambda_{3})h(\lambda_{4})\label{eq:isingT}
\end{equation}
and you see that the $h$s (that contain the $K$-dependencies) can
be completely separated from the $g$s. We will pick up from that
fact again in section \ref{sec:ising-matchgates}.

\subsection{\label{sub:mps-error-analysis}Error analysis}

Having found the form of the tensors, we can now follow the calculation
outlined in section \ref{sec:calc-part-func}. The results for the
Ising model will be presented at the end of this chapter. For now,
let us only analyze the errors that come from approximating the largest
eigenvalue and corresponding eigenvector by a MPS with limited bond
dimension.

\begin{figure}
\begin{centering}
\includegraphics[width=1\textwidth]{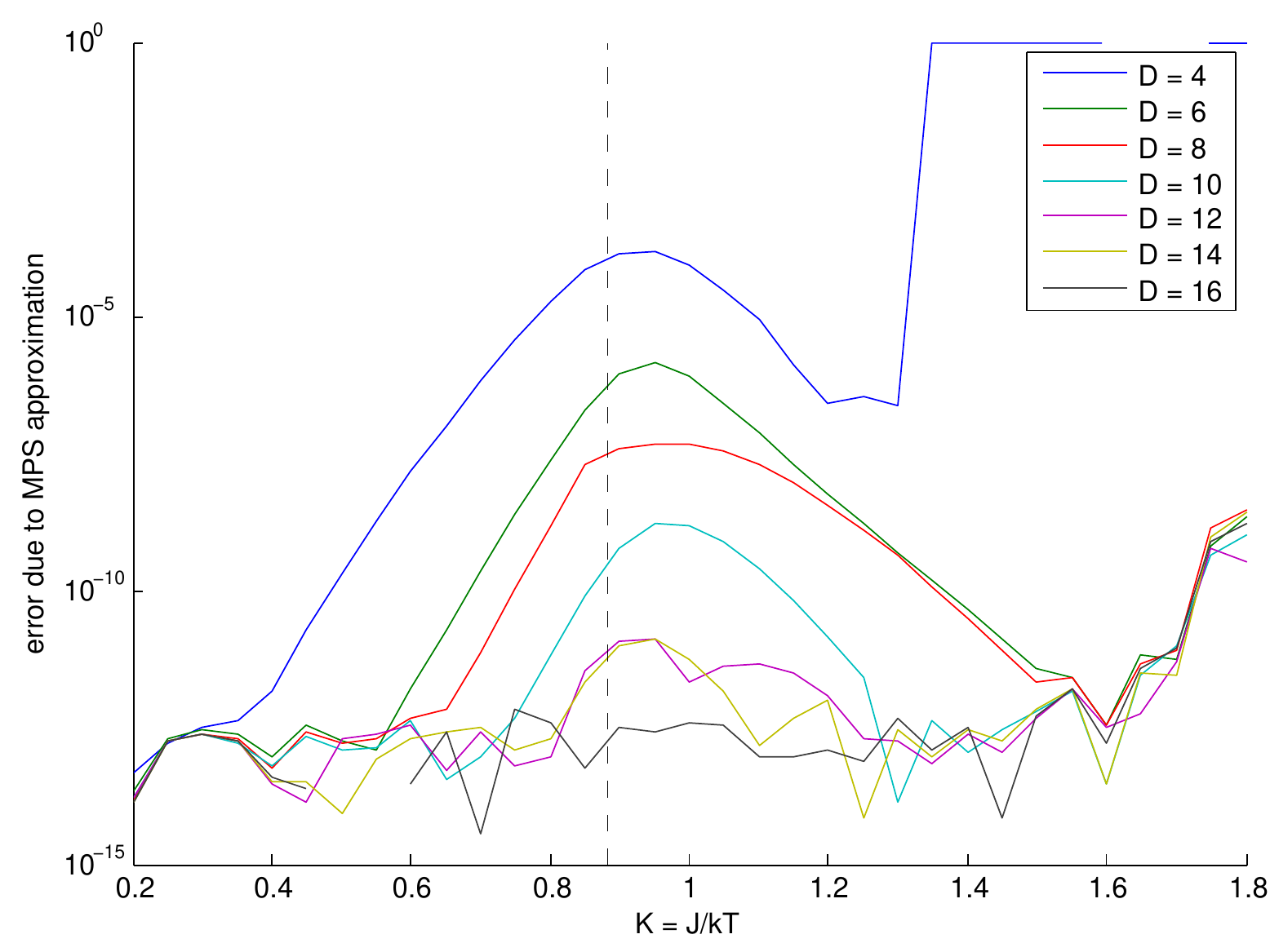}
\par\end{centering}

\caption{\label{fig:mps-error}Error made in the calculation of the mutual
information by using a MPS with bond dimension $D$ rather than the
exact eigenvector. A system with 16 rows and open boundary conditions
was studied. The dashed black line marks the location of the phase
transition in the thermodynamic limit.}
\end{figure}

Since the focus of this work is not on Matrix Product States and their
properties, let us only pick out one specific setting to discuss the
general features. Figure \ref{fig:mps-error} shows the error made
by using a MPS of bond dimension $D$ rather than the exact eigenvector
of the transfer matrix, for a system of 16 rows, where it is still
possible to calculate that eigenvector exactly. To avoid any other
error sources, it is of course necessary that we also carry out the
summation of \eqref{eq:I_strip} exactly. This is indeed possible
after the simplifications described in section \ref{sub:strip-simplifications}.

We notice two things: First of all, the errors show a very systematic
behaviour, and have a pronounced maximum (note the logarithmic scale).
The maximum occurs where the system becomes critical, and this is
exactly the behaviour we would expect from Matrix Product States.

Second, the quality of the MPS approximation depends strongly on the
bond dimension. For the system studied, a MPS with bond dimension
of $D=16$ is very close to the exact eigenvector, and for most temperatures
the only difference is numerical noise. However, almost any choice
of $D$ results in an error that will be lower than the statistical
error we will introduce by Monte Carlo sampling. Since a smaller $D$
results in faster Monte Carlo updates, we will choose a relatively
modest $D$, in order to be able to do as many Monte Carlo steps as
possible and minimize the total error.

Even if it will be the Monte Carlo sampling that decides our errors,
let us dedicate the next two sections to methods for the exact calculation
of Ising partition functions.

\section{\label{sec:ising-matchgates}Matchgates}

Let us remember the tensors \eqref{eq:isingT} from the preceding
section. We can rewrite them in matrix form, with $(\lambda_{1},\lambda_{2})$
being the column indices (corresponding to {}``incoming'' indices)
and $(\lambda_{3},\lambda_{4})$ being the row ({}``outgoing'')
indices. We have $T=G\cdot H_{3}(K_{3})\cdot H_{4}(K_{4})$ with
\[
G=\left(\begin{array}{cccc}
1 &  &  & 1\\
 & 1 & 1\\
 & 1 & 1\\
1 &  &  & 1
\end{array}\right)\,,\: H(K)=\left(\begin{array}{cc}
\cosh K\\
 & -\sinh K
\end{array}\right)\;,
\]
\[
H_{3}(K_{3})=H(K_{3})\otimes I\,,\: H_{4}(K_{4})=I\otimes H(K_{4})\;.
\]
We will now interpret these tensors/matrices as {}``gates'' acting
on some quantum state. The gates turn out to have a special form,
they are \emph{matchgates }\cite{valiant2002quantum}. As the name
{}``gate'' suggests, the context in which this has mostly been considered
are quantum circuits, the gates in which are usually unitary. This
also provides the connection to what has become known as fermionic
linear optics \cite{knill2001fermionic,terhal2002classical,bravyi2002fermionic,bravyi2005lagrangian,jozsa2008matchgates,van2009quantum,jozsa2010matchgate}.
We will work along similar lines, and map the tensors/matrices/gates
to fermionic operators. However, the actual mapping is significantly
different from what has been done in the literature -- in particular,
our operators are non-unitary.

\subsection{\label{sub:fermionic-representation}Fermionic representation}

So, let us now try to express the gates in terms of fermionic creation
and annihilation operators. Without loss of generality, let us restrict
ourselves to just two modes, which we can label $1$ and $2$, with
occupation numbers $\lambda_{1}$ and $\lambda_{2}$. The state space
is Fock space, states are given as 
\[
|\psi\rangle=\sum_{\lambda_{1},\lambda_{2}}\psi_{\lambda_{1},\lambda_{2}}(a_{1}^{\dagger})^{\lambda_{1}}(a_{2}^{\dagger})^{\lambda_{2}}|\Omega\rangle.
\]
To determine the components, say $\psi_{00}$ or $\psi_{01}$, we
can project

\begin{eqnarray*}
\psi_{00} & = & \langle\Omega|a_{1}a_{1}^{\dagger}a_{2}a_{2}^{\dagger}|\psi\rangle\\
\psi_{01} & = & \langle\Omega|a_{1}a_{1}^{\dagger}a_{2}|\psi\rangle\\
 &  & \textrm{etc.}
\end{eqnarray*}
Let us start by looking at the entangling gate $G$. In terms of fermionic
operators, this can be written as the operator $\hat{O}=I+a_{1}a_{2}+\ldots=b_{1}^{\dagger}b_{1}$
in a transformed basis given by
\[
\begin{alignedat}{1}b_{1} & =(+a_{1}+a_{2}-a_{1}^{\dagger}+a_{2}^{\dagger})/\sqrt{2}\\
b_{1}^{\dagger} & =(-a_{1}+a_{2}+a_{1}^{\dagger}+a_{2}^{\dagger})/\sqrt{2}.
\end{alignedat}
\]
Going back to the old basis, the bond-strength gates are just single-mode
gates of the form
\begin{equation}
H(K)=\exp(-K)aa^{\dagger}+\sinh(K)\mathbb{I}\label{eq:g}
\end{equation}
where $a$ is either of $a_{1}$ or $a_{2}.$ A gate is required for
each bond, i.e. at a typical site in the bulk there are two gates:
one for the horizontal bond and one for the vertical bond.

The essential thing is that all of these gates are only quadratic
in the fermionic operators. In the next section we will see how this
allows us to efficiently describe the application of the gates using
only a polynomially-sized covariance matrix, rather than a full exponentially-sized
state vector.

How do we handle the boundary conditions needed for our partial partition
functions? In the end there turns out to be a pretty simple way to
handle them: what we do is fix the relative orientations of boundary
spins by using infinite bond strengths. That is, if we want to implement
a boundary configuration in which two neighbouring boundary spins
are pointing in the same direction, we connect them by an infinitely
strong ferromagnetic bond. Otherwise, we use an infinitely strong
antiferromagnetic bond. This fixes the boundary configuration up to
a possible inversion of all the spins at once. However, because of
the inversion symmetry of the whole problem, both these configurations
will result in identical values of the partition function. Therefore,
we simply do the calculation with the spins only fixed among each
other, and divide the result by a factor of two. The gates remain
in fact well-defined even for infinitely strong couplings, if we renormalize
them -- since these infinitely strong bonds are not relevant for the
partition function, this does not cause any problem.

If one prefers, one can even work without infinitely strong antiferromagnetic
bonds at all: Rather than fixing the relative orientations of the
boundary spins, we can achieve the same effect by fixing all boundary
spins to point in the same direction, which only requires infinitely
strong ferromagnetic couplings. We then simply invert the couplings
that connect some of the boundary spins with the spins in the interior.

\subsection{\label{sub:calc-part-matchgates}Calculating the partition function}

\begin{figure}
\begin{centering}
\includegraphics[width=0.5\columnwidth]{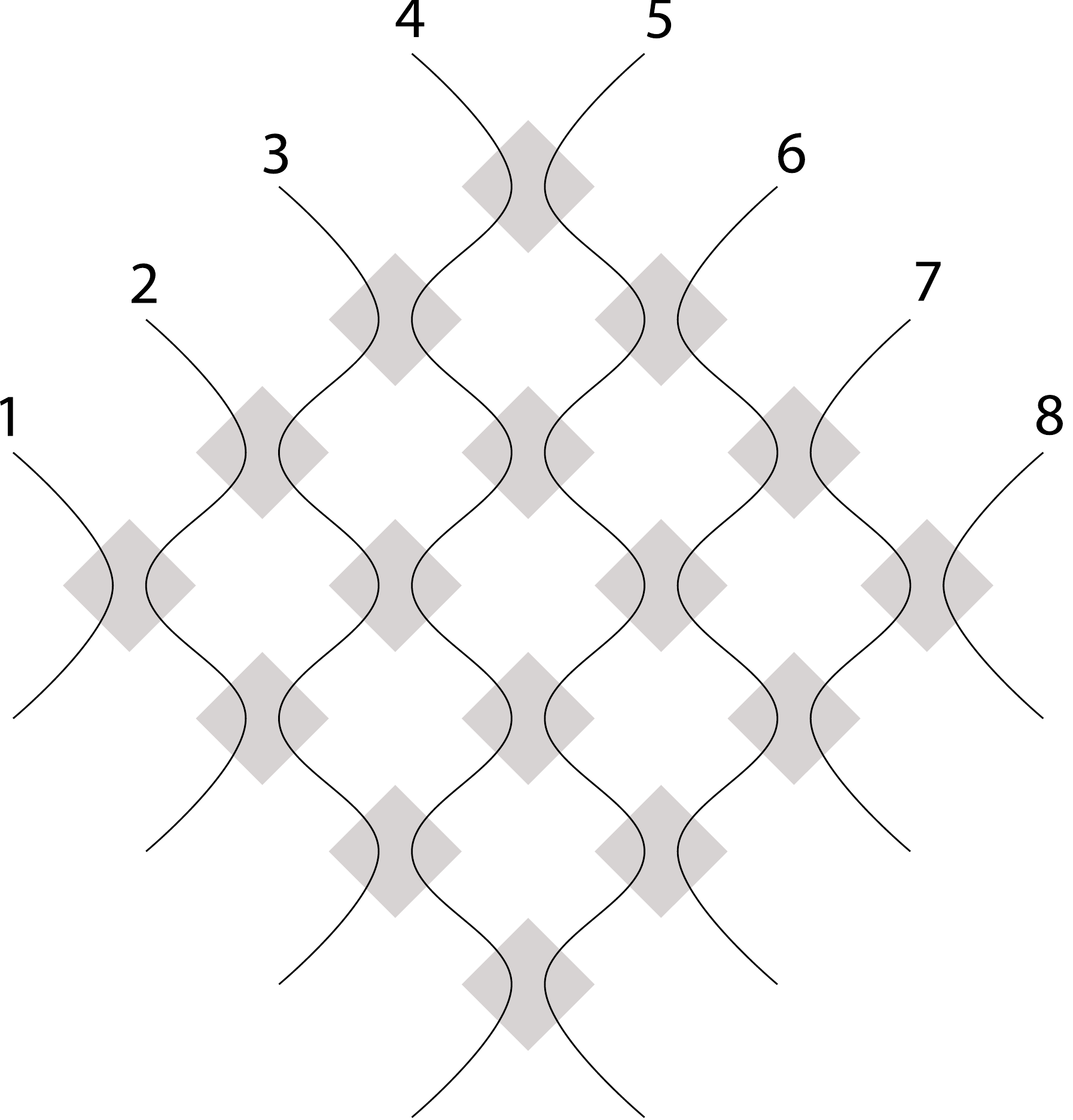}
\par\end{centering}

\caption{\label{fig:fermionic-modes}The 8 fermionic modes required for a $4\times4$
lattice}
\end{figure}
Now, we know how to express the tensors as fermionic operators, but
we still need to calculate the tensor network contraction to find
the partition function. So how do we do that? We start from a state
$|\Omega\rangle$ and begin applying the operators, and finally project
onto $|\Omega\rangle$ again. However, why can this be done efficiently?
Let us first see how we can identify modes such that the only thing
that acts on them are the quadratic operators found in the previous
section. Figure \ref{fig:fermionic-modes} shows how to do that in
a square lattice by having the modes run at an angle of 45 degrees
to the rows and columns of the lattice. The gates act on nearest neighbours
and there is no problem in keeping track of signs due to fermionic
anticommutation rules.

It is now really essential that the gates are only quadratic in the
fermionic operators, because that means we do not need to work with
a general fermionic state, but only with a fermionic Gaussian state.
You can find more details about this class of states e.g. in reference
\cite{bravyi2005lagrangian}. In particular, what we will do is the
following: Instead of working with an exponentially large state vector,
we work only with its \emph{covariance matrix}, which contains all
the information if we apply at most quadratic operators. This covariance
matrix can then be updated using Wick's theorem \cite{wick1950evaluation},
and we have all the pieces needed to calculate the partition function
efficiently, i.e. with polynomial rather than exponential complexity.
Let us see how these pieces indeed require only polynomial resources:

\subsubsection*{Covariance matrix}

The covariance matrix contains the expectation values for all quadratic
expressions of creation and annihilation operators of fermionic modes.
We can order the operators for example as $(c_{i})_{i=1,2,\ldots,2n}=(a_{1},a_{2},\ldots,a_{n},a_{1}^{\dagger},a_{2}^{\dagger},\ldots,a_{n}^{\dagger})$.
A mapping to Majorana fermions \cite{majorana1937teoria} is also
possible, but not necessary. The covariance matrix is then $C=(\langle c_{i}c_{j}\rangle)_{i,j=1,2,\ldots,2n}$
where $\langle\cdot\rangle=\langle\psi|\cdot|\psi\rangle$ is the
expectation value with respect to the current state $|\psi\rangle$
of the system. Applying a gate operator $\hat{O}$ means transforming
$|\psi\rangle$ into $\hat{O}|\psi\rangle$, or transforming the expectation
value as $\langle\hat{O}\cdot\hat{O}\rangle$.

\subsubsection*{Wick's Theorem}

So, what we have to do is update the covariance matrix by applying
some fermionic operators, thereby inducing a transformation of the
covariance matrix. To calculate the transformed expectation values
in the covariance matrix, we can use Wick's theorem

\[
\langle c_{u}c_{v}c{}_{w}\cdots\rangle=\mathrm{Pf}(\Gamma)\quad\mathrm{where}\quad\Gamma=(\langle c_{i}c_{j}\rangle)_{i,j=u,v,w,\ldots}
\]
i.e. what you have to calculate is the Pfaffian of the matrix restricted
to rows and columns $u,v,w,\ldots$ . In particular for the case of
just 4 fermionic operators (which turns out to be all that is actually
needed)

\[
\left\langle abcd\right\rangle =\left\langle ab\right\rangle \left\langle cd\right\rangle -\left\langle ac\right\rangle \left\langle bd\right\rangle +\left\langle ad\right\rangle \left\langle bc\right\rangle .
\]
What is that Pfaffian in general? We will give a proper definition
and discuss it in more detail in section \ref{sub:matching-as-pfaffian}.
This is already a hint that the method presented in the next section
does in fact make use of the same intrinsic properties of the Ising
model, even though the actual manifestation will look pretty different.

\section{\label{sec:ising-fkt}The FKT method }

In this section, we will revisit a long-established method used to
calculate Ising partition functions, which can be formulated without
any understanding of the (partial) partition functions as tensor network
contractions. While the method itself is old, we will see that it
can also be adapted to deal with the fixed boundary conditions needed
for our partial partition functions. Maybe even more importantly,
we will also see how it can be used in such a way that changing of
the boundary conditions requires only a small amount of new computation,
just like with the MPS method discussed in the previous chapter.

\subsection{Ising partition function as a dimer covering}

So how does this method calculate the partition functions? Let us
start by following the treatment of chapter V of \cite{mccoy1973tdim}.
For simplicity of exposition, let us work in the homogenous case,
but it should be obvious that the following can straightforwardly
be generalized to interactions of varying strengths, just as in all
the other methods we have discussed. As $s_{i}$, and consequently
also $s_{i}s_{j}$ can only take on the values $+1$ and $-1,$ it
holds that $\exp(Ks_{i}s_{j})=\cosh K+s_{i}s_{j}\sinh K=\cosh K(1+s_{i}s_{j}\tanh K)$
and consequently we can write the partition function as

\[
Z=\sum_{\{s_{k}\}}\exp(K\sum_{\left\langle ij\right\rangle }s_{i}s_{j})=(\cosh K)^{N_{b}}\sum_{\{s_{k}\}}\prod_{\left\langle ij\right\rangle }(1+s_{i}s_{j}\tanh K)
\]
where $N_{b}$ is the number of bonds in the lattice. Expanding the
product gives us (exponentially) many terms, which have some factors
of $1$ and some factors of the type $s_{i}s_{j}\tanh K$, and in
fact any $s_{k}$ can turn up at most in a power equal to the coordination
number of the lattice, so for a quadratic lattice up to $4$. More
interesting though is that terms that have an odd number of any $s_{k}$
will vanish once the sum over this $s_{k}$ is carried out. The only
remaining terms are those that contain even powers of every $s_{k}$
appearing in them, and those will pick up a factor of $2$ from each
sum over some $s_{k},$ so a total factor of $2^{N_{s}}$, where $N_{s}$
is the number of sites in our lattice.

We want to represent every such term by {}``marking'' the bonds
appearing in it in the lattice. The condition that we can only have
an even power of $s_{k}$ means we have a configuration where at each
site an even number of marked bonds meet. This leads to a configuration
consisting of one or more closed loops of connecting bonds, and we
get the product of as many $\tanh K$ as there are bonds in the loop(s).
The loops may intersect but cannot share segments. In the literature
you will sometimes also read that these allowed configurations are
called polygon configurations (if you draw the graph using straight
edges, closed loops are polygons).

So all in all we have
\[
Z=2^{N_{s}}(\cosh K)^{N_{b}}\sum_{\textrm{c}}\prod_{\langle ij\rangle\in c}(\tanh K)
\]
where we sum over all configurations $c$ fulfilling the above conditions,
with the suitable weight given as the product over the weight of all
the bonds in the configuration. You can indeed notice that the above
formula can be generalized to lattices where each bond has, in general,
a different strength $K_{ij}$:
\begin{equation}
Z=2^{N_{s}}\prod_{\langle ij\rangle}(\cosh K_{ij})\sum_{\textrm{l}}\prod_{\langle ij\rangle\in l}(\tanh K_{ij}).\label{eq:fktZ}
\end{equation}
What we now want to do is calculate this weighted sum over closed
loop configurations. Let us now follow the presentation of \cite{fisher1966dspim},
but with updated terminology. We will map each loop configuration
to a {}``perfect matching'', i.e. a set of edges such that each
vertex is part of exactly one edge in the set. This sort of matching
is exactly what gave the name to the matchgates that were mentioned
in the previous section.

\begin{figure}
\begin{centering}
\includegraphics[width=0.8\columnwidth]{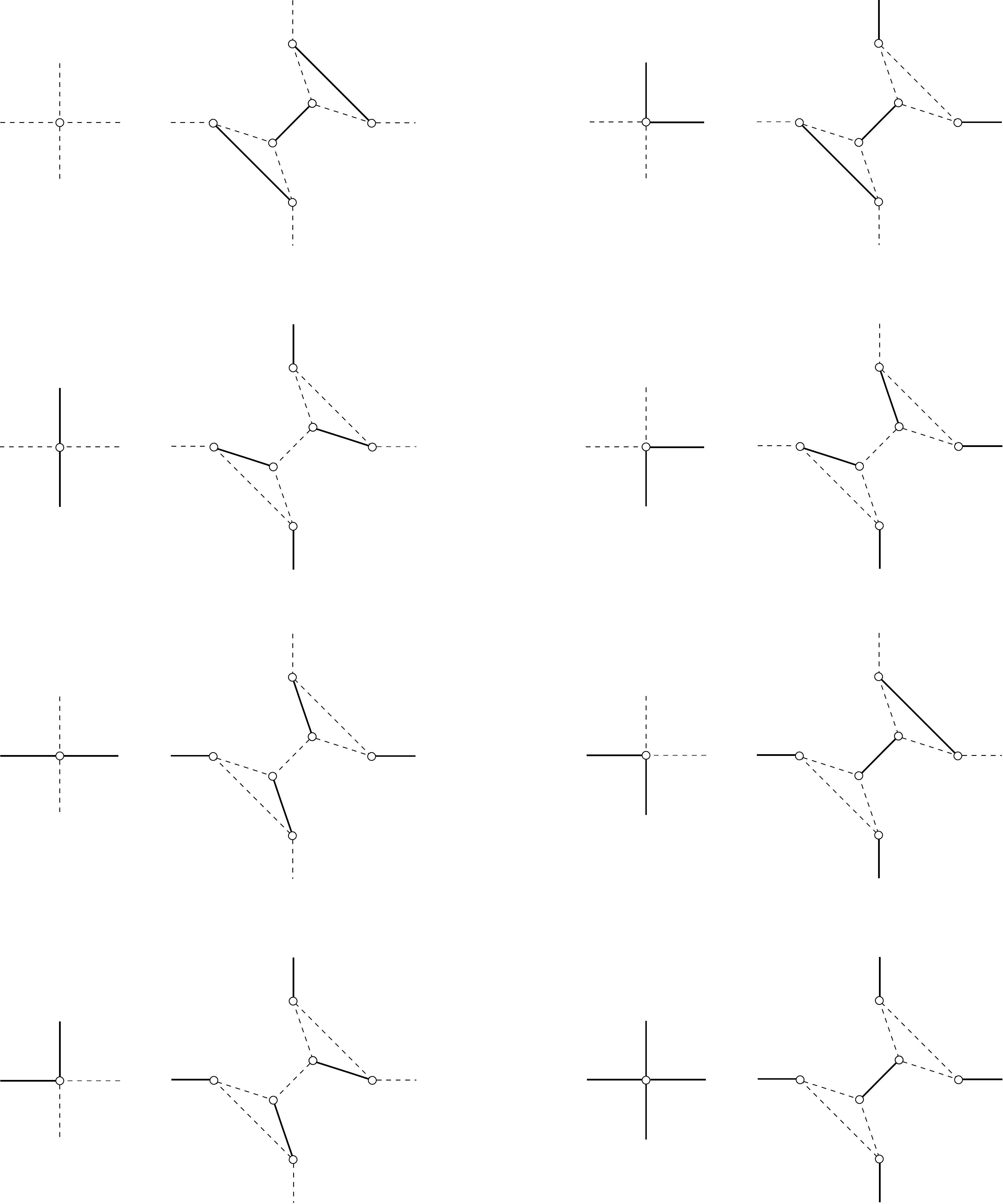}
\par\end{centering}

\caption{\label{fig:dimers}Even numbers of bonds correspond to perfect matchings
of subgraphs. Adapted from \cite{fisher1966dspim}.}
\end{figure}

The mapping to perfect matchings can be achieved by extending the
lattice; let us illustrate it for a lattice where each vertex has
a coordination number of four, i.e. the square lattice. The nicest
way is to replace each vertex by six new ones and map the possibilities
of an even number of bonds meeting there, of which there are 
\[
{4 \choose 0}+{4 \choose 2}+{4 \choose 4}=1+6+1=8,
\]
to the eight possible perfect matchings as shown in figure \ref{fig:dimers}.
In order not to change the weight of each contribution in the partition
function, it is necessary that all the extra bonds introduced must
have a weight of unity.

\subsection{\label{sub:matching-as-pfaffian}Dimer covering calculated as a Pfaffian}

\begin{figure}
\begin{centering}
\includegraphics[width=0.85\columnwidth]{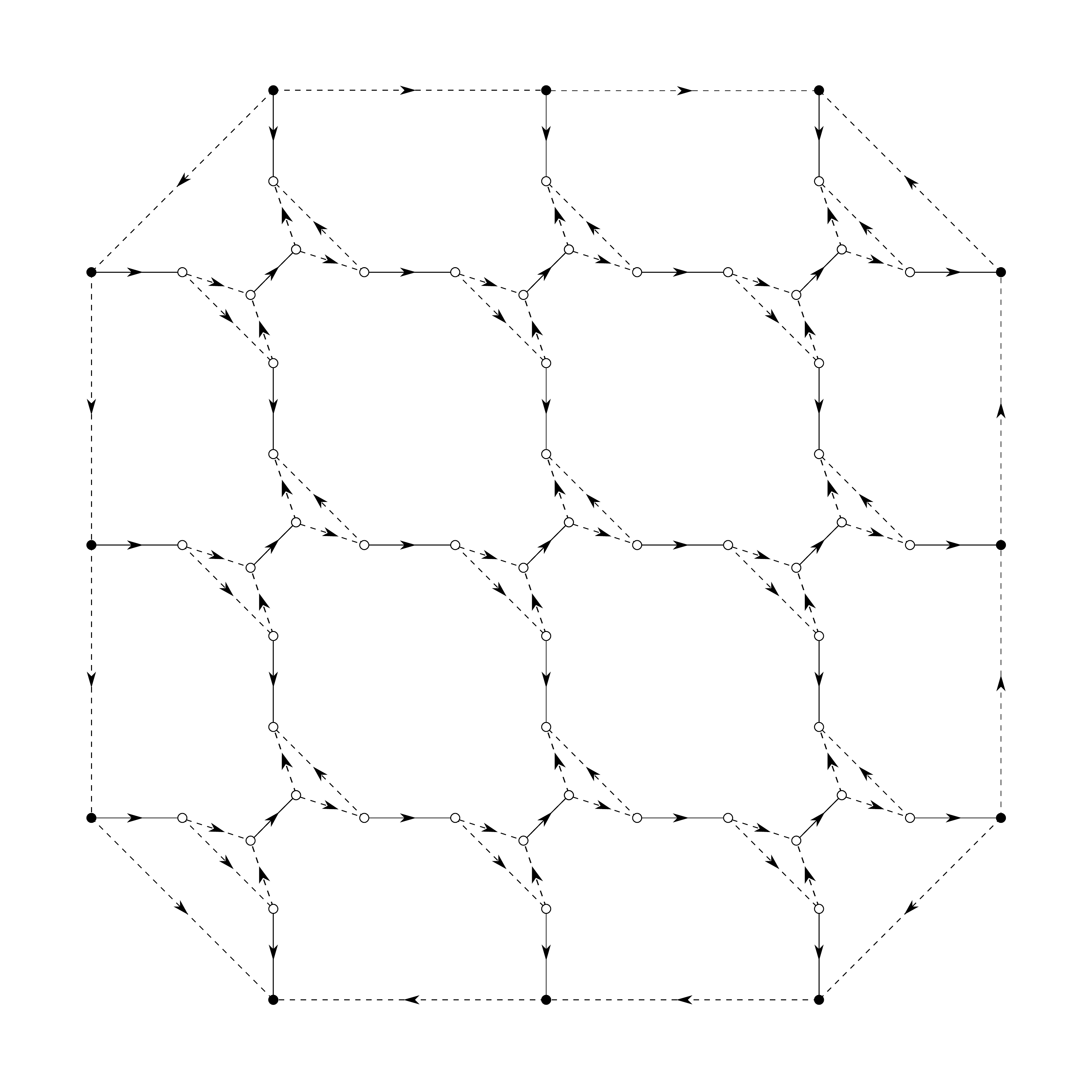}
\par\end{centering}

\caption{\label{fig:oriented-extended-lattice}Oriented extended lattice. Adapted
from \cite{fisher1966dspim}.}
\end{figure}

Using this construction, we can build up the extended lattice, as
is shown in figure \ref{fig:oriented-extended-lattice}. Now, the
essential ingredient is what is sometimes called the FKT theorem after
Fisher, Kasteleyn, and Temperley \cite{fisher1961statistical,kasteleyn1961statistics,temperley1961dimer}.
It tells us that we can calculate the partition function, i.e. the
sum over all the weighted dimer coverings, or perfect matchings, as
the Pfaffian of a suitably constructed matrix. Crucially however,
it must be possible to orient the edges of the lattice graph in a
specific way, such that every possible loop contains an odd number
of bonds oriented in clockwise direction, as indicated by the arrows
in figure \ref{fig:oriented-extended-lattice}. Such an orientation
can always be found for planar lattices, but in general not for arbitrary
lattices. This is what ultimately restricts this method in the same
way that the mapping to free fermions and matchgates is restricted.
Let us now take a closer look at this Pfaffian, which also occurred
in section \ref{sub:calc-part-matchgates}.

\subsubsection*{What is a Pfaffian?}

The Pfaffian of a $2n\times2n$ antisymmetric matrix $A=(a_{kl})_{1\leq k,l\leq2n}$
is conventionally defined as
\begin{equation}
\mathrm{Pf}(A)=\frac{1}{2^{n}\cdot n!}\sum_{\sigma\in S_{2n}}\mathrm{sgn}(\sigma)a_{\sigma_{1}\sigma_{2}}a_{\sigma_{3}\sigma_{4}}\cdots a_{\sigma_{2n-1}\sigma_{2n}}\label{eq:def-pfaffian}
\end{equation}
where the sum is over all possible permutations $\sigma$ of $(1,2,\ldots,2n).$
It can however also be written as a sum over all partitions of $(1,2,\ldots,2n)$
into $n$ pairs $\{(p_{1},\tilde{p}_{1}),(p_{2},\tilde{p}_{2}),\ldots,(p_{n},\tilde{p}_{n})\}=:p$.
Let us call the set of all such partitions $P$. We then get
\begin{equation}
\mathrm{Pf}(A)=\sum_{p\in P}\mathrm{sgn}(p)a_{p_{1}\tilde{p_{1}}}a_{p_{2}\tilde{p}_{2}}\cdots a_{p_{n}\tilde{p}_{n}}\label{eq:pfaffian}
\end{equation}
where $\mathrm{sgn}(p)$ is the same as $\mathrm{sgn}(\sigma$) for
one of the corresponding permutations $\sigma$, one of which would
be

\begin{center}
\begin{tabular}{|c||c|c|c|c|c|c|c|}
\hline 
$i$ & 1 & 2 & 3 & 4 & $\ldots$ & $2n-1$ & $2n$\tabularnewline
\hline 
$\sigma_{i}$ & $p_{1}$ & $\tilde{p}_{1}$ & $p_{2}$ & $\tilde{p}_{2}$ & $\ldots$ & $p_{n}$ & $\tilde{p}_{n}$\tabularnewline
\hline 
\end{tabular} .
\par\end{center}

There are $n!$ ways to order the pairs, and two possibilities to
order each of the $n$ pairs, so that the factor in the definition
\eqref{eq:def-pfaffian} is exactly canceled, and we might really
say \eqref{eq:pfaffian} is the more natural definition. It is pretty
obvious that equation \eqref{eq:pfaffian} provides the required connection
to the weighted perfect matchings, which are exactly such partitions
into pairs. Note that it is crucial that we can assign the correct
signs, which means this only works if it is possible to find a suitable
orientation of the lattice.

\subsubsection*{Relation to determinant}

We realize that the matrix of which we need to calculate the Pfaffian
is essentially just a weighted oriented adjacency matrix, and its
dimension is therefore just the number of sites. But the definitions
of the Pfaffian still seem to involve exponentially many terms, so
how can it be calculated efficiently? For this, we can use the identity
$\mathrm{Pf}(A)^{2}=\det(A)$, for we know efficient methods to evaluate
determinants. Efficient means in time polynomial rather than exponential
in the size of the matrix. To be precise, \eqref{eq:pfaffian} consists
of $(2n)!/(2^{n}\cdot n!)$ terms, just as the usual definition of
a determinant of a ($2n\times2n$) matrix does:

\[
\det(A)=\sum_{\sigma\in S_{2n}}\mathrm{sgn}(\sigma)a_{1,\sigma_{1}}a_{2,\sigma_{2}}\cdots a_{2n,\sigma_{2n}}
\]
but this can nevertheless be computed in time $O(n^{3})$, e.g. by
doing a LU decomposition (see below).

\subsubsection*{\label{sub:reduced-determinant}Efficient evaluation of the determinant}

So the evaluation of a determinant is already efficient, but for the
final calculation of the mutual information we actually have to repeatedly
calculate determinants of still pretty big matrices where only very
few entries (those related to the boundary conditions) change, while
the {}``bulk'' remains unchanged. In this situation, we can do better
still, as the determinant of a matrix can be split into factors

\begin{align}
\det\left(\begin{array}{cc}
A & B\\
C & D
\end{array}\right) & =\det\left[\left(\begin{array}{cc}
A & 0\\
C & \mathbb{I}
\end{array}\right)\left(\begin{array}{cc}
\mathbb{I} & A^{-1}B\\
0 & D-CA^{-1}B
\end{array}\right)\right]\nonumber \\
 & =\det(A)\det(D-CA^{-1}B),\label{eq:det-formula}
\end{align}
so if just the $D$ block, assumed to be relatively small, changes,
we just have to calculate the determinant of the big block $A$ once,
and find once the (again relatively small) matrix $CA^{-1}B$, and
each time we want the determinant of the changed full matrix we can
then get away with calculating a determinant of a matrix of only this
small size.

\subsubsection*{Numerical computations of the determinants}

Since the calculation of determinants is the essential computational
step, let us spend a moment discussing it: The standard way to calculate
a determinant numerically, is to do a LU decomposition of the matrix
in question. This decomposes a matrix $A$ into a product of two matrices
$L$ and $U$, $A=LU$, such that $L$ is a lower triangular matrix
and $U$ is an upper triangular one. The determinant of a triangular
matrix is simply the product of its diagonal entries, so the determinants
of $L$ and $U$ can be easily computed, and the determinant of $A$
is then given as $\det(A)=\det(L)\det(U)$ due to the multiplicativity
property of the determinant. In fact, a typical LU decomposition even
produces an $L$ with unit diagonal, such that $\det(L)=1$ and therefore
$\det(A)=\det(U)$, and we just need to multiply the diagonal elements
of $U$.

Especially for the full matrix (but also for the reduced ones, depending
on their size) the determinant is a very big number. This is not so
surprising: If all the elements on the diagonal of $U$ are greater
than 1, the number will grow exponentially with the matrix size. If
they are all smaller than 1, the determinant will become exponentially
small. While these conditions may not occur often with more general
matrices, they are pretty typical for the ones we have to deal with,
and neither exponentially small nor exponentially large numbers can
be handled particularly well in the usual floating point representation
\cite{ieee754} -- you can lose precision, or, in the worst case,
numbers can even become too large or small to be represented at all
(\emph{overflow} or \emph{underflow}). However, we can work perfectly
well with the logarithms of these numbers, and therefore we can simply
use a custom routine for the determinant calculation that does just
that.

\subsubsection*{Bond strengths and boundary conditions}

As mentioned at the beginning of this section, one other thing that
goes beyond the established FKT algorithm is the handling of the fixed
boundary conditions. In fact, it turns out to be pretty straightforward
once you realize that we you again employ the procedure described
at the end of \ref{sub:fermionic-representation}, and use bonds of
infinite strength to fix the boundary in itself (and take care of
the extra factor of two introduced). It turns out that we can again
avoid any possible problems with infinities, because only the hyperbolic
tangens of the bond strength is relevant, and that is simply $1$
for an infinite bond strength. Seemingly divergent terms like the
$\cosh K_{ij}$-terms in \eqref{eq:fktZ} will always cancel for the
relevant partition functions.

\section{Results}

The preceding sections contain all the information needed to successfully
implement the calculations algorithmically, e.g. as Matlab programs.
It is a bit of a challenge to get all the different normalization
factors correct, but in the end of course all the algorithms give
the same results, so that there is little to show or discuss here,
and we can simply use whatever algorithm seems most appropriate for
a given problem geometry.

In particular, we want to work with systems that are as large as possible
and therefore require Monte Carlo sampling. For the geometry introduced
in figure \ref{fig:systems-borders} the MPS approximation method
turns out to be most convenient in practical implementation, and as
we have seen in section \ref{sub:mps-error-analysis}, the error is
typically limited by the statistical Monte Carlo error rather than
the approximation of the eigenvector.

\begin{figure}
\begin{centering}
\includegraphics[width=0.4\textwidth]{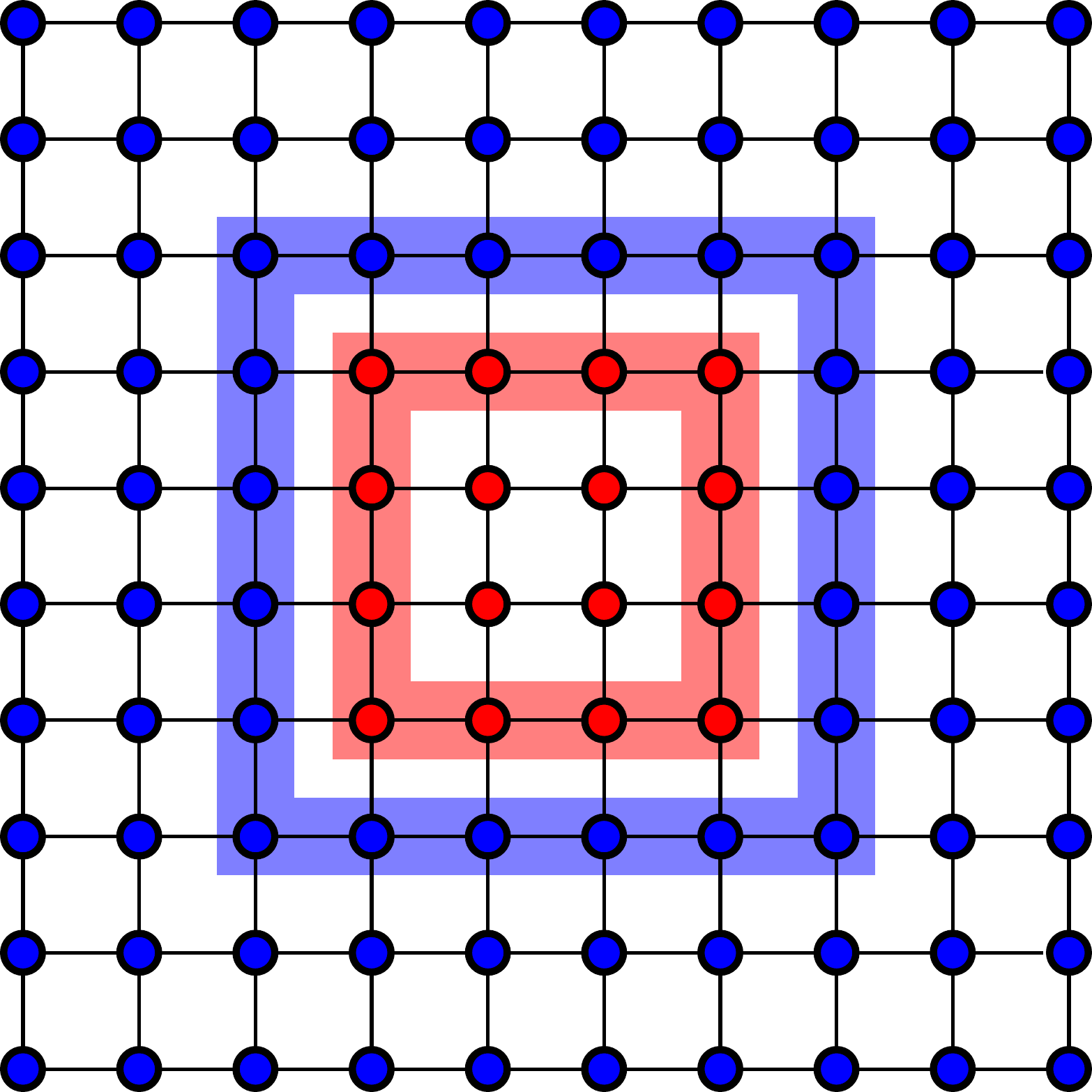}
\par\end{centering}

\caption{\label{fig:nested-systems-borders}Nested geometry}
\end{figure}

One geometry is however a bit limited, so is there a second choice
of geometry that would be interesting? In fact, there might be an
even more natural one: Not just one system that is cut in the middle,
but one system that it is fully embedded in a bigger one. This is
illustrated in figure \ref{fig:nested-systems-borders}. For such
a geometry the MPS method is less convenient, which is why the results
presented are obtained by the FKT method (with Monte Carlo sampling
of the boundary configurations again). The matchgate method is of
course also suitable, but its Matlab implementation turned out to
be somewhat slower in practice.

\begin{figure}
\includegraphics[width=1\textwidth]{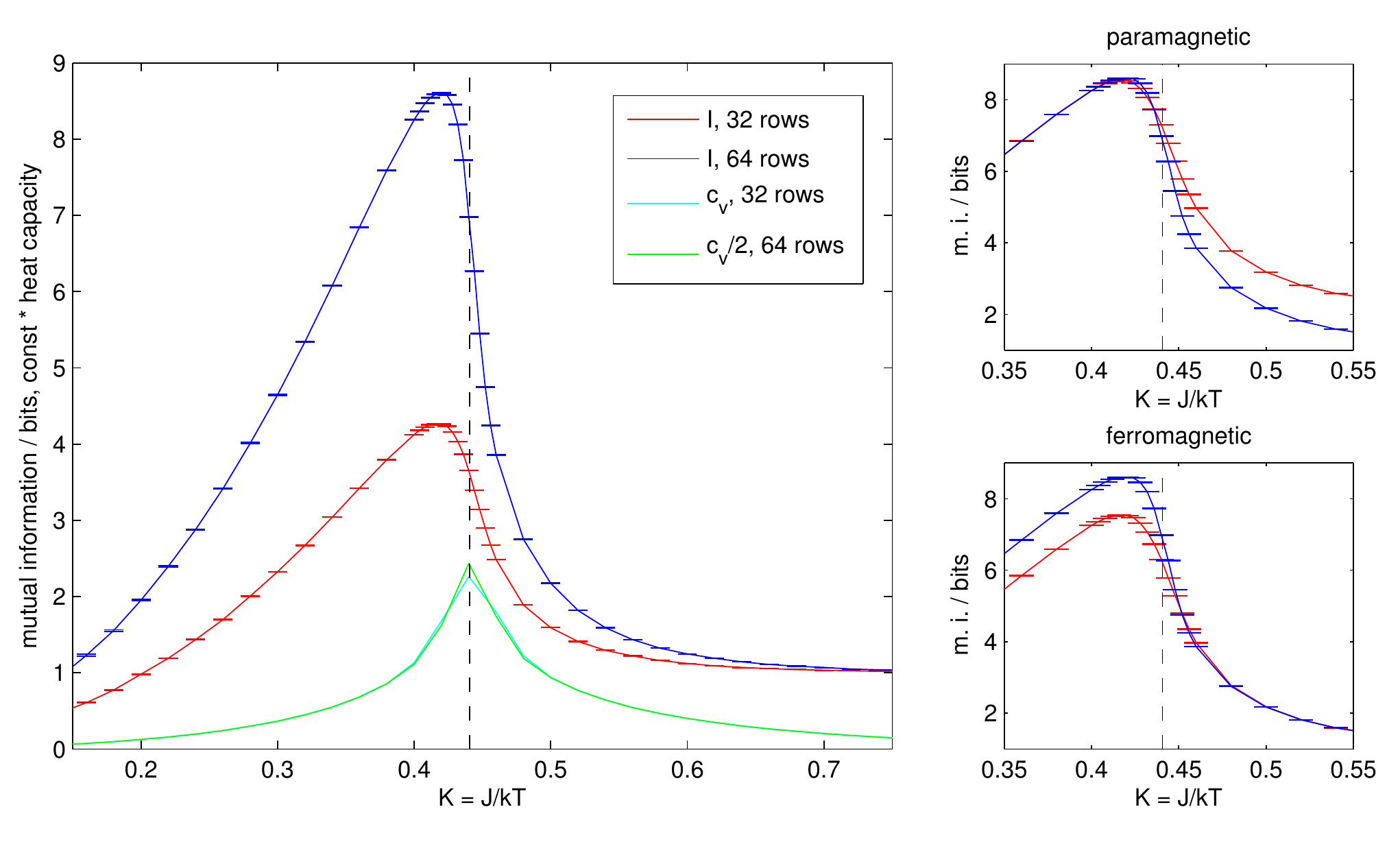}

\caption{\label{fig:ising-pbc-results}Results for the cylindrical geometry}
\end{figure}

\begin{figure}
\includegraphics[width=1\textwidth]{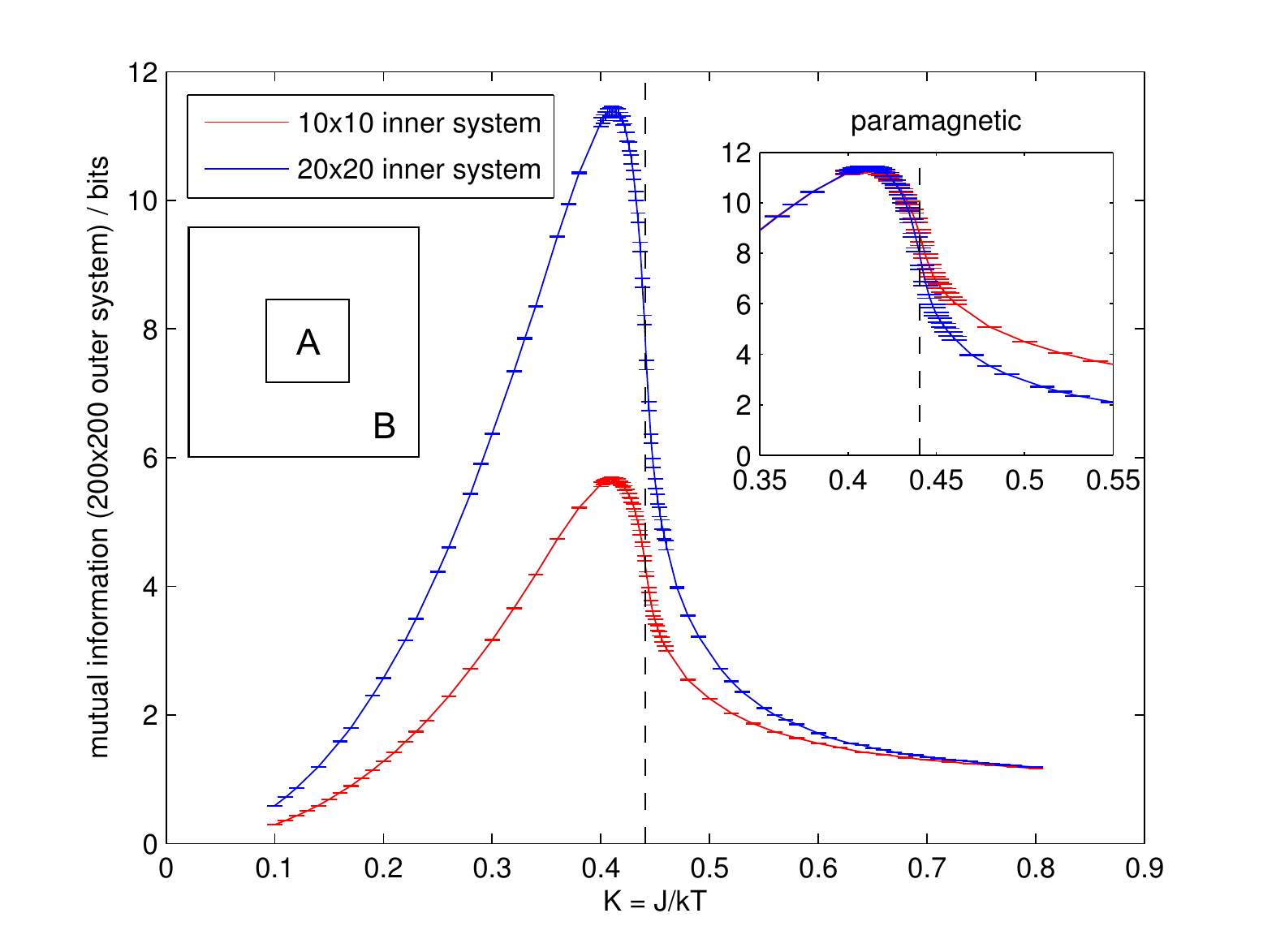}

\caption{\label{fig:ising-nested-results}Results for the nested geometry}
\end{figure}

So let us take a look at the results, shown in figures \ref{fig:ising-pbc-results}
and \ref{fig:ising-nested-results} for the two respective geometries:
Figure \ref{fig:ising-pbc-results} shows the results for the geometry
introduced in figure \ref{fig:systems-borders}, of a strip that is
cut in the middle. The strip is assumed to be infinitely long in horizontal
direction, which is another advantage of the MPS method, avoiding
any finite-size effects in that direction. Also, we can most easily
employ periodic boundary conditions in the vertical direction, which
makes for a much larger effective system than open boundary conditions,
and means that we are working with a cylinder rather than a strip.
This is in fact also possible with for example the FKT method, but
increases the number of Pfaffians that have to be calculated there.

Let us start by taking a look at the general features of mutual information:
We notice that it goes to zero for high temperatures. That is exactly
what we would have expected: at high temperatures the spins are all
randomly aligned, and if you look at the spins in one of the subsystems,
this does not give you any information about those in the other subsystem.

At low temperatures, the mutual information, calculated using a base-2
logarithm, becomes 1, i.e. 1 bit. Again, this is exactly what should
happen: All the spins will be aligned either up or down. In any case,
looking at one part of the system tells us that the spins will be
aligned in the same way in the other part, which is exactly 1 bit
of information.

In between those two limits, there is a phase transition. Its location
in the thermodynamic limit is known exactly, and indicated by the
dashed black line at $K=\mathrm{arcsinh(1)/2}=\ln(1+\sqrt{2})/2\approx0.44$.
Clearly something is happening around there, again as we would have
hoped. However, the naive reasoning may have gone like this: We know
that at criticality the correlation length, which describes the spatial
behaviour of the correlation function, diverges. That is why maybe
the most natural thing would have been to expect a maximum of the
mutual information exactly at the critical point. However, the maximum
lies within the high-temperature (paramagnetic) phase. Still, there
is something happening at the phase transition: It appears to be an
inflection point of the curve and the point of the largest (negative)
slope. In fact this slope, i.e. the first derivative of the mutual
information, will obviously diverge in the thermodynamic limit. You
can see that for example by comparing the results for two different
numbers of rows (circumferences of the cylinder) that are shown in
the figure.

To be specific, results are shown for both 32 and 64 rows. Let us
however first take a look at the heat capacity, which can be straightforwardly
calculated as a suitable second derivative of the logarithm of the
full partition function
\[
c_{v}=K^{2}\frac{\partial^{2}}{\partial K^{2}}\ln Z
\]
It has been scaled by some arbitrary factors to be nicely visible
in the figure and is plotted in cyan and green, respectively. You
can see that it has a sharp peak near the location of the phase transition
in the thermodynamic limit. This is already a first indicator that
our system sizes are sufficiently large to make useful statements.

Let us however focus on the plots of the mutual information now. It
turns out that it shows a remarkably simple scaling behaviour with
the system size, scaling linearly with it in the paramagnetic regime.
In the ferromagnetic regime, you first need to subtract 1, the value
of the mutual information in the low-temperature limit, and then you
recover exactly the same linear scaling. This can be seen in the right
parts of figure \ref{fig:ising-pbc-results}, where the data for 32
rows have been scaled as described and are contrasted with the data
for 64 rows. You can see that these scalings indeed work very well;
there appear to be very little residual finite size effects. Also,
you notice that not both of the scalings can work at the phase transition
-- unless this is indeed a point at which the slope becomes infinite
in the thermodynamic limit.

The error bars might deserve an additional explanation: They show
the Monte Carlo errors, calculated by a binning procedure as described
at the end of section \ref{sec:MC-sampling}. The bond dimension of
the MPS was chosen $D=8$, such that the errors due to this approximation
are lower than the ones introduced by Monte Carlo sampling even at
the critical point.

Let us now take a look at the second geometry, where the two parts
of the system are nested within each other. The results are shown
in figure \ref{fig:ising-nested-results}, and are now obtained by
using the FKT method as described in section \ref{sec:ising-fkt}.
We find pretty much the same behaviour as in the cylindrical geometry,
which serves as a nice confirmation of our results. In particular,
we even find the same scaling in the paramagnetic phase. This is shown
in the inset of the figure, where again the data for 32 rows has simply
been scaled by a factor of 2. The ferromagnetic phase appears to be
a bit more complicated in this geometry, and does not exhibit a straightforward
scaling behaviour any more. Again the error bars show the Monte Carlo
errors; remember that the FKT method itself is numerically exact.

\section{Fortuin-Kasteleyn clusters}

Can we understand why the mutual information has a maximum not at
the phase transition, but within the high-temperature phase?

What follows is certainly not a full explanation, but a motivation
which you may or may not find convincing: For spin systems like the
Ising model, it is sometimes useful to consider clusters of aligned
spins. For example, single spin updates can become very inefficient
in Monte Carlo sampling methods, when the typical cluster size diverges
at the critical point. What one therefore does instead is work with
Fortuin-Kasteleyn (FK) clusters \cite{fortuinkasteleyn}. Those are
clusters of spins pointing in the same direction, but not the simple
geometric clusters that would arise from combining all neighbouring
aligned spins. Instead, we only connect spins that are pointing in
the same direction with a probability $1-\exp(\beta\Delta E)$ where
$\Delta E$ is the energy difference between two aligned and unaligned
spins, and $\beta=1/k_{B}T$ the inverse temperature. This procedure
results in clusters which can be treated independently from each other,
which then allows for efficient Monte Carlo updates that flip the
whole clusters. If one did not use this procedure, but always connected
aligned spins, you would form clusters even at infinitely high temperature
($\beta=0$), simply because neighbouring spins will sometimes happen
to randomly point in the same direction.

So how can we make any argument about the behaviour of mutual information
by looking at clusters? It is true that spins within a cluster are
always perfectly correlated. So whenever a cluster lies in both parts
between which we calculate the mutual information, it represents some
information that is accessible in one system about the other one.
It is not obvious how one would exactly quantify that amount of information,
since we do not know how far any given cluster will extend into the
other part; but we can nevertheless think of some interesting quantities:
Namely, we can implement a Monte Carlo simulation that identifies
FK clusters, using for example the Swendsen-Wang algorithm \cite{swendsenwang}.
We can then ask how many of these clusters lie in both parts of the
system; in other words, how many clusters would be cut if we were
to cut the system in half. We can also ask for a slightly different
thing, namely, in how many pieces each of this clusters would disintegrate.
Since it might turn into more than two pieces, this is a slightly
different measure.

\begin{figure}
\centering{}\includegraphics[width=0.8\columnwidth]{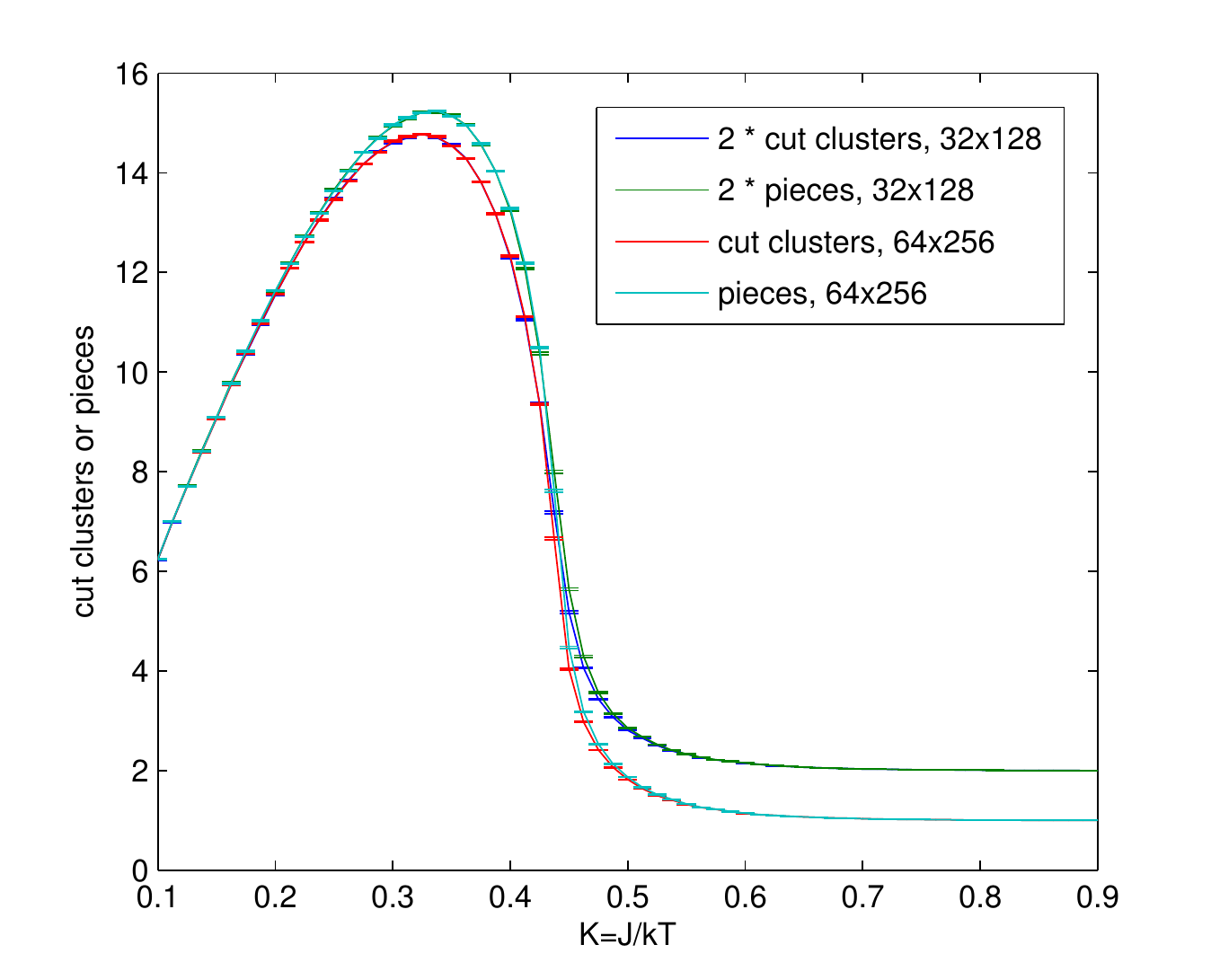}\caption{\label{fig:clusters}FK clusters being cut (cylindrical geometry).
The simulations were done using Swendsen-Wang cluster updates, which
also allow the immediate identification of clusters that are being
cut. Code from the ALPS project \cite{ALPS,ALPSscheduler} was used
in the simulations.}
\end{figure}

Take a look at figure \ref{fig:clusters} for the results. We again
looked at the same cylindrical geometry as for the results in figure
\ref{fig:ising-pbc-results}, so as to allow for greatest comparability.
Instead of a cylinder that is infinitely long in horizontal direction,
we just used a very elongated one, such that all relevant clusters
were contained within it. Again, let us look at the cases of 32 and
64 rows circumference. You see that we recover many of the behaviours
of mutual information: At very high temperature, there are no clusters
to be cut. At low temperature, there is just one big cluster. In particular,
also the scaling behaviour is exactly the same as for the mutual information:
For high temperature, we find a perfectly linear scaling, no matter
which of the two slightly different quantities {}``clusters cut''
or {}``resulting cluster pieces'' we study. For low temperature,
we again need to take the {}``residual'' contribution of the one
big cluster into account.

We also see that these quantities show the same behaviour as the mutual
information near the phase transition: There is again an inflection
point at the critical temperature, with diverging slope. And again
we have a maximum in the paramagnetic phase, and now we can give an
interpretation of this maximum in terms of clusters: The phase transition
may be the point where the cluster size diverges, but if our measure
is instead how many clusters we can cut (or into how many pieces we
can cut them), it is clear that the maximum of such a quantity will
not be at the phase transition, but at higher temperature, where there
are more clusters that can be cut -- but also not at very high temperature,
because there the clusters become too small, until they just consist
of individual spins.

So, if we accept that the {}``cut clusters'' quantities capture
some of the essential properties of mutual information, I believe
we have found a good intuition for why it behaves the way it does.

\section{Related studies in the Ising model}

At this point, some other studies made in the Ising model deserve
mentioning \cite{stephanshannon,renyi2dising}. They study entropies
of the probability distributions of boundary configurations in the
2-dimensional Ising model in a cylindrical geometry. This definition
of entropy is actually very similar to the term \eqref{eq:ising-ent},
which appears as an important part in our calculations of the mutual
information, and they are able to establish very nice scaling properties
and relations to conformal field theories. However, there does not
seem to be a way to actually connect this with our results concerning
the mutual information.

\chapter{\label{chap:More-lattice-spin}More lattice spin models}

In this chapter, we will take a look at a few models which may be
understood as generalizations of the Ising model. The essential generalization
is that the spins can now take more than two different values; you
might imagine it for example as a spin which can point in more different
directions than just up or down. Two different generalizations of
the Hamiltonian then lead to either the Potts models or the clock
models.

\section{Potts models}

Let us start with the Potts models \cite{Potts51,potts1952some,pottsrmpwu},
defined by the Hamiltonian
\[
H=-J\sum_{\langle ij\rangle}\delta(s_{i},s_{j})\,.
\]
If the $s_{k}$ only take two different values, this is indeed identical
to the Ising model \eqref{eq:isinghamiltonian}, up to a scaling of
the coupling constant by factor of two (and an irrelevant constant
energy shift). Now however we will let them take $q$ different values.
The critical temperatures for Potts models are known exactly, just
like for the Ising model. They are $T_{c}(q)=\ln(1+\sqrt{q})$ \cite{pottsrmpwu}.

\begin{figure}
\begin{centering}
\includegraphics[width=0.75\textwidth]{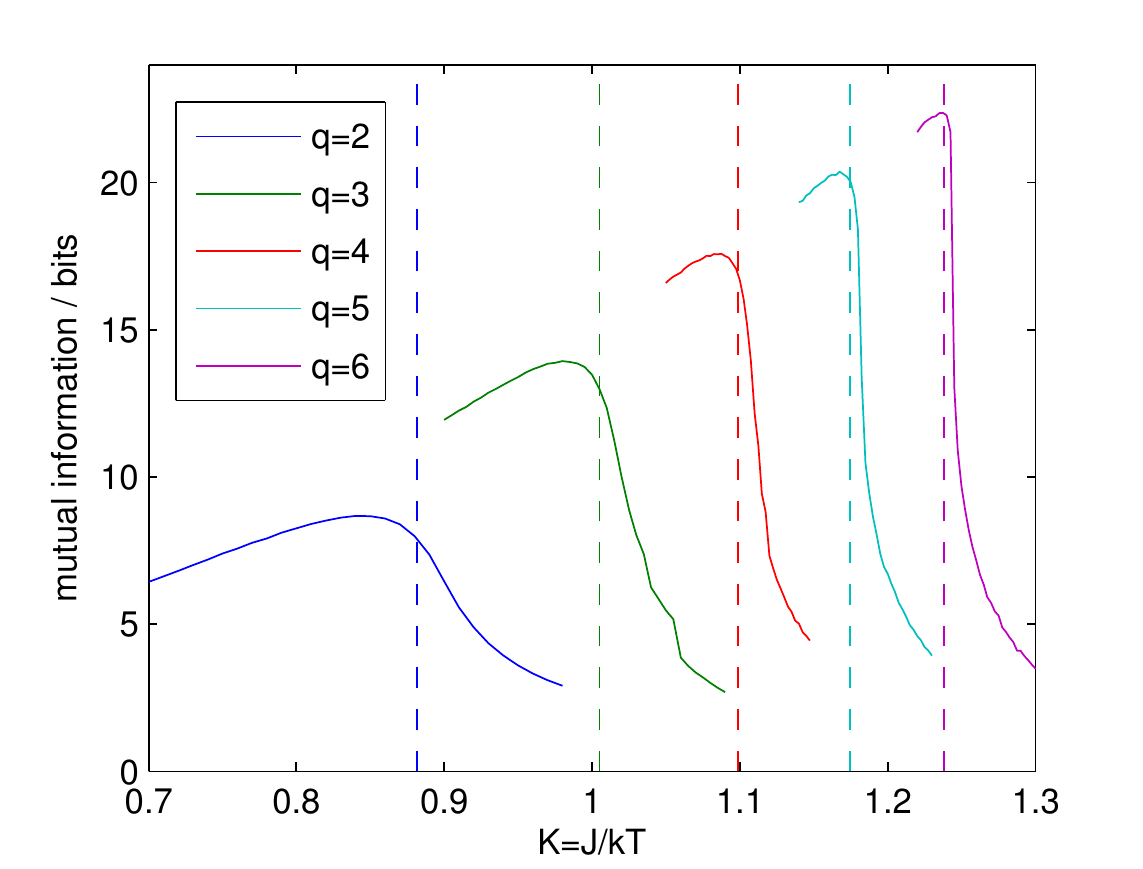}
\par\end{centering}

\caption{\label{fig:potts-q}Potts models}
\end{figure}

While we no longer have methods for exact solutions, like they were
available in the Ising model, we can still numerically calculate the
mutual information using the method explained in section \ref{sec:calc-part-func}.
Figure \ref{fig:potts-q} shows the results for $N=32$ rows, up to
$q=6$. It should be noted that for larger $q$, the calculations
become harder: We need to work with MPOs and MPSs with larger bond
dimensions, and there is a bigger space of boundary states that needs
to be sampled.

So what do we see? In short, pretty much the same behaviour as in
the Ising model: A maximum in the high-temperature phase near the
critical point, which again seems to be an inflection point with diverging
slope. The maximum becomes larger with larger $q$, which also makes
sense of course -- there can indeed now be a bigger amount of correlation,
or shared information if you will. It is known that for $q>4$ the
phase transition should change from a second-order phase transition
to a first-order one. However, there is nothing notably changing in
our results.

\section{Clock models}

\begin{figure}
\begin{centering}
\includegraphics[width=0.75\textwidth]{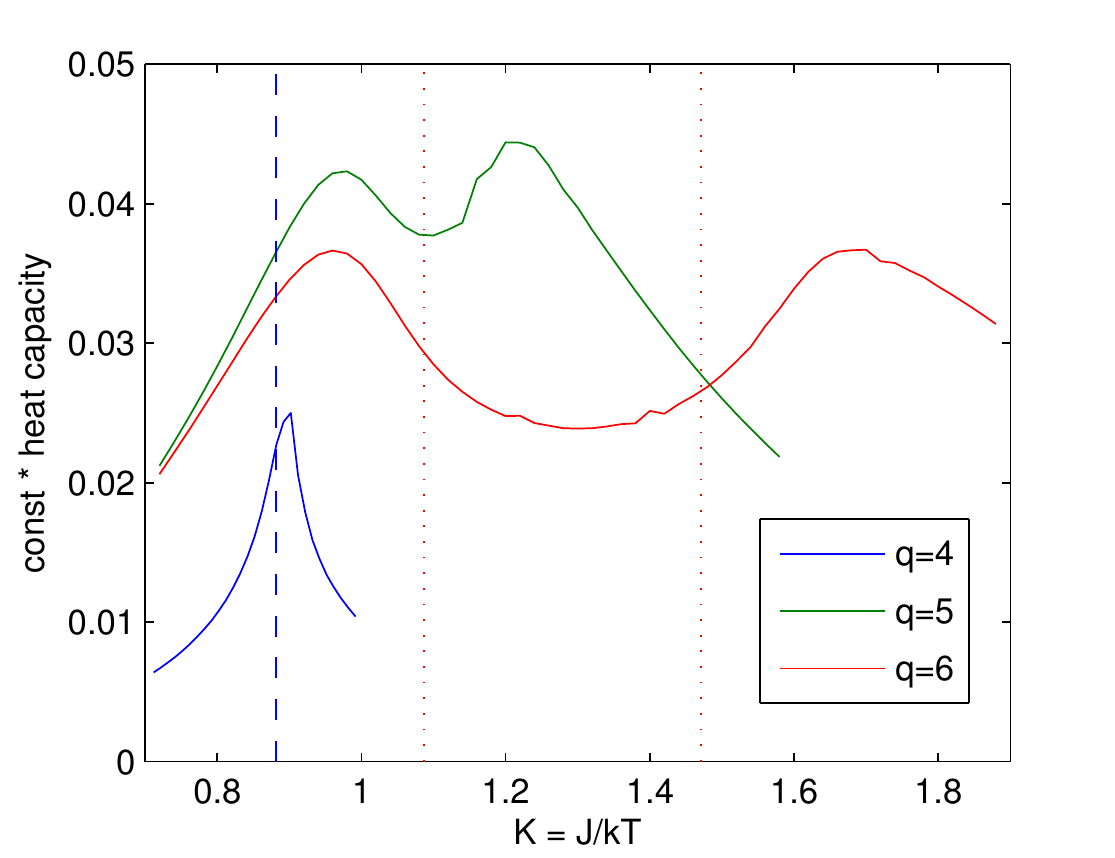}
\par\end{centering}

\caption{\label{fig:clock-hc}Clock models: heat capacity}
\end{figure}

The previous section described what is now known as the standard Potts
models. These are however not the only possible generalizations of
the Ising model. In fact, even Potts himself originally studied something
a bit different, which is now usually called planar Potts model, vector
Potts model, or clock model. Let us stick to that last name for clarity.
Again, we have $q$ possible states for each site/spin, and let us
now explicitly understand them as (planar) angles $s_{k}=2\pi k/q$
where $k\in\{0,1,\ldots,q-1\}$. The Hamiltonian of the $q$-state
clock model can then be defined as 
\[
H=-J\sum_{\langle ij\rangle}\cos(s_{i},s_{j})\:.
\]
For $q>3$, the clock models are different from, and rather more complex
than the Potts models; and not much seems to be known about their
phase transitions. Therefore, let us first take a look at the heat
capacity, which we can also easily extract from our calculation of
partition functions, just as we did before, for figure \ref{fig:ising-pbc-results}.
The heat capacities behave pretty much like has been found in \cite{clock456,clock456erratum,clock6}.
Figure \ref{fig:clock-hc} shows our results for the heat capacities
for $q=4,5,6$. For $q=4$, the heat capacity has a pronounced maximum
that is a good match for the location of the phase transition, which
is known exactly in this special case \cite{betts1964exact,clock4}
and indicated as the vertical dashed blue line. For $q>4$ however,
the clock model apparently has not just one, but two phase transitions.
There are in fact also two maxima of the heat capacity, but these
are not very sharp. They also do not match very well the estimated
locations of the phase transitions; the dotted red lines show the
estimates of \cite{clock6} for $q=6$.

\begin{figure}
\begin{centering}
\includegraphics[width=0.75\textwidth]{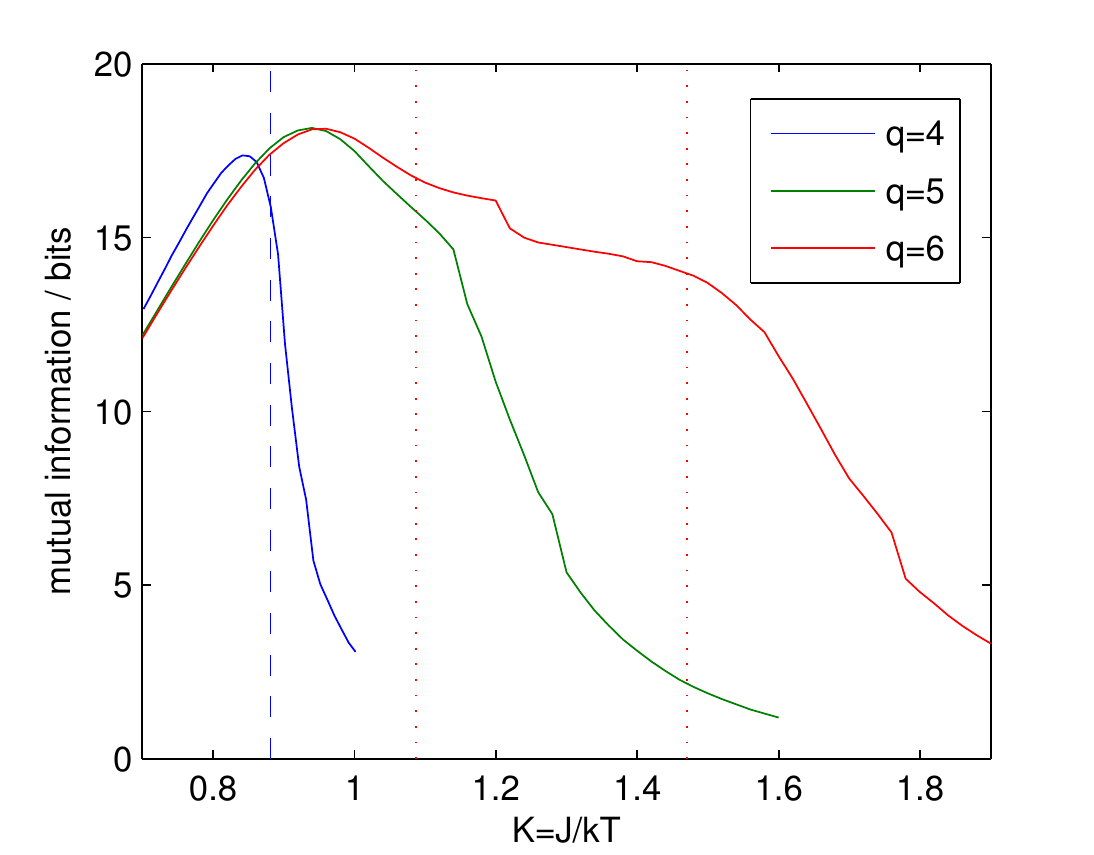}
\par\end{centering}

\caption{\label{fig:clock-mi}Clock models: mutual information}
\end{figure}

Correspondingly, we also find a rather straightforward behaviour for
the mutual information in the $q=4$ case, see figure \ref{fig:clock-mi}
-- it seems essentially the same again as for Ising and standard Potts
models, with a maximum near and the inflection point at the phase
transition. This is actually not so surprising because this case can
be related to regular Ising models \cite{betts1964exact,clock4}.

For $q=5$ and 6, the models clearly become more complicated. Since
the different phases in the models are not very well understood, it
does not seem possible to say much about the behaviour of the mutual
information either. Again, there appear to be inflection points close
to the suspected locations of the phase transitions. They now seem
to be accompanied by some sort of slight {}``kinks''. 

\section{Quantum spin models}

As has already been mentioned in section \ref{sec:what-done}, there
have been some independent investigations in quantum spin models \cite{Hastings10,Melko10,Singh11}.
These works made use of so-called replica tricks to allow for calculation
of integer Rényi entropies in an adapted Quantum Monte Carlo simulation.
As far as the results can be compared, they find some similar features
as we did: a maximum within the high-temperature phase, an inflection
point at criticality. However, their results are critically limited
by the fact that they have to calculate mutual information according
to \eqref{eq:mi-definition-shannon} while using Rényi entropies,
which is not well-justified from an information-theoretic perspective;
there is also no smoothing, which is generally necessary to give a
clear meaning to Rényi entropies, as was discussed briefly in section
\ref{sec:mi-renyi}. Therefore I will not try to make any further
connections between the results.

\chapter{\label{chap:lmg}The Lipkin-Meshkov-Glick model}

In this chapter we will start to consider a very different kind of
model. It can also be understood as a model of interacting spins.
But rather than having only local interactions as in the previous
chapters, these spins will now all be interacting with each other.
Such a model is therefore also called a fully-connected model (the
graph which describes the interactions as edges between the nodes
representing the spins is fully connected) or sometimes also a collective
model.

The model studied in this chapter is the Lipkin-Meshkov-Glock model,
which can be seen as the fully-connected version of an Ising-type
model and which has received the most attention of all fully-connected
models; a subset of the results of this chapter has already been published
as \cite{wilms2012finite}. The chapter after this one will then deal
with some more general fully-connected models.

\section{Motivation}

Most readers will probably agree that the assumption of having local
interactions is a very natural one, and might therefore doubt if it
even makes sense to consider fully-connected models. So where does
the Lipkin-Meshkov-Glick (LMG) model come from, and is it even relevant
for anything other than maybe a very small system?

The model got its name when it was studied by Lipkin, Meshkov, and
Glick in the 1960s \cite{Lipkin65,Meshkov65,Glick65}. Stigler's law
of eponymy \cite{Stigler80} applies, meaning it may be named after
them, buy they are not its original discoverers: As they mention themselves
in their first paper, it had certainly been discussed before that,
e.g. in the (otherwise unpublished) PhD thesis of Stavros Fallieros
\cite{Fallieros59}.

Even though Lipkin, Meshkov, and Glick were working in the field of
nuclear physics, and thinking about the spin-1/2 particles of neutrons
and protons, they did not think about a model with spin-spin interactions
in the same sense as e.g. in the Ising model, with a {}``magnetic''
(ferromagnetic or antiferromagnetic) interaction between spins. Indeed,
in a nucleus that is not a very relevant interaction -- the relevant
interaction is instead the strong nuclear interaction. However, in
order to arrive at the LMG model the exact form of the interaction
is not actually relevant. What is relevant is the fact that one has
a system of fermions that arrange themselves in shells; the analogy
to spin-1/2 systems comes about via fermions that can be in either
of two possible shells, thereby forming effective two-level systems.
Assuming some natural symmetries then leads to the LMG Hamiltonian.

Since the precise form of the interaction is not important in arriving
at the LMG Hamiltonian, it turns out to work as a simple model Hamiltonian
in quite a number of other systems as well. Typically, these are systems
of fermions that form shells, like the nucleons in a nucleus do, but
the interaction itself can be very different \cite{Maruhn10}. A prime
example of such systems are metal clusters, but the same is also true
for droplets of helium-3. There are also connections to, for example,
Bose-Einstein condensates \cite{cirac1998quantum}, or circuit quantum
electrodynamics \cite{larson2010circuit}.

Another strong motivation to pick the LMG model for further study
is the fact that in recent years, its ground state entanglement properties
have received a lot of attention \cite{Vidal04_1,Dusuel04_3,Dusuel05_2,Wichterich10,Latorre05_2,Barthel06_2,Vidal07,Orus08_2}.
We can therefore hope to be able to study how these relate to what
we will find at finite temperature. In particular, the model has a
phase transition both at zero and at finite temperature. In section
\ref{sec:mi-scaling-vs-qpt} we will see how in fact the mutual information
shows a comparable scaling behaviour to the ground state entanglement
entropy, in the form of a logarithmic divergence in both cases.

\section{\label{sec:lmg-ham-symmetries}Hamiltonian and symmetries}

We will use the LMG Hamiltonian in the following form:
\begin{equation}
H=-\frac{1}{4N}\left(\sum_{i,j}\sigma_{x}^{(i)}\sigma_{x}^{(j)}+\gamma\sum_{i,j}\sigma_{y}^{(i)}\sigma_{y}^{(j)}\right)+\frac{h}{2}\sum_{i}\sigma_{z}^{(i)}\label{eq:lmggeneral}
\end{equation}
which shows very clearly how it can be understood as a system of $N$
two-level systems -- spin-1/2 particles, if you like. In this language,
every spin interacts with every other one. Let it be noted that this
is not quite the most general form, but the form in which it has generally
been studied recently. In particular, note that there is a ferromagnetic
coupling between spins. In the next few sections we will deal, for
simplicity of exposition and readability, with the special case of
$\gamma=0$, and only afterwards go on to consider the effects of
a nonzero $\gamma$.

The essential property of this Hamiltonian is that all the interactions
are exactly identical, which is a symmetry that makes this system
much more accessible than a general system of $N$ spins. One can
see this by introducing the total spin $\boldsymbol{S}=(S_{x},S_{y},S_{z})$,
component-wise:
\begin{eqnarray*}
S_{x} & = & \sum_{i}\sigma_{x}^{(i)}/2\\
S_{y} & = & \sum_{i}\sigma_{y}^{(i)}/2\\
S_{z} & = & \sum_{i}\sigma_{z}^{(i)}/2
\end{eqnarray*}
and then one sees that this symmetry implies that the Hamiltonian
commutes with the magnitude of the total spin, $[H,\boldsymbol{S}^{2}]=0$.
We can in fact rewrite the Hamiltonian, up to constants, as
\[
H=-\frac{1}{N}\left(S_{x}^{2}+\gamma S_{y}^{2}\right)+hS_{z}\;.
\]
The symmetry implies the following: If you think of the Hamiltonian
of $N$ spins-1/2 as a $\mbox{{\ensuremath{2^{N}\times2^{N}}}}$ matrix,
this matrix can be brought into block-diagonal form, where the blocks
correspond to different representations of the total spin, indexed
by the magnitude of the total spin.

Why is this such a useful property? The reason is that this hugely
reduces the complexity of numerically approaching the problem, because
we no longer need to handle a Hamiltonian that is exponentially large
in the number of spins. Instead we only need to deal with blocks of
the size of the angular momentum representations. The possible values
of total spin lie between $s=0$ or $s=1/2$ and $s=N/2$. If $N$
is even, only integer values of $s$ appear; otherwise, only half-integer
ones. The size of such a representation is $(2s+1)\times(2s+1)$,
and there are only $\left\lfloor N/2\right\rfloor $ different possible
values of $s$. The exponential size of the full Hamiltonian is reproduced
by the multiplicity with which these representations appear, but these
are just numbers that can be calculated rather easily and of course
be handled much more efficiently.

It will turn out that calculating mutual information is still far
from easy, but let us get a little more familiar with the model first.

\section{Phase diagram}

Rather than just the one parameter $K=J/k_{B}T$ of e.g. the classical
Ising model on a lattice we now have several independent parameters.
Therefore it seems appropriate to spend some time to get familiar
with the phase space of the model, in particular since I am not aware
of any suitable existing review. It will be convenient to set $k_{B}=1$
(i.e., expressing temperature in units of $k_{B}$), and sometimes
to work with the inverse temperature $\beta=1/T$.

\subsection{\label{sub:lmg-mean-field}Mean-field theory}

Let us continue to interpret the model as $N$ interacting spin-1/2
particles and stick, for now, with the $\gamma=0$ case of the Hamiltonian
\eqref{eq:lmggeneral}, i.e.
\[
H=-\frac{1}{4N}\sum_{i,j}\sigma_{x}^{(i)}\sigma_{x}^{(j)}+\frac{h}{2}\sum_{i}\sigma_{z}^{(i)}\;.
\]
We have a ferromagnetic interaction, and if we just apply our intuition
from spin models on a lattice, we suspect (correctly) that there will
again be an ordered (ferromagnetic) phase and a disordered (paramagnetic)
phase. In fact, the average magnetization in $x$-direction $m_{x}=\left\langle \sigma_{x}\right\rangle $
is a good order parameter. It is zero in the disordered phase and
finite in the ordered one. As before, we will be mostly interested
in what happens at and near the phase transition.

To find the location of this phase transition, we just need to find
where the order parameter $m_{x}$ goes to zero. We can calculate
this using mean field theory (see also \cite{Botet82,Botet83}; similar
approaches can be found in \cite{Dutta11} and \cite{Quan09}): We
rewrite $\sigma_{x}^{(i)}=m_{x}+(\sigma_{x}^{(i)}-m_{x})$ and make
the replacement in the Hamiltonian, ignoring quadratic and higher
terms in the {}``deviation'' $(\sigma_{x}^{(i)}-m_{x})$. We get
\[
H=-\boldsymbol{h}_{\mathrm{eff}}\cdot\boldsymbol{S}+H_{0}
\]
where $\boldsymbol{h}_{\mathrm{eff}}=(m_{x},0,h)$ and $H_{0}=Nm_{x}^{2}/4$
(an added scalar constant that does however depend on $m_{x}$). Now,
by using the self-consistent condition 
\[
m_{x}\overset{!}{=}\left\langle \sigma_{x}\right\rangle =\mathrm{Tr}(\sigma_{x}\mathrm{e}^{-H/T})/Z=m_{x}\tanh(\sqrt{m_{x}^{2}+h^{2}}/(2T))/\sqrt{m_{x}^{2}+h^{2}}
\]
with $Z=\mathrm{Tr}\,\mathrm{e}^{-H/T}$ we can solve this for $m_{x}$
and determine the solution with the lowest free energy
\[
F=-T\ln Z=Nm_{x}^{2}/4-N\ln(2\cosh(\sqrt{m_{x}^{2}+h^{2}}/(2T)))\,.
\]
It is then possible to determine the temperature at which there is
no longer a nonzero solution for the magnetization (which is exactly
the transition temperature). This can easily be done numerically,
but in this case it is even possible to give an analytical expression
for this critical temperature as a function of the magnetic field
$h$:
\[
T_{c}(h)=h/(2\mathrm{\, arctanh}(h))\,.
\]

\subsection{\label{sub:lmg-blocks}Finite-$N$ numerical treatment}

As has been mentioned before, the Hilbert space of a system consisting
of $N$ spins-1/2 has dimension $2^{N}$, which usually limits the
study of such systems to a few tens of spins at most. But due to the
symmetry $[H,\boldsymbol{S}^{2}]=0$, the Hamiltonian becomes block
diagonal in the total spin basis. For the numerical calculations,
we need to know the multiplicities with which the blocks corresponding
to the different values of total spin occur. We can get these from
the theory of addition of angular momenta, which tells us that there
are $d_{s}^{N}$ distinct ways of obtaining a total spin $s$ when
combining $N$ spins 1/2, with

\[
d_{s}^{N}=\begin{pmatrix}N\\
N/2-s
\end{pmatrix}-\begin{pmatrix}N\\
N/2-s-1
\end{pmatrix}=\frac{2s+1}{N/2+s+1}\begin{pmatrix}N\\
N/2-s
\end{pmatrix}.
\]
There is one possible problem with the numerical evaluation of these
multiplicities, namely the factorials contained in the definition
of the binomial coefficient. Of course, one can immediately get rid
of all the canceling factors, but even then, some pretty large numbers
remain, and in fact they can quickly become larger than the largest
floating point number that can be represented e.g. in usual IEEE754
64-bit arithmetic \cite{ieee754}.

In fact, for our purposes the above formula can just about still be
evaluated, but for future problems it might also be useful to be aware
that the multiplicities can be generated recursively according to
the following scheme, which is perfectly stable:

\begin{center}
\begin{tabular}{|c||c|c|c|c|c|c|c|c|c|}
\hline 
$N\backslash s$  &  & 0  & $\frac{1}{2}$  & 1  & $\frac{3}{2}$  & 2  & $\frac{5}{2}$  & 3 & \ldots{}\tabularnewline
\hline 
\hline 
1  & 0  & 0  & 1  & 0  &  &  &  &  & \tabularnewline
\hline 
2  & 0  & 1  & 0  & 1  & 0  &  &  &  & \tabularnewline
\hline 
3  & 0  & 0  & 2  & 0  & 1  & 0  &  &  & \tabularnewline
\hline 
4  & 0  & 2  & 0  & 3  & 0  & 1  & 0  &  & \tabularnewline
\hline 
\,5\,  & \,0\,  & \,0\,  & \,5\,  & \,0\,  & \,4\,  & \,0\,  & \,1\,  & \,0\, & \ldots{}\tabularnewline
\hline 
\end{tabular}
\par\end{center}

Here, each line is obtained from the one above by summing the multiplicities
from the columns to the left and to the right (but not directly above).

One can of course check that

\[
\sum_{s=[S]}^{S}(2s+1)d_{s}^{N}=2^{N},
\]
where $S=N/2$ is the maximum spin, and $[S]$ is the minimum spin,
namely $[S]=0$ if $N$ is even and $[S]=1/2$ if $N$ is odd. The
sum over $s$ then runs over integers (half-integers) if $N$ is even
(odd).

We can now efficiently diagonalize the Hamiltonian in the separate
blocks, keeping track of their multiplicities, and by exponentiation
of the blocks obtain the density matrix which we are interested in.
Exponentiation of the blocks can be done either via diagonalizing
them, or via direct matrix exponentiation algorithms (see e.g. \cite{moler1978nineteen,moler2003nineteen}).
In our setting, it does not really matter which option one chooses.
They all work well, and for the more involved calculation of the mutual
information later on, this step will only take a negligible amount
of computation time anyway.

In fact, at this point it should also be noted that the LMG model
possesses additional symmetries, specifically the so-called spin-flip
symmetry $[H,\prod_{j}\sigma_{j}^{z}]=0$. This does in fact further
reduce the size of the blocks, roughly by half, but we will mostly
ignore it in order to give general results for any Hamiltonian satisfying
only $[H,\boldsymbol{S}^{2}]=0$. Of course, whenever such a symmetry
exists, it is in general worthwhile implementing it to speed up the
algorithm.

Because we will use it later, let us formally write down our spin-conserving
Hamiltonian in block-diagonal form as

\begin{eqnarray}
H & = & \sum_{s=[S]}^{S}\,\sum_{i=1}^{d_{s}^{N}}H_{i}^{(s)},\nonumber \\
 & = & \sum_{s=[S]}^{S}\,\sum_{i=1}^{d_{s}^{N}}\,\sum_{m=-s}^{s}\,\sum_{m'=-s}^{s}\, h_{m',m}^{(s)}\,\ket{s,m'}_{i}\,\,{}_{i}\bra{s,m},\qquad\label{eq:def1}
\end{eqnarray}
where $i$ labels the $d_{s}^{N}$ degenerate subspaces of spin $s$,
and where the notations of the sums over $s$ have already been introduced
above. Furthermore, we introduced eigenstates $\ket{s,m}_{i}$ of
operators $\mathbf{S}^{2}$ and $S_{z}$ with eigenvalues $s(s+1)$
and $m$ respectively. The matrix elements $h_{m',m}^{(s)}$ are independent
of $i$ so that, for each $s$, one has $d_{s}^{N}$ copies of the
same matrix to diagonalize. Once the diagonalizations are performed,
one can write

\begin{equation}
H=\sum_{s=[S]}^{S}\,\sum_{i=1}^{d_{s}^{N}}\,\sum_{\alpha=1}^{2s+1}E_{\alpha}^{(s)}\ket{s;\alpha}_{i}\,\,{}_{i}\bra{s;\alpha},\label{eq:diag}
\end{equation}
where eigenvalues $E_{\alpha}^{(s)}$ of $H_{i}^{(s)}$ are independent
of $i$ and where the corresponding eigenvector $\ket{s;\alpha}_{i}$
is given by

\begin{equation}
\ket{s;\alpha}_{i}=\sum_{m=-s}^{s}a_{\alpha;m}^{(s)}\ket{s,m}_{i},\label{eq:evec}
\end{equation}
with coefficients $a_{\alpha;m}^{(s)}\in\mathbb{C}$ independent of
$i$. The partition function is then

\begin{eqnarray}
Z & = & \mathrm{Tr}\:{\rm e}^{-\beta H},\nonumber \\
 & = & \sum_{s=[S]}^{S}\,\sum_{i=1}^{d_{s}^{N}}\,\sum_{\alpha=1}^{2s+1}\mathrm{e}^{-\beta E_{\alpha}^{(s)}},\nonumber \\
 & = & \sum_{s=[S]}^{S}\, d_{s}^{N}\left[\sum_{\alpha=1}^{2s+1}\,\mathrm{e}^{-\beta E_{\alpha}^{(s)}}\right]=\sum_{s=[S]}^{S}\, d_{s}^{N}Z^{(s)},\label{eq:lmgZn}
\end{eqnarray}
where $Z^{(s)}=\mathrm{Tr}\,\mathrm{e}^{-\beta H_{\mathrm{ref}}^{(s)}}$
is the partition function associated to any of the $H_{i}^{(s)}$
(here we choose a reference index $i=\mathrm{ref}$).

Calculating expectation values of observables $A$ as $\mathrm{\langle A\rangle=Tr}\:\rho A$
with $\rho={\rm e}^{-\beta H}/Z$ is a simple generalization of the
above formula, as long as the observable can be decomposed in the
same block structure, which is however the case for all the observables
we are interested in.

One additional remark: we will later be comparing finite-temperature
properties to ground-state properties. It is obvious that our diagonalization
procedure allows us to find the ground state(s) by just choosing the
state(s) with the lowest energy. In fact, it is known that this state
is always part of the (non-degenerate) maximum-spin sector, due to
the ferromagnetic nature of the interaction \cite{Dusuel04_3}, so
that it is sufficient to look at just that sector.

\subsection{\label{sub:LMG-Order-parameter}Order parameter}

\begin{figure}
\includegraphics[width=1\textwidth]{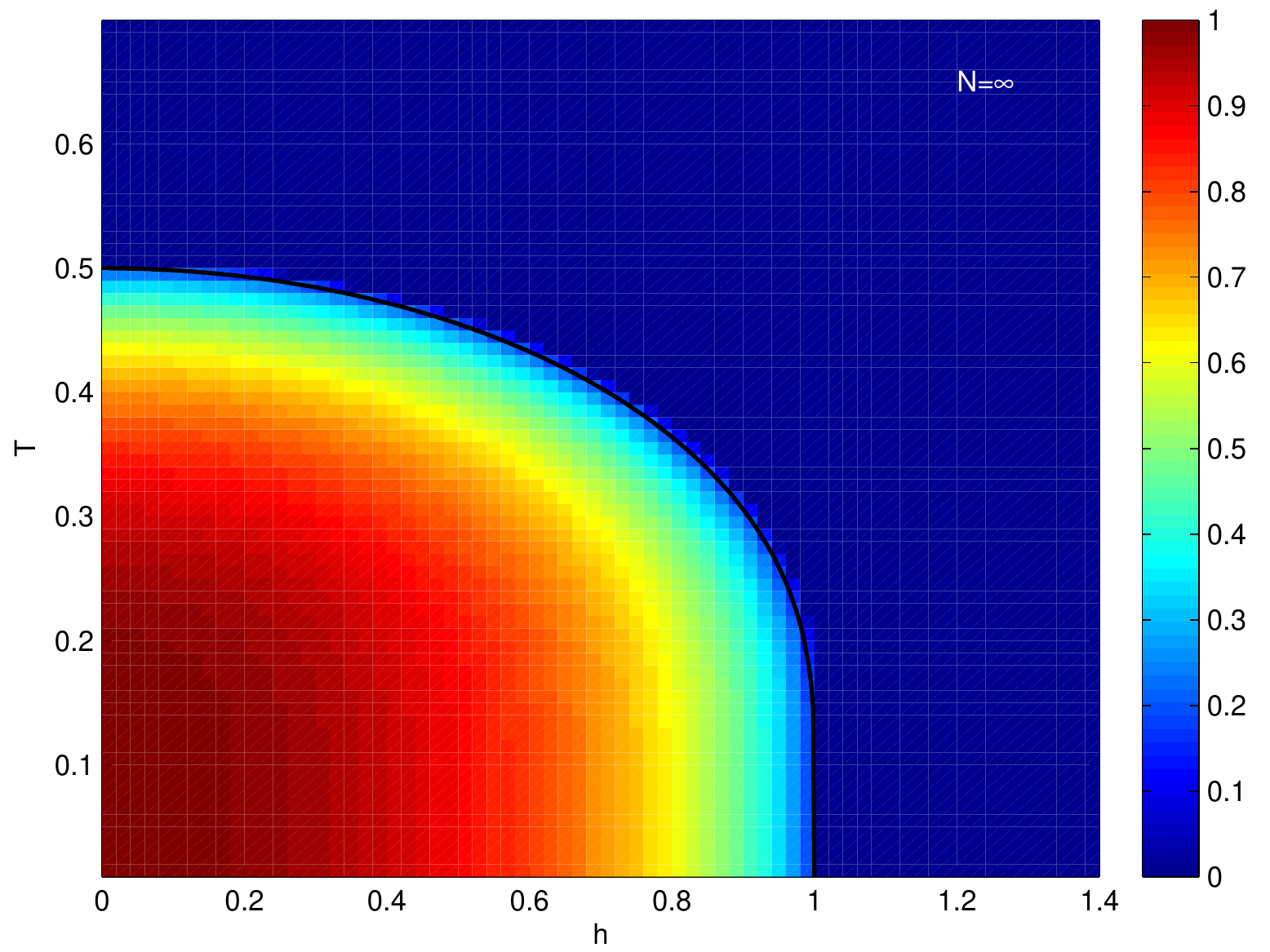}\\

\includegraphics[width=1\textwidth]{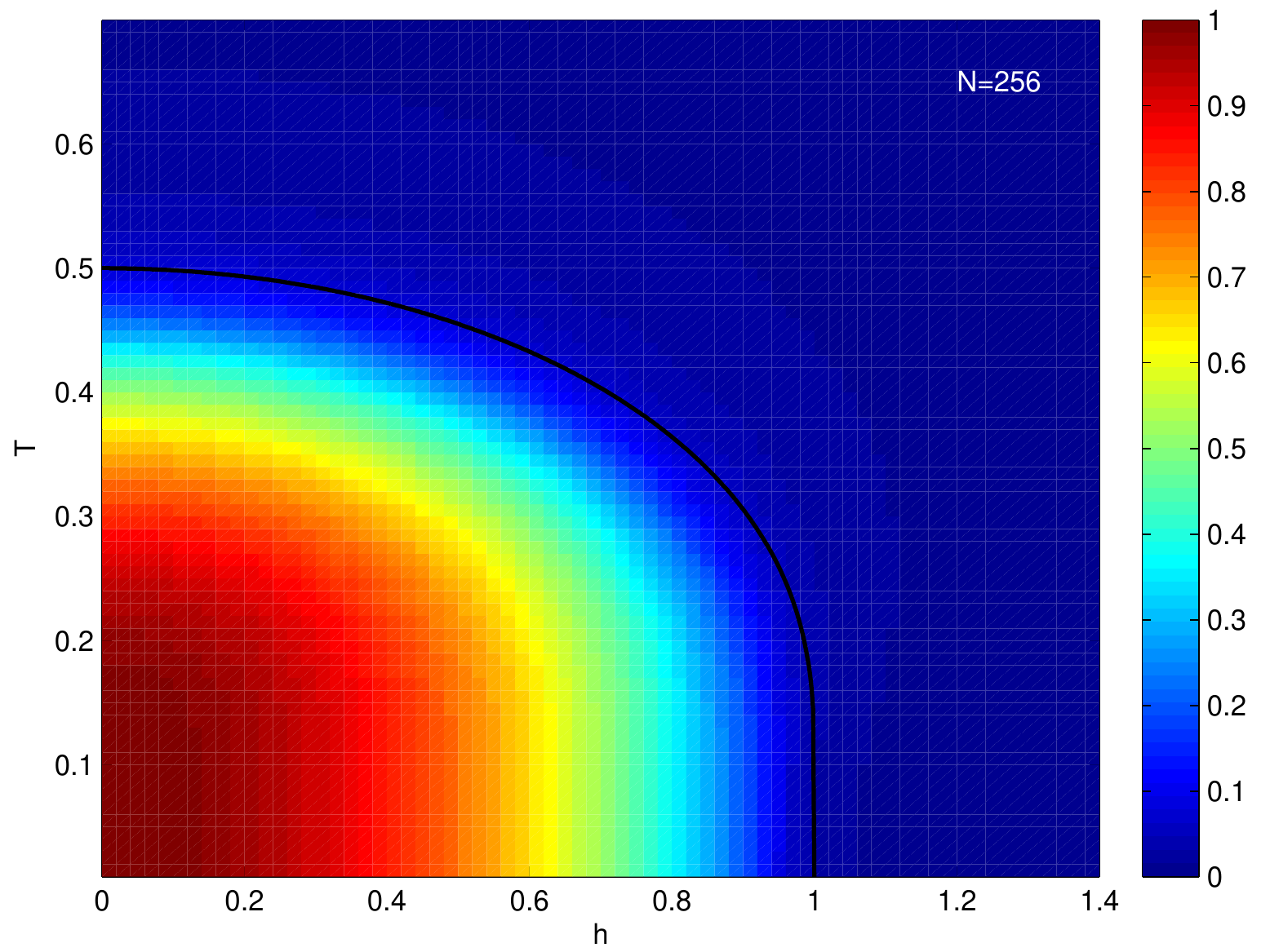}

\caption{\label{fig:lmg_op_256}Order parameter $m_{x}$ for the LMG model,
mean-field theory (top) and $N=256$ (bottom). The black line denotes
the location of the phase transition.}
\end{figure}

Now that we understand how to do the calculations, let us get a better
feeling for the LMG model by looking explicitly at the aforementioned
order parameter. It clearly differentiates the two phases, as can
be seen from the upper part of figure \ref{fig:lmg_op_256} (obtained
using the mean-field theory of section \ref{sub:lmg-mean-field}):
There is the ordered phase for low field and temperature, and the
disordered phase elsewhere.

The lower part of figure \ref{fig:lmg_op_256} then shows what happens
if we do the calculation not by using mean-field theory, but working
with a fixed number $N$ of spins, using the approach of section \ref{sub:lmg-blocks}.
In fact, we numerically calculate the order parameter as $m_{x}=2\sqrt{\langle S_{x}^{2}\rangle}/N$
in order to avoid any problems due to symmetry-breaking. 

For $N=256$ as chosen in the figure we get the same characteristic
picture as in the thermodynamic limit, with of course some inevitable
finite-size corrections. We can conclude that $N=256$ already gives
us a very good idea even if we declare ourselves interested in the
behaviour in the thermodynamic limit.

There are two interesting extreme cases of the LMG model that we will
specially consider in the following: The left edge of the phase space,
$h=0$, is what we will call the classical limit: Since there is no
transverse field, we can work with classical spins (in the $\sigma_{x}$-basis).
The bottom edge, where $T\to0$, is the zero-temperature limit. The
$T=0$ case will usually not be explicitly shown in the plots since,
as mentioned, it requires a different (if simpler) algorithm and would
just smoothly add another line of barely visible pixels anyway.

\begin{figure}
\includegraphics[width=1\textwidth]{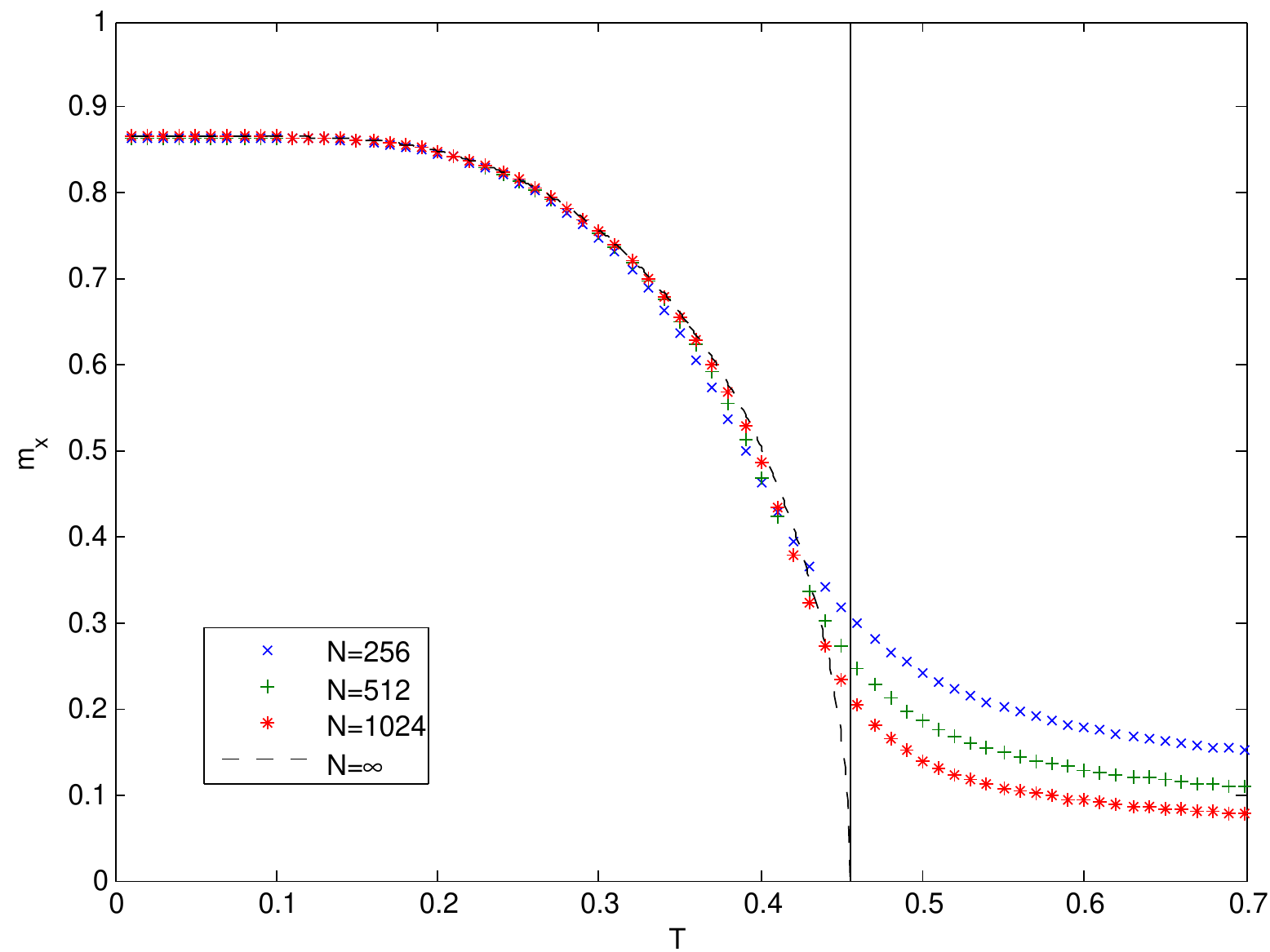}

\caption{\label{fig:lmg_op_T}Order parameter $m_{x}$ as a function of temperature
for $h=1/2$. Different $N$ are shown together with the mean-field
theory valid in the thermodynamic limit (dashed black line). The critical
temperature for $h=1/2$ is also marked by the vertical solid black
line. }
\end{figure}

Let us also look at some cross-sections of these plots, in figure
\ref{fig:lmg_op_T}. You can see that (as one would expect) for increasing
$N$ the thermodynamic limit is approximated better and better. $h=1/2$
was chosen since it represents a somewhat more general case than e.g.
$h=0$ (which looks very similar, only with the transition at $T_{c}=1/2$).

\subsection{Magnetic susceptibility}

\begin{figure}
\includegraphics[width=1\textwidth]{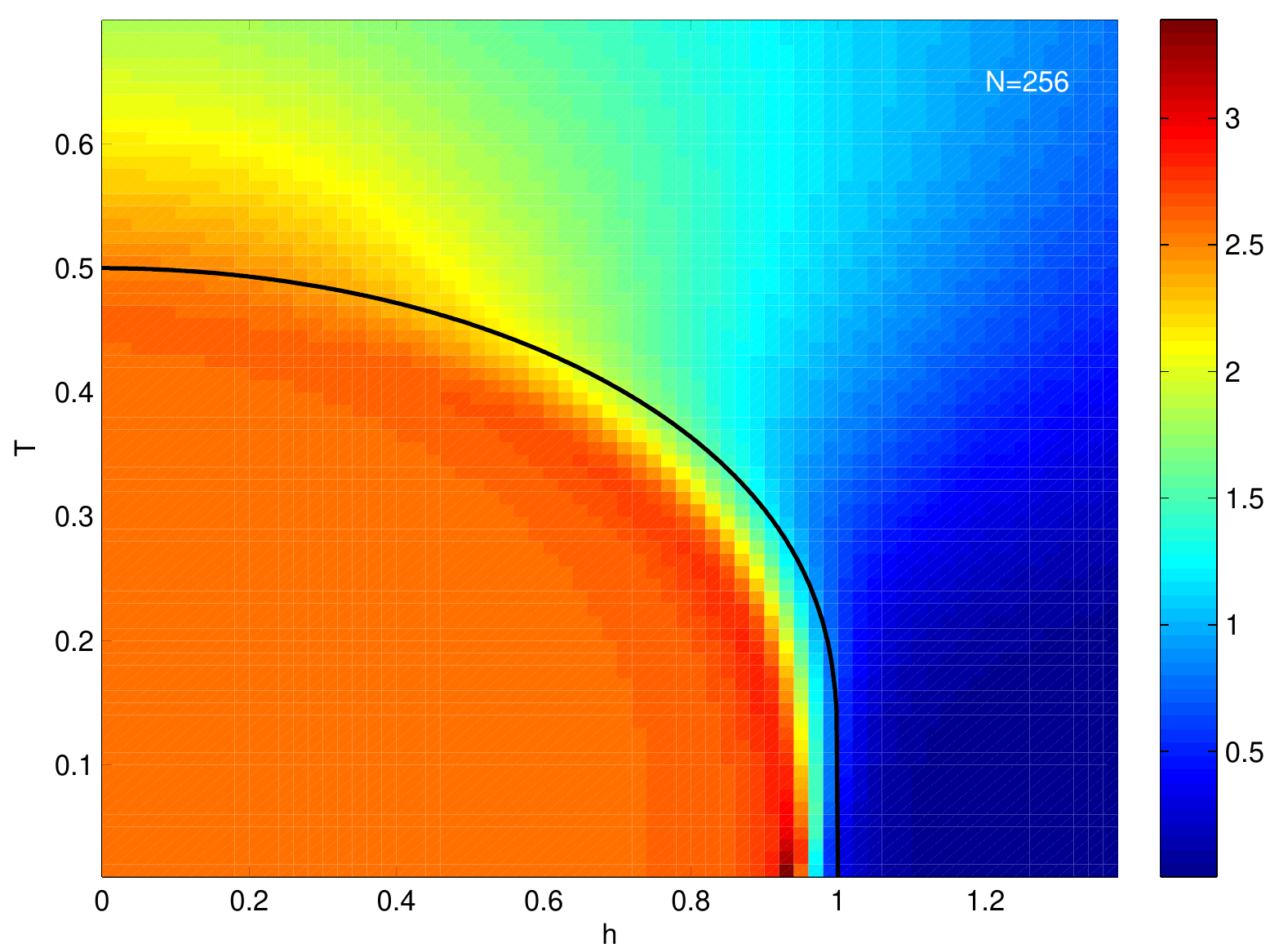}

\caption{\label{fig:lmg_su_256}Magnetic susceptibility}
\end{figure}
The order parameter gives us a good idea about the two distinct phases,
the ordered (ferromagnetic) and disordered (paramagnetic) one. We
will however be most interested in the phase transition between those.
So are there quantities that show a more {}``interesting'' behaviour
at the phase transition than just going to zero as the order parameter
does? Maybe the most obvious choice is the magnetic susceptibility,
which is just the derivative of our order parameter with respect to
the magnetic field, $\partial m_{x}/\partial h$. In the thermodynamic
limit, this will diverge at the phase transition. At finite $N$,
there is still a pronounced maximum near the phase transition, as
can be seen in figure \ref{fig:lmg_su_256}. While it would still
be possible to produce mean-field plots for classical thermodynamic
quantities such a the susceptibility, from now on we will mostly stick
to finite-$N$ plots, since these will be the only thing that we will
be able to produce for the more interesting quantities we will study
later, at least in the full phase space. 

One more thing to say about the susceptibilities: It would also be
possible to use not the magnetization in $x$-direction, but rather
the magnetization in the direction of the field (or maybe even the
total magnetization). These will all result in slightly different
plots, but we will not go into that any further, since this is of
little relevance to the quantities we will want to study later anyway.

\subsection{Heat capacity}

\begin{figure}
\includegraphics[width=1\textwidth]{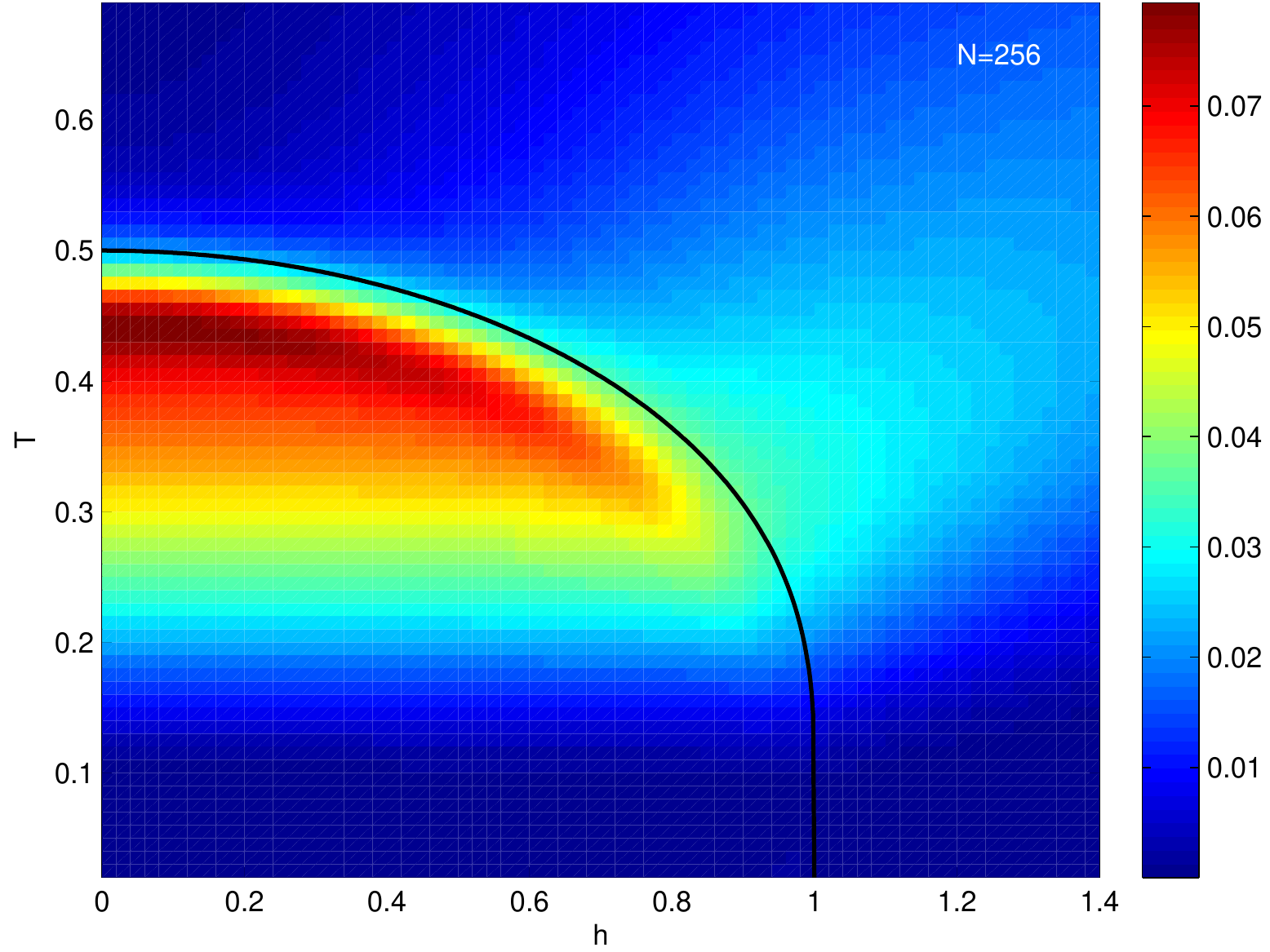}

\caption{\label{fig:lmg_hc_256}Heat capacity}
\end{figure}
In order to complete the survey of all the {}``standard'' quantities
that one thinks of in relation to the study of phase transitions,
let us also take a quick look at the heat capacity. In particular,
this was also our choice in the classical Ising model on a lattice
(where there was no obvious choice of external field, with respect
to which one would derive).

The study of the heat capacity is in some way complementary to the
study of the order parameter and related magnetic susceptibilities:
rather than derive with respect to the magnetic field to arrive at
magnetic susceptibilities, we derive with respect to the temperature.

This is somehow evident in the fact that, as figure \ref{fig:lmg_hc_256}
shows, in a colour plot the heat capacity seems to work better as
an indicator for the {}``fully thermal'' transition at low fields,
and not work quite as well where the external field is relevant. It
should however be noted that even there the heat capacity still has
a clear maximum near the phase transition; it is just not easily visible
in this figure that is trying to show the full transition line at
once.

\section{\label{sec:lmg-op-fss}Finite size scaling properties}

We said we wanted to study the behaviour at (or nearby) the phase
transition, so what is it that one studies quantitatively in such
a situation?

\begin{figure}
\includegraphics[width=1\textwidth]{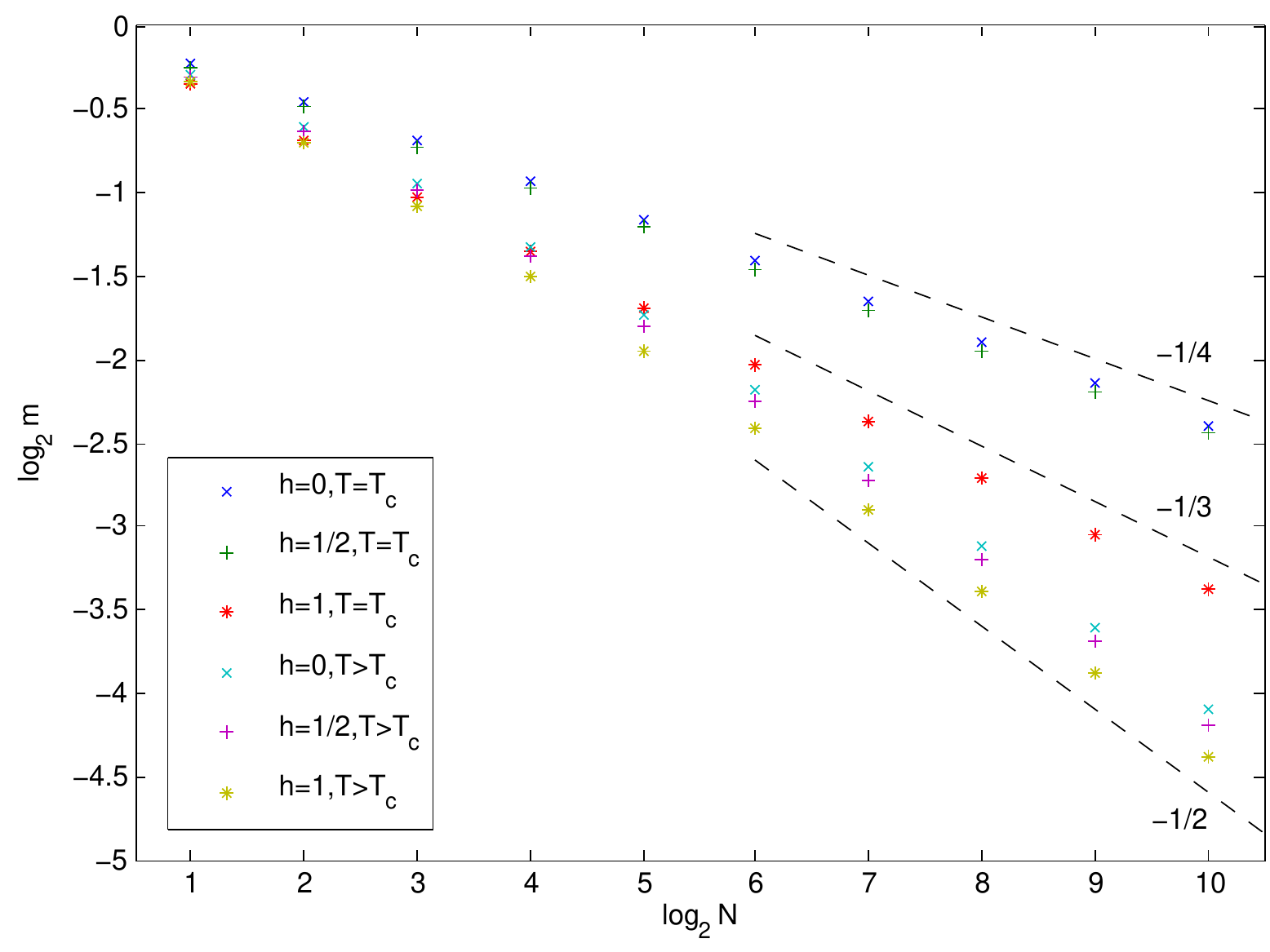}\caption{\label{fig:lmg_op_N}Order parameter as a function of $N$. For the
$T>T_{c}$ a $T$ of 0.7 was chosen in all cases; the dependence on
the exact value of $T$ is very weak there.}
\end{figure}

We are dealing with a second-order phase transition, and that implies
that there will be critical exponents to be found. So where would
we look for such critical exponents? The obvious choice is the order
parameter: we can examine how it tends to zero either as a function
of $(T-T_{c})$, or as a function of the number of spins $N$, at
the critical point. The former was already displayed in figure \ref{fig:lmg_op_T},
and when studying this in more detail one can indeed see that it matches
the known critical behaviour: the order parameter goes to zero as
$(T-T_{c})^{\alpha}$ with the known critical exponent $\alpha=1/2$
\cite{Botet83}.

What about critical exponents associated to the behaviour as a function
of $N$? As we can also see in figure \ref{fig:lmg_op_N}, within
the disordered phase, i.e. for $T>T_{c}$, the order parameter (which
is zero in the thermodynamic limit) vanishes as $N^{-1/2}$. You would
probably not even call that a critical exponent, but it is important
to know in order to compare it to the critical behaviour: At $h<1$,
where the phase transition occurs at a nonzero temperature $T_{c},$
the order parameter instead behaves as $N^{-1/4}$. As you can see,
the exact value of $h$ is irrelevant for the asymptotic behaviour.

However, the case $h=1$ is special: Here the critical temperature
is zero; we are dealing with a quantum phase transition. First of
all, this means that a different numerical approach is required. As
described in section \ref{sub:lmg-blocks} however, we can also easily
enough access the ground state and calculate its properties, such
as the expectation values needed for the order parameter. What we
can see is a different critical exponent, with the order parameter
going to zero as $N^{-1/3}$.

We will be interested in studying such scaling behaviour also for
the mutual information, and in particular again contrast finite with
zero temperature in section \ref{sec:mi-scaling-vs-qpt}.

\section{\label{sec:lmg-ent-measures}Entanglement measures}

\begin{figure}[t]
\includegraphics[width=1\textwidth]{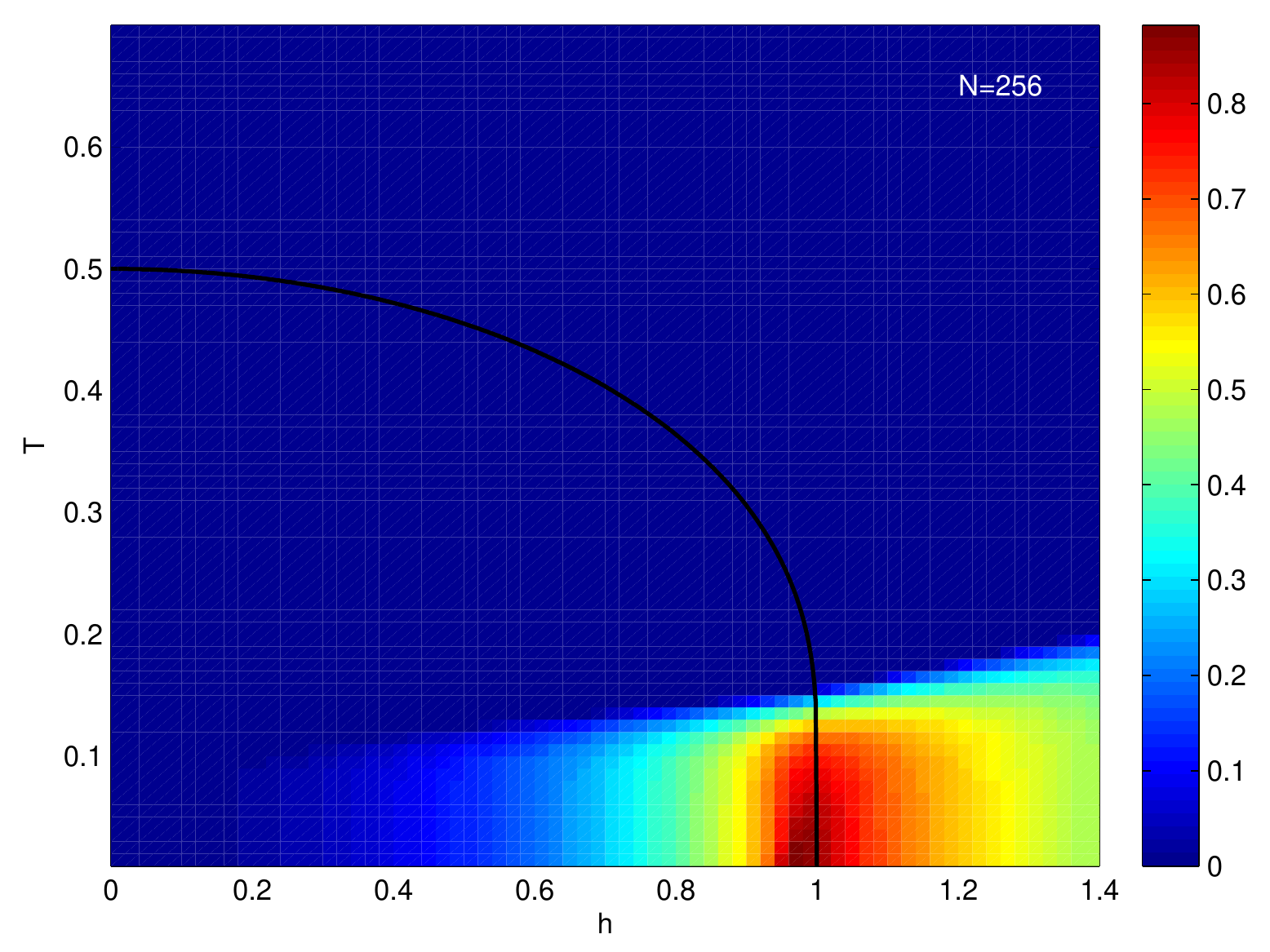}

\caption{\label{fig:lmg_conc_256}Concurrence}
\end{figure}
Let us stay with the topic of zero versus finite temperature for a
bit: Part of the motivation for looking at mutual information was
that it is the generalization of entanglement entropy, which proved
useful for pure states, and in particular also for the study of the
ground state of the LMG model. While entanglement entropy was not
the only entanglement measure that proved interesting there, a general
problem with entanglement measures is that many of them only work
well for pure states. An exception that remains well-defined for mixed
states and can also still be calculated efficiently is concurrence
\cite{Wootters98}, which has indeed already been examined for this
model, both at zero \cite{Vidal04_1,Dusuel04_3,Dusuel05_2} and at
non-zero temperature \cite{Matera08}. However, take a look at figure
\ref{fig:lmg_conc_256}: As you can see, it works well in the vicinity
of the quantum phase transition, but has no interesting features at
all for most of the finite-temperature transition. 

Can we explain why concurrence works so badly? I believe the answer
lies in the argument presented in the first chapter: concurrence may
qualify as an entanglement measure, but it is not a good measure of
all many-body correlations. In fact, it can be calculated from the
reduced density matrix of just two spins, and the calculation can
even be reduced to a formula involving only expectation values of
expressions at most quadratic in spin operators. It is therefore also
much easier to calculate than mutual information will prove to be
-- but even if it seems to work well for the quantum phase transition,
it is of little use to us if it does not exhibit any interesting behaviour
at finite temperature as well.

Among the other quantities that have been studied in the zero-temperature
case for the LMG model, there is one more where the definition straightforwardly
generalizes to the finite-temperature case, namely, (logarithmic)
negativity \cite{Wichterich10}. However, the calculation is again
significantly more involved, and there is probably little chance of
achieving it for thermal states.

While they do not actually qualify as entanglement measures, this
might still be a good place to mention \emph{fidelity} and related
quantities, which have also been studied as markers for phase transitions,
both quantum and thermal. Let me just point to the references \cite{Kwok08,Ma08,Quan09}.

\section{\label{sec:lmg_mi_general}Mutual information}

\begin{figure}
\includegraphics[width=1\textwidth]{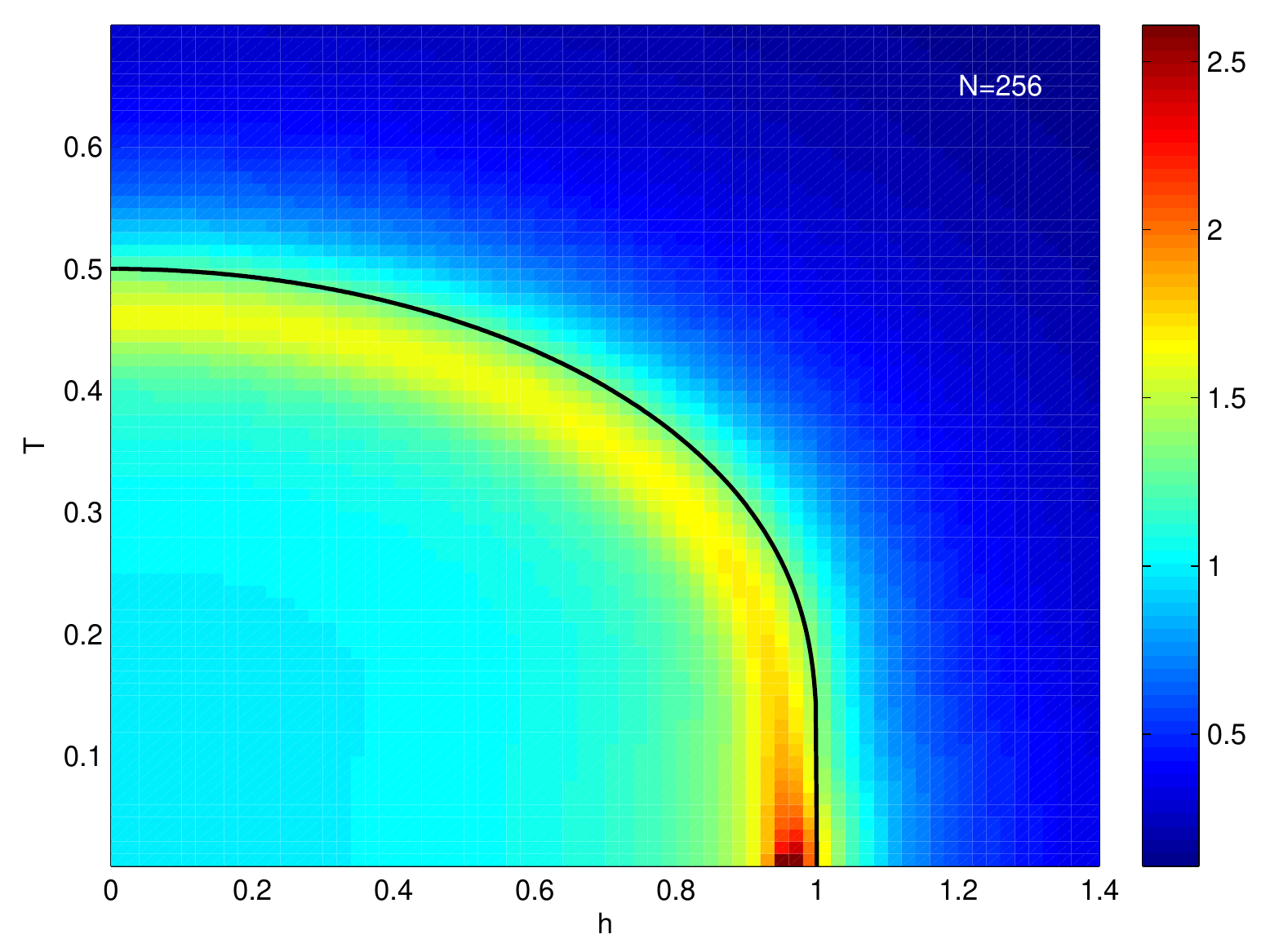}

\caption{\label{fig:lmg_mi_256}Mutual information}
\end{figure}
Now let us finally look at mutual information, as promised. Figure
\ref{fig:lmg_mi_256} shows the results; the numerical algorithm will
be described in section \ref{sec:numerical_algorithm_lmg}. Section
\ref{sec:lmg-analytical} will show how an analytical approach is
possible in the case of vanishing magnetic field $h=0,$ the left
edge of the plot. As mentioned, this case will also be called the
classical limit in the following, since there are no non-commuting
observables any more, and everything can be expressed using classical
(thermal) probability distributions.

We see that the result is rather convincing: the maximum of the mutual
information follows exactly the phase transition. Certainly this is
a much more convincing plot than the one shown for concurrence in
the previous section, further supporting our arguments from chapter
\ref{chap:mutual-information} for studying mutual information as
a measure of correlations. I also think it is a much nicer plot than
either susceptibility (figure \ref{fig:lmg_su_256}) or heat capacity
(figure \ref{fig:lmg_hc_256}).

\begin{figure}
\includegraphics[width=1\textwidth]{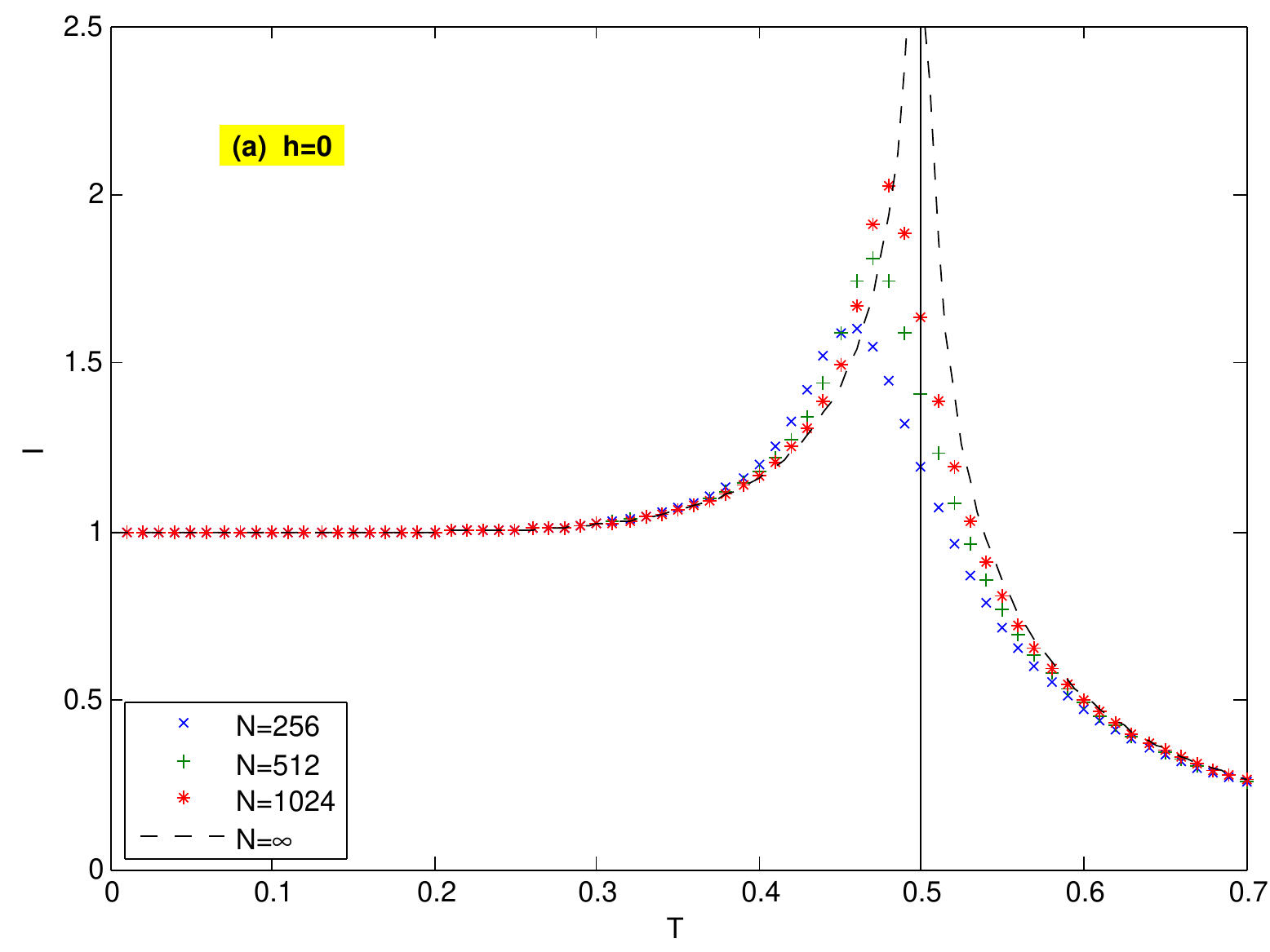}

\includegraphics[width=1\textwidth]{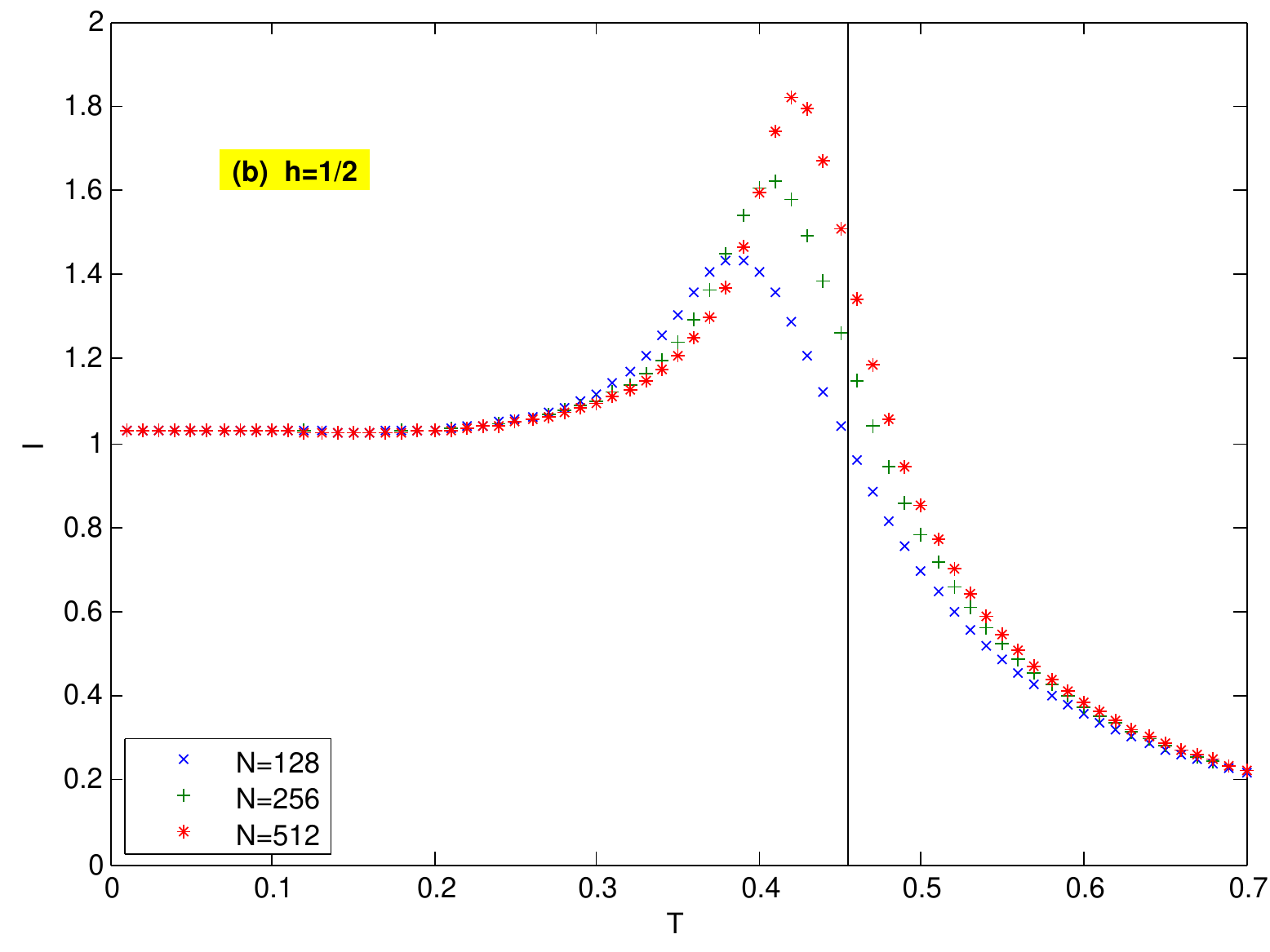}

\caption{\label{fig:lmg_mi_T}Mutual information as a function of temperature
for both the classical limit $h=0$ and for $h=1/2$. The critical
temperatures are marked by the vertical solid black lines. }
\end{figure}

Similarly to what we did in figure \ref{fig:lmg_op_T}, we can again
plot some cross-sections to get a better feeling for the behaviour
of the mutual information. Figure \ref{fig:lmg_mi_T} shows cross-sections
for constant magnetic field. Part (a) is for the special case of $h=0$
(which we now call the classical limit), and part (b) shows the rather
generic case of $h=1/2$.

As you can see, the two cases are actually not very different, but
studying the classical limit has two distinct advantages: For one,
the numerics actually become much simpler than what is described in
section \ref{sec:numerical_algorithm_lmg} for the general case: The
Hamiltonian becomes proportional to $S_{x}^{2}$, and we know that
everything becomes diagonal in the $S_{x}$-basis, so we just need
to deal with classical probability distributions rather than density
matrices. In fact, there are just $N+1$ different possible eigenvalues
and their degeneracies are easily calculated numerically as binomial
coefficients. Of course, we actually need something a bit more complicated
than that, splitting the whole system into subsystems $A$ and \textbf{$B$},
but it can be written down straightforwardly and easily implemented
numerically (which was in this case done in Mathematica, because that
allows to calculate binomial coefficients even where it becomes problematic
with IEEE754 64-bit arithmetic \cite{ieee754}, which was important
for the large-$N$ studies in section \ref{sec:mi-scaling-vs-qpt}).

The second advantage is that in this setting, it also becomes possible
to do analytical calculations extending to the thermodynamic limit
$N\to\infty$, the result of which is also shown in figure \ref{fig:lmg_mi_T}(a).
This is however far more challenging, as outlined in section \ref{sec:lmg-analytical}.
The general idea is to approximate the binomial coefficients (along
the lines of Stirling's formula), replace sums by integrals, and approximate
those integrals. For the $N=\infty$ plot in figure \ref{fig:lmg_mi_T}(a),
however, numerical evaluation of the resulting integrals was in fact
sufficient.

\section{\label{sec:mi-scaling-vs-qpt}Scaling behaviour and comparison to
the QPT}

Can we extract more predictions about the mutual information from
the analytical treatment of the classical case in section \ref{sec:lmg-analytical}?
In fact, after a lot of calculation we find a very nice critical scaling
of the mutual information (compare equation \eqref{eq:Icrit}):
\[
I(T_{\mathrm{c}})\propto\frac{1}{4}\log_{2}N.
\]
We can call the coefficient $1/4$ a critical exponent, just like
the exponents we found when studying the order parameter in section
\ref{sec:lmg-op-fss}.

\begin{figure}
\includegraphics[width=1\textwidth]{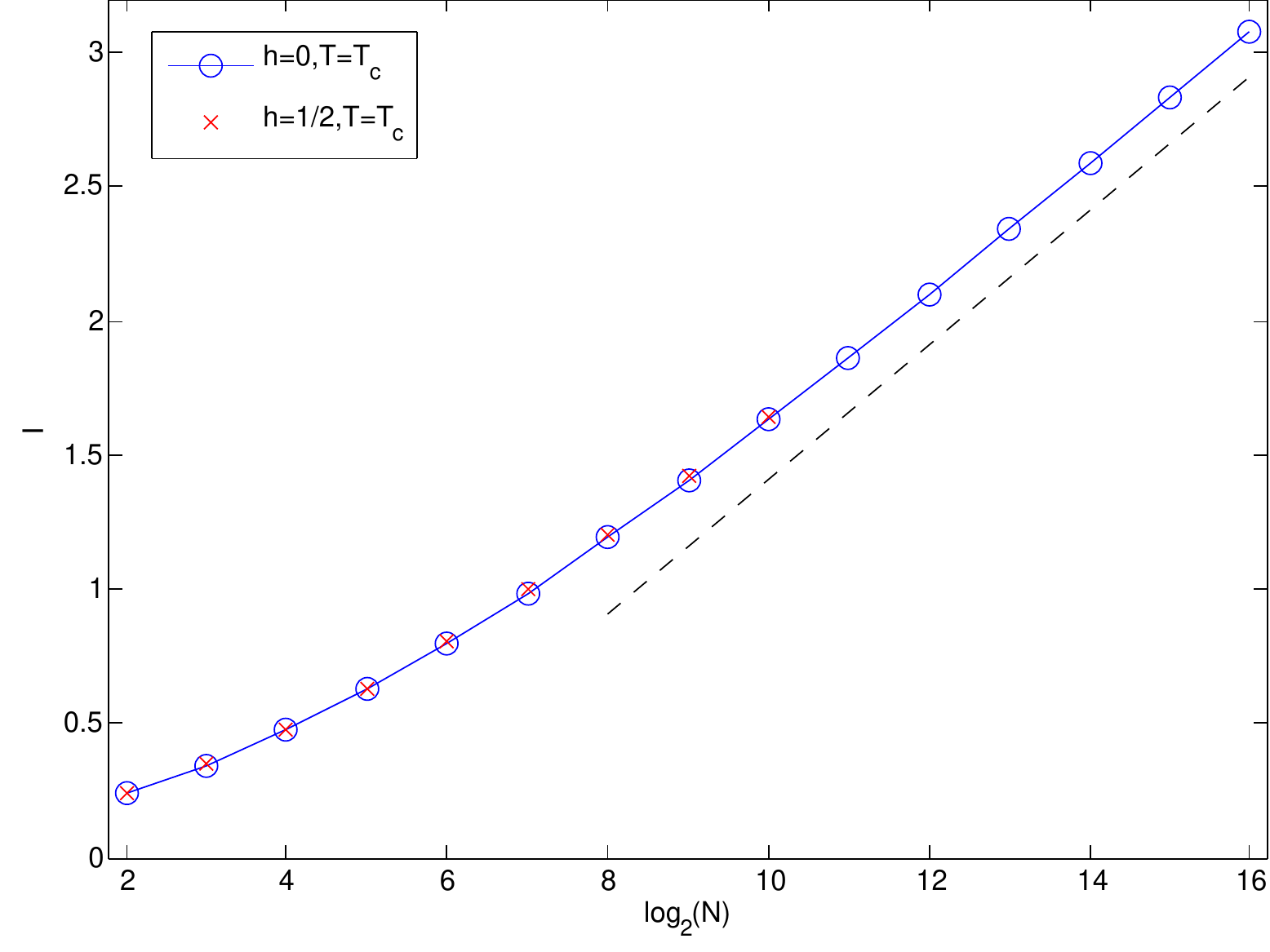}\caption{\label{fig:lmg_mi_N}Mutual information at criticality as a function
of $N$. The dashed line has a slope of exactly $1/4$.}
\end{figure}

Let us first check if our numerical simulations match the analytical
calculation. Figure \ref{fig:lmg_mi_N} shows that this is indeed
very much the case. In particular the agreement for $h=0$ is very
good for the larger $N$ we can access numerically in this case, as
explained in the previous section. For a non-zero $h$, like the $h=1/2$
displayed here, we cannot access quite such large system sizes, but
as you can see, the mutual information is almost identical for those
system sizes that we can access.

Of course this changes considerably if we get very close to the quantum
phase transition at $h=1$. Remember that there also the order parameter
showed a different critical exponent, of $-1/3$ rather than $-1/4$.
Do you want to predict what happens for the mutual information?

Since at zero temperature, mutual information is just twice the entanglement
entropy, its scaling behaviour has already been carefully studied
in \cite{Latorre05_2,Barthel06_2,Vidal07,Orus08_2}, and in fact it
has been found analytically that the entanglement entropy behaves
like $1/6\log_{2}N$, and the mutual information therefore as $1/3\log_{2}N$%
\footnote{The initial numerical study \cite{Latorre05_2} mistakenly identified
the scaling behaviour of the entanglement entropy as $1/3\log_{2}N$.
The later references contain the correct result.%
}. This parallel between the behaviour of the order parameter and the
mutual information does in fact seem extraordinarily nice.

\section{\label{sec:ising-to-xy}From Ising to XY}

So far, we considered the special case of a zero $\gamma$ in \eqref{eq:lmggeneral}.
What happens for nonzero $\gamma$ (which we could now maybe call
an anisotropy parameter)? This corresponds to going from a pure Ising-
to an XY-type model of competing $x$- and $y$-interactions; there
is now no classical limit anymore, but our numerical procedure as
described in section \ref{sec:numerical_algorithm_lmg} remains applicable.

\begin{figure}
\includegraphics[width=1\textwidth]{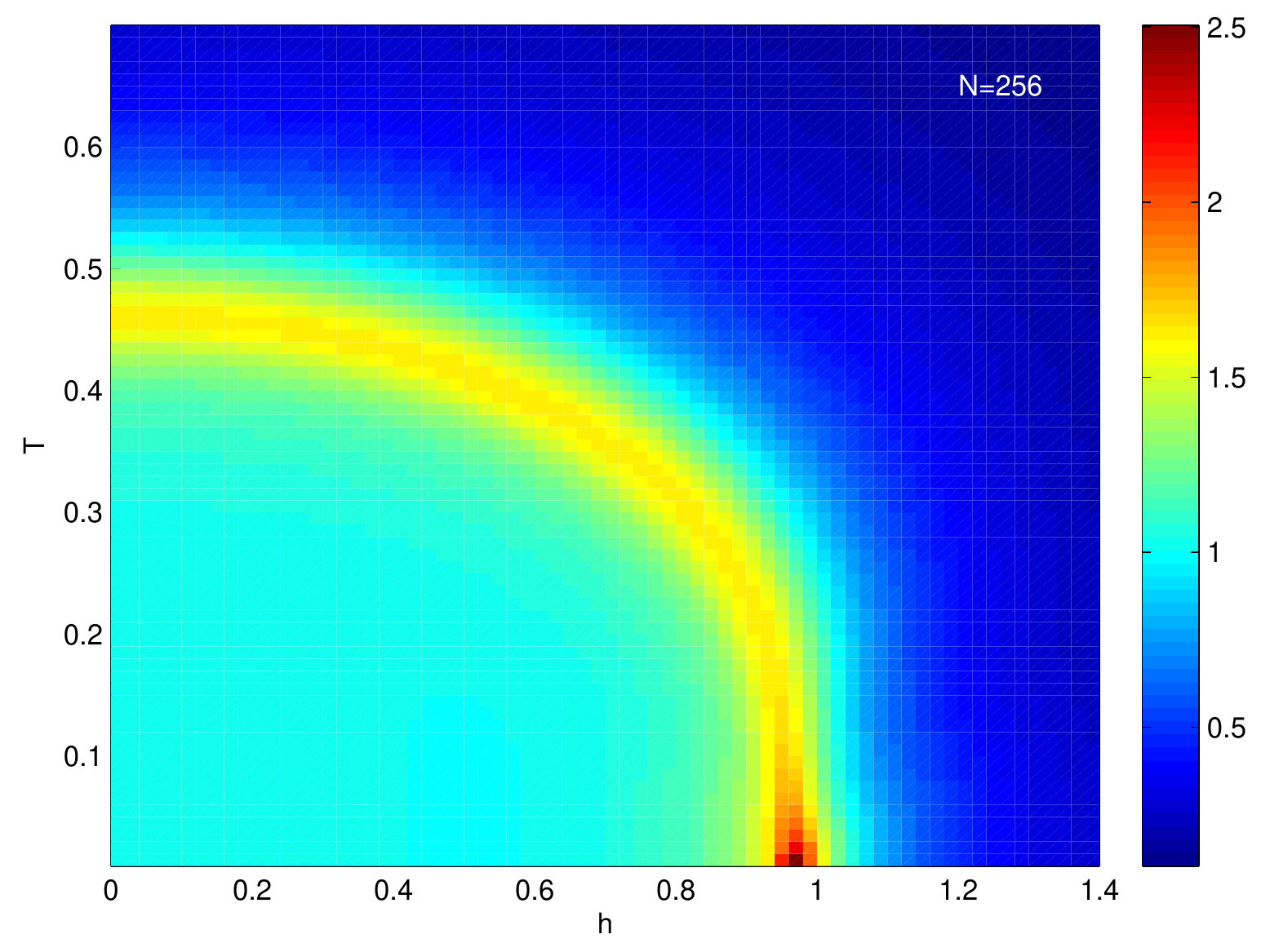}

\caption{\label{fig:lmg_mi_256_gamma025}Mutual information, for $\gamma=1/4$}
\end{figure}

\begin{figure}
\includegraphics[width=1\textwidth]{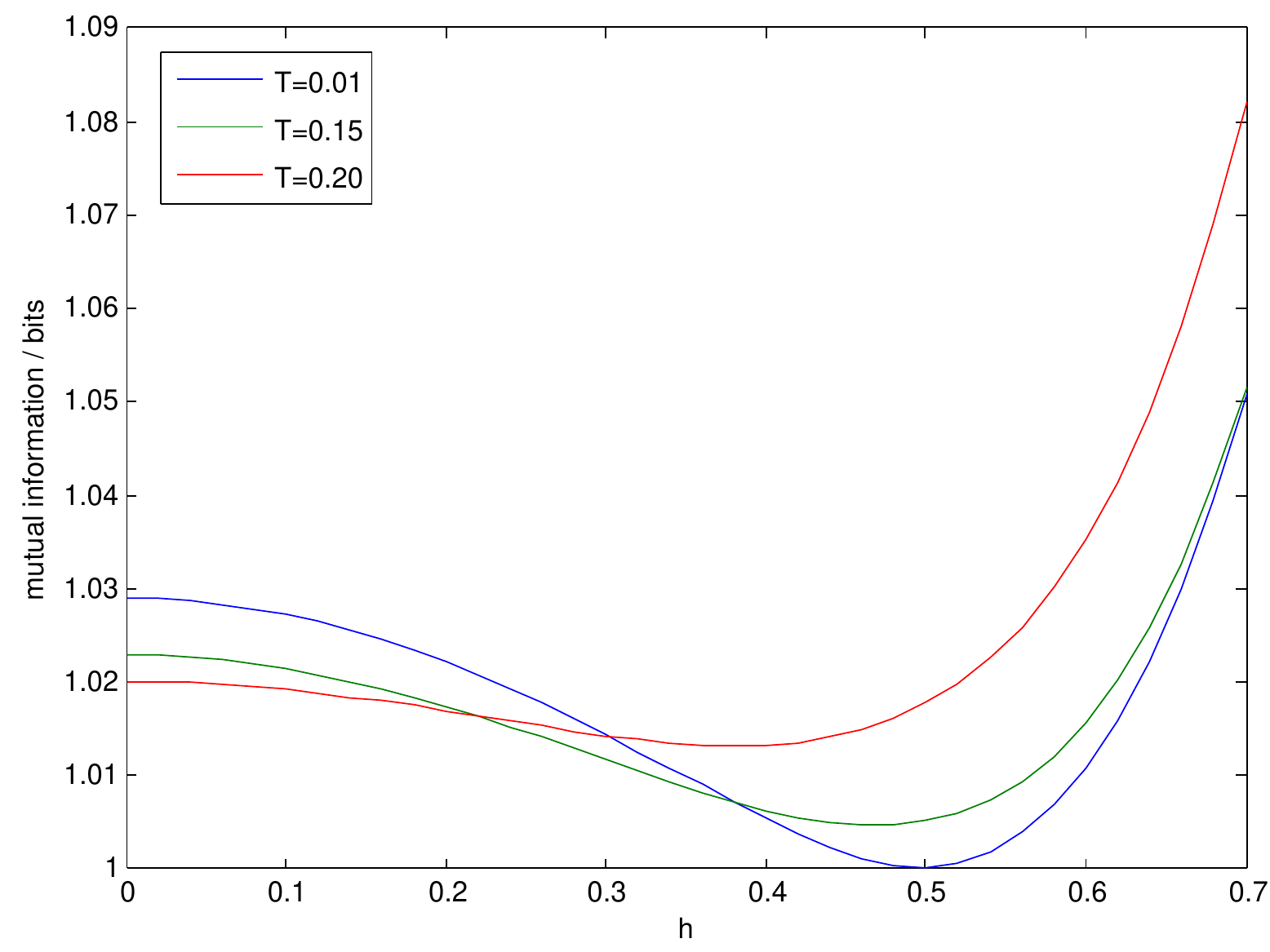}

\caption{\label{fig:lmg_kurmann}Mutual information for the case $\gamma=1/4$
as a function of magnetic field, for several {}``low'' temperatures}
\end{figure}

Figure \ref{fig:lmg_mi_256_gamma025} shows some results for $\gamma=1/4$,
and in fact not much seems to have changed. However, there is one
rather nice additional feature: If you look carefully at low temperatures
and around $h=1/2$, you notice that the mutual information seems
to be a bit lower there than at both lower and higher $h$. This is
much better visible in figure \ref{fig:lmg_kurmann}, which plots
exactly the same data, but this time only a few relevant cross-sections
at fixed low values of $T$.

In fact, this behaviour is a signature of the two-fold degenerate
and separable ground state at $h=\sqrt{\gamma}$ \cite{Dusuel05_2,Vidal07,kurmann1982},
so for our choice of $\gamma=1/4$ this occurs at $h=1/2$. We can
call this a {}``Kurmann point'', and see that the mutual information
(where we put in absolutely no knowledge of such behaviour) gives
us a good idea of how this effect still remains visible at non-zero
temperatures.

\begin{figure}
\includegraphics[width=1\textwidth]{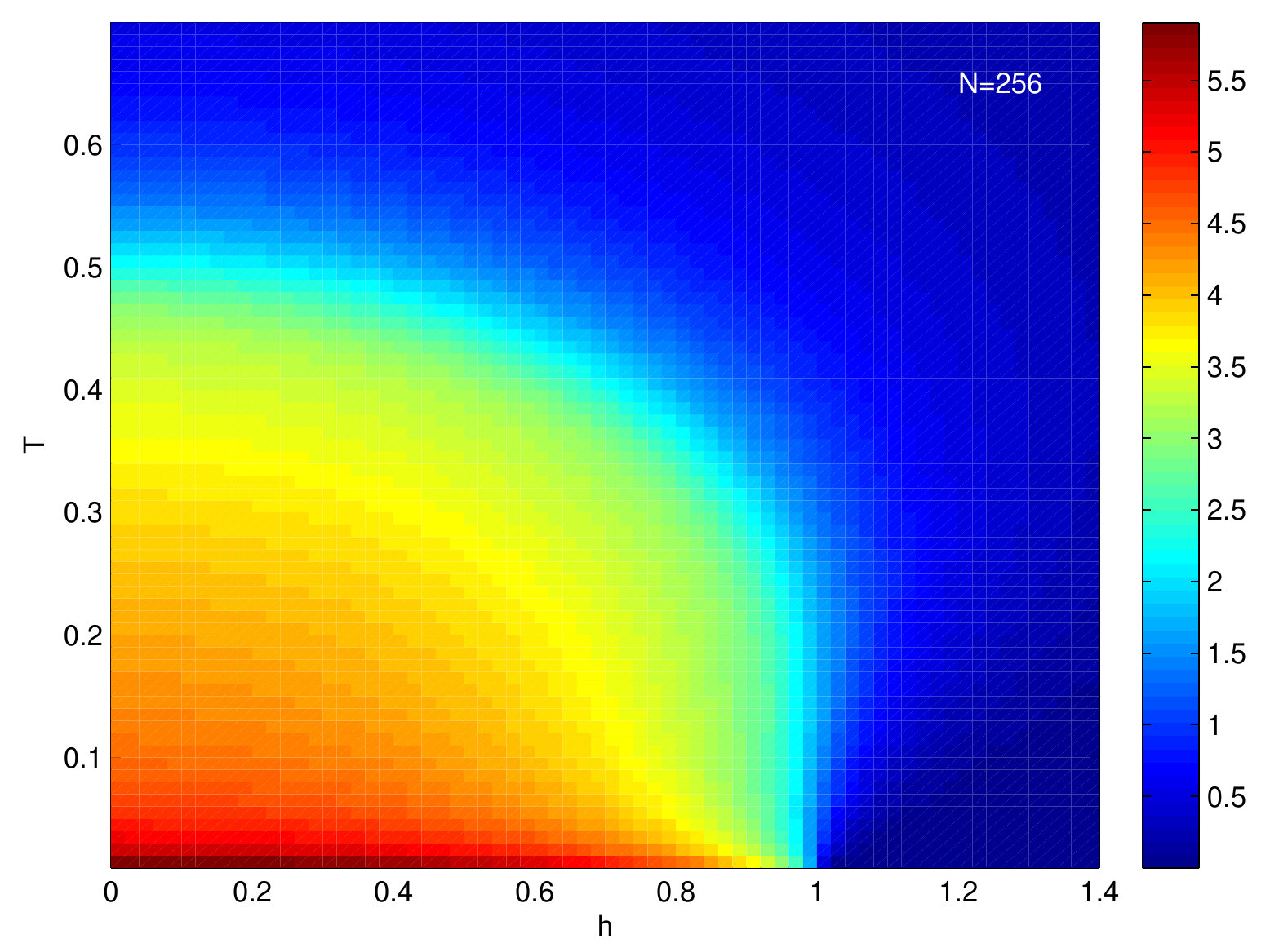}

\caption{\label{fig:lmg_mi_256_gamma1}Mutual information, for $\gamma=1$}
\end{figure}

What happens if we go all the way towards the case of an interaction
that is isotropic in $x$ and $y$? The results are shown in figure
\ref{fig:lmg_mi_256_gamma1}. Clearly, this is quite a different picture
now! The essential difference is that the Hamiltonian now has an additional
symmetry; it is isotropic in the $x$-$y$-plane. This changes the
universality class, and explains why $\gamma=1$ is qualitatively
so different from the $\gamma=0$ and $\gamma=1/4$ cases (both of
which are anisotropic and generally pretty similar). This has been
studied in the ground state case in the aforementioned references
\cite{Vidal04_1,Dusuel04_3,Dusuel05_2,Wichterich10,Latorre05_2,Barthel06_2,Vidal07,Orus08_2}.
It will however be a challenge to make any analytical statements about
the thermal case, especially as there is now no classical limit any
more.

\begin{figure}
\includegraphics[width=1\textwidth]{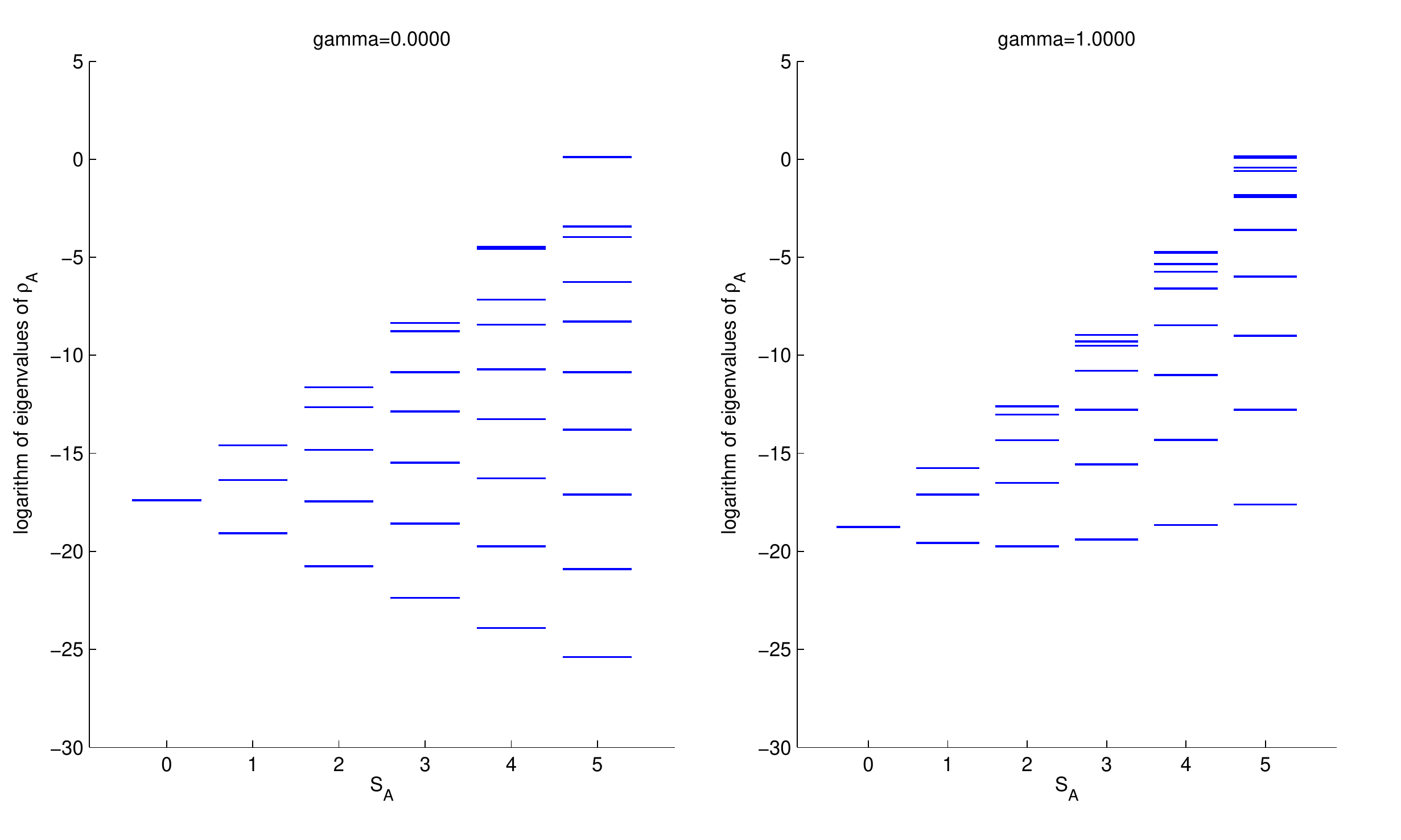}

\caption{\label{fig:lmg_ent_spectra}Entanglement spectra for $\gamma=0$ and
$\gamma=1$ contrasted, for $N=20,$ $h=1/2$, $T=0.2$}
\end{figure}

Let us also use this opportunity to point out that our methods give
us access not only to the mutual information, but really to the full
spectrum of the reduced density matrices from which it is calculated.
These {}``entanglement spectra'', where the eigenvalues of the reduced
density matrices are studied, typically sorted by a good quantum number
such as the total angular momentum in the subsystem, have received
some attention in recent years \cite{li2008entanglement,calabrese2008entanglement,fidkowski2010entanglement,turner2010entanglement}.
Unfortunately, there does not currently seem to be much theory that
we can compare the entanglement spectra of our models to. Nevertheless,
figure \ref{fig:lmg_ent_spectra} presents some finite-temperature
entanglement spectra which show that not only the mutual information,
but also the spectra themselves are significantly different between
the $\gamma=0$ and $\gamma=1$ cases.

\section{\label{sec:numerical_algorithm_lmg}Algorithm for calculating mutual
information in spin-conserving Hamiltonians}

Let us now give the algorithm we used for numerically calculating
the mutual information, as also presented in \cite{wilms2012finite}.
It works for any Hamiltonian fulfilling the essential property $[H,\boldsymbol{S}^{2}]=0$.
In section \ref{sub:lmg-blocks}, we saw how to calculate the partition
function using the block decomposition of such a Hamiltonian while
keeping track of multiplicities. With all the spin quantum numbers,
there will now be a lot of instances of the letter $S$ around. To
avoid confusion, let us therefore use a different letter for the entropies,
namely, $\mathcal{E}.$ Just like we did for the partition function
before, in equation \eqref{eq:lmgZn}, we can now write the (total)
entropy as

\begin{equation}
\mathcal{E}_{AB}=-\mathrm{Tr}[\rho\log_{2}\rho]=\sum_{s=[S]}^{S}\, d_{s}^{N}\mathcal{E}_{AB}^{(s)}\,,\label{eq:EAB}
\end{equation}
where $\rho=\tfrac{1}{Z}\mathrm{e}^{-\beta H}$ is the density matrix
and where we defined $\mathcal{E}_{AB}^{(s)}=-\mathrm{Tr}\left[\rho_{\mathrm{ref}}^{(s)}\log_{2}\rho_{\mathrm{ref}}^{(s)}\right]$
and $\rho_{\mathrm{ref}}^{(s)}=\frac{1}{Z}\mathrm{e}^{-\beta H_{\mathrm{ref}}^{(s)}}$.

\subsection*{Computation of $\mathcal{E}_{A}$ and $\mathcal{E}_{B}$}

The much more challenging part is the calculation of entropies of
the subsystems in our definition of the mutual information \eqref{eq:mi-definition-shannon}.
We split the system into two subsystems $A$ and $B$ containing $L=\tau N$
and $N-L=(1-\tau)N$ spins respectively, and aim to compute the entropies
of the reduced density matrices $\rho_{A}=\mathrm{Tr}_{B}\rho$ and
$\rho_{B}=\mathrm{Tr}_{A}\rho$.

Again, the essential idea is to work with a block decomposition of
the Hamiltonian, and density matrix. However, what makes this more
challenging now is that we have to be able to take partial traces,
which means we need a basis with a tensor product structure. Therefore,
we cannot straightforwardly work with the total spin. However, what
we can do instead is combine the spins in $A$ and $B$ separately,
to total spins which we can label $s_{1}$ and $s_{2}$. The most
straightforward thing would then be to work in the basis of states
\[
\ket{s_{1},m_{1}}\otimes\ket{s_{2},m_{2}}
\]
where the $m$s label e.g. the $z$-component of the respective spins.
In this basis, taking partial traces is straightforward (although
we would still need to keep track of all multiplicities). However,
this is not the most efficient approach, because this forces us to
work with much bigger blocks which arise in the tensor product basis. 

It is more efficient to couple $s_{1}$ and $s_{2}$ to a total spin
$s$ again and work with the density matrix in the blocks of fixed
$s.$ In order to calculate partial traces, we will then need to work
with Clebsch-Gordan coefficients describing the angular momentum coupling.
However, this can be done rather efficiently. While this basic idea
is simple enough, the formulas will unfortunately start to look rather
messy, due to the fact that not every $s_{1}$ and $s_{2}$ can be
coupled to every $s$. Let us start with the Hamiltonian, which reads

\begin{eqnarray*}
H & = & \sum_{s=[S]}^{S}\,\sum_{s_{1}=[S_{1}]}^{S_{1}}{\phantom{\sum}\hspace{-7mm}}'\,\sum_{s_{2}=[S_{2}]}^{S_{2}}{\phantom{\sum}\hspace{-7mm}}'\,\sum_{i_{1}=1}^{d_{s_{1}}^{L}}\,\sum_{i_{2}=1}^{d_{s_{2}}^{N-L}}H_{i_{1},i_{2}}^{(s_{1},s_{2};s)}\\
 & = & \sum_{s=[S]}^{S}\,\sum_{s_{1}=[S_{1}]}^{S_{1}}{\phantom{\sum}\hspace{-7mm}}'\,\sum_{s_{2}=[S_{2}]}^{S_{2}}{\phantom{\sum}\hspace{-7mm}}'\,\sum_{i_{1}=1}^{d_{s_{1}}^{L}}\,\sum_{i_{2}=1}^{d_{s_{2}}^{N-L}}\sum_{m=-s}^{s}\,\sum_{m'=-s}^{s}\,\\
 &  & h_{m',m}^{(s)}\,\ket{s_{1},s_{2};s,m'}_{i_{1},i_{2}}\,\,{}_{i_{1},i_{2}}\bra{s_{1},s_{2};s,m},
\end{eqnarray*}
where $\ket{s_{1},s_{2};s,m}_{i_{1},i_{2}}$ denotes an eigenstate
of operators $\mathbf{S}_{1}^{2},\mathbf{\, S}_{2}^{2},\,\mathbf{S}^{2}$
and $S_{z}$ with eigenvalues $s_{1}(s_{1}+1)$, $s_{2}(s_{2}+1)$,
$s(s+1)$ and $m$ respectively. Index $i_{1}$ ($i_{2}$) labels
the $d_{s_{1}}^{L}$ ($d_{s_{2}}^{N-L}$) degenerate subspaces of
spin $s_{1}$ ($s_{2}$) that can be built from $L$ ($N-L$) spins
1/2. For a given $s$, primed sums are restricted to values of $s_{1}$
and $s_{2}$ that can add to a total spin $s$. That is to say, one
must fulfill the inequalities $|s_{1}-s_{2}|\leqslant s\leqslant s_{1}+s_{2}$,
so one can alternatively write:

\[
\sum_{s=[S]}^{S}\,\sum_{s_{1}=[S_{1}]}^{S_{1}}{\phantom{\sum}\hspace{-7mm}}'\,\sum_{s_{2}=[S_{2}]}^{S_{2}}{\phantom{\sum}\hspace{-7mm}}'\,=\sum_{s_{1}=[S_{1}]}^{S_{1}}\,\sum_{s_{2}=[S_{2}]}^{S_{2}}\,\sum_{s=|s_{1}-s_{2}|}^{s_{1}+s_{2}}.
\]
We have denoted as \mbox{$S=N/2$}, $S_{1}=L/2$ and $S_{2}=(N-L)/2$
the maximum spins of the whole system and of each of the subsystems.
As before, minimum spins are denoted with square brackets. Note that
the degeneracy $d_{s}^{N}$ of the spin-$s$ sector can be recovered
from

\[
\sum_{s_{1}=[S_{1}]}^{S_{1}}{\phantom{\sum}\hspace{-7mm}}'\,\sum_{s_{2}=[S_{2}]}^{S_{2}}{\phantom{\sum}\hspace{-7mm}}'\, d_{s_{1}}^{L}d_{s_{2}}^{N-L}=d_{s}^{N}.
\]
The matrix elements $h_{m',m}^{(s)}$ are the same as in formula \eqref{eq:def1},
and do not depend on $s_{1}$, $s_{2}$, $i_{1}$ or $i_{2}$. Therefore,
all Hamiltonians $H_{i_{1},i_{2}}^{(s_{1},s_{2};s)}$ have the same
eigenvalues $E_{\alpha}^{(s)}$ (which are the same as in equation
\eqref{eq:diag}), and corresponding eigenvectors

\[
\ket{s_{1},s_{2};s;\alpha}_{i_{1},i_{2}}=\sum_{m=-s}^{s}a_{\alpha;m}^{(s)}\ket{s_{1},s_{2};s,m}_{i_{1},i_{2}}.
\]
Once again, coefficients $a_{\alpha;m}^{(s)}\in\mathbb{C}$ are independent
of $s_{1}$, $s_{2}$, $i_{1}$ and $i_{2}$, and are the same as
in equation \eqref{eq:evec}.

Now, the density matrix $\rho=\frac{1}{Z}\mathrm{e}^{-\beta H}$ reads

\begin{eqnarray*}
\rho & = & \frac{1}{Z}\sum_{s_{1}=[S_{1}]}^{S_{1}}\,\sum_{s_{2}=[S_{2}]}^{S_{2}}\,\sum_{s=|s_{1}-s_{2}|}^{s_{1}+s_{2}}\,\sum_{i_{1}=1}^{d_{s_{1}}^{L}}\,\sum_{i_{2}=1}^{d_{s_{2}}^{N-L}}\,\sum_{\alpha=1}^{2s+1}\,\mathrm{e}^{-\beta E_{\alpha}^{(s)}}\ket{s_{1},s_{2};s;\alpha}_{i_{1},i_{2}}\,\,{}_{i_{1},i_{2}}\bra{s_{1},s_{2};s;\alpha}\\
 & = & \frac{1}{Z}\sum_{s_{1}=[S_{1}]}^{S_{1}}\,\sum_{s_{2}=[S_{2}]}^{S_{2}}\,\sum_{s=|s_{1}-s_{2}|}^{s_{1}+s_{2}}\,\sum_{i_{1}=1}^{d_{s_{1}}^{L}}\,\sum_{i_{2}=1}^{d_{s_{2}}^{N-L}}\,\sum_{\alpha=1}^{2s+1}\,\sum_{m=-s}^{s}\,\sum_{m'=-s}^{s}\,\\
 &  & \mathrm{e}^{-\beta E_{\alpha}^{(s)}}{a_{\alpha;m}^{(s)}}^{\!\!\!*}a_{\alpha;m'}^{(s)}\ket{s_{1},s_{2};s,m'}_{i_{1},i_{2}}\,\,{}_{i_{1},i_{2}}\bra{s_{1},s_{2};s,m}.\qquad
\end{eqnarray*}
where $a^{*}$ denotes the complex conjugate of $a$.

Now we have to decompose each basis state $\ket{s_{1},s_{2};s,m'}_{i_{1},i_{2}}$
into the tensor product state basis using Clebsch-Gordan coefficients

\begin{eqnarray*}
 &  & \ket{s_{1},s_{2};s,m'}_{i_{1},i_{2}}=\sum_{m_{1}=-s_{1}}^{s_{1}}\,\sum_{m_{2}=-s_{2}}^{s_{2}}C_{s_{1},s_{2};s}^{m_{1},m_{2};m}\ket{s_{1},m_{1}}_{i_{1}}\otimes\ket{s_{2},m_{2}}_{i_{2}}\\
\end{eqnarray*}
with obvious notations. The only pairs $(m_{1},m_{2})$ that contribute
to the above sum are such that $m_{1}+m_{2}=m$, since the Clebsch-Gordan
coefficients vanish otherwise. Introducing the shorthand notation

\[
\sum_{\mathrm{all}}\,=\,\sum_{s_{1}=[S_{1}]}^{S_{1}}\,\sum_{s_{2}=[S_{2}]}^{S_{2}}\,\sum_{s=|s_{1}-s_{2}|}^{s_{1}+s_{2}}\,\sum_{i_{1}=1}^{d_{s_{1}}^{L}}\,\sum_{i_{2}=1}^{d_{s_{2}}^{N-L}}\,\sum_{\alpha=1}^{2s+1}\,\sum_{m=-s}^{s}\,\sum_{m'=-s}^{s}\,\sum_{m_{1}=-s_{1}}^{s_{1}}\,\sum_{m_{2}=-s_{2}}^{s_{2}}\,\sum_{m_{1}'=-s_{1}}^{s_{1}}\,\sum_{m_{2}'=-s_{2}}^{s_{2}},
\]
we get

\begin{equation}
\rho=\frac{1}{Z}\sum_{\mathrm{all}}\mathrm{e}^{-\beta E_{\alpha}^{(s)}}{a_{\alpha;m}^{(s)}}^{\!\!\!*}a_{\alpha;m'}^{(s)}C_{s_{1},s_{2};s}^{m_{1},m_{2};m}C_{s_{1},s_{2};s}^{m_{1}',m_{2}';m'}\ket{s_{1},m_{1}'}_{i_{1}}\,\,{}_{i_{1}}\bra{s_{1},m_{1}}\otimes\ket{s_{2},m_{2}'}_{i_{2}}\,\,{}_{i_{2}}\bra{s_{2},m_{2}}.
\end{equation}
It is now possible to perform the partial traces. Let us focus on
computing $\rho_{A}=\mathrm{Tr}_{B}\rho$. This partial trace will
enforce $m_{2}'=m_{2}$ in $\sum_{\mathrm{all}}$. Furthermore, the
index $i_{2}$ disappears from the quantity to be summed over, so
that $\sum_{i_{2}=1}^{d_{s_{2}}^{N-L}}$ simply yields a factor $d_{s_{2}}^{N-L}$.
At the end of the day, one finds

\begin{eqnarray}
\rho_{A} & = & \frac{1}{Z}\sum_{s_{1}=[S_{1}]}^{S_{1}}\,\sum_{i_{1}=1}^{d_{s_{1}}^{L}}\,\sum_{m_{1}=-s_{1}}^{s_{1}}\,\sum_{m_{1}'=-s_{1}}^{s_{1}}{r_{A}}_{m_{1}',m_{1}}^{(s_{1})}|s_{1},m_{1}'\rangle_{i_{1}}\,\,{}_{i_{1}}\langle s_{1},m_{1}|\quad\mbox{with}\nonumber \\
{r_{A}}_{m_{1}',m_{1}}^{(s_{1})} & = & \sum_{s_{2}=[S_{2}]}^{S_{2}}\, d_{s_{2}}^{N-L}\,\sum_{s=|s_{1}-s_{2}|}^{s_{1}+s_{2}}\,\sum_{m=\min_{m}}^{\max_{m}}\,\left(\sum_{\alpha=1}^{2s+1}\,\mathrm{e}^{-\beta E_{\alpha}^{(s)}}{a_{\alpha;m}^{(s)}}^{\!\!\!*}a_{\alpha;m+m_{1}'-m_{1}}^{(s)}\right)\nonumber \\
 &  & C_{s_{1},s_{2};s}^{m_{1},m-m_{1};m}\, C_{s_{1},s_{2};s}^{m_{1}',m-m_{1};m+m_{1}'-m_{1}}\qquad\quad\label{eq:reduceddensity}
\end{eqnarray}
with

\begin{eqnarray*}
{\min}_{m} & = & \max(-s,-s+m_{1}-m_{1}',-s_{2}+m_{1}),\\
{\max}_{m} & = & \min(s,s+m_{1}-m_{1}',s_{2}+m_{1}).
\end{eqnarray*}
These limits come from the fact that not all values of $m$ are allowed
in between $-s$ and $s$, but only those that fulfill \mbox{$-s_{2}\leqslant m-m_{1}\leqslant s_{2}$}
and \mbox{$-s\leqslant m+m_{1}'-m_{1}\leqslant s$}. Let us remark
that, here, we have got rid of the sum over $m_{2}$ and kept the
sum over~$m$. One could have of course done the opposite, resulting
in a somewhat different implementation.

Expression \eqref{eq:reduceddensity} of $\rho_{A}$ allows one to
numerically compute $\mathcal{E}_{A}$. The partial entropy $\mathcal{E}_{B}$
can be computed similarly. We already saw $\mathcal{E}_{AB}$ in formula
\eqref{eq:EAB}. Once we have all of these, the mutual information
$I$ is simply obtained from its definition \eqref{eq:mi-definition-shannon}.

It should be noted that the above algorithm requires the numerical
computation of many Clebsch-Gordan coefficients. To see why that is
important, let us estimate the computational complexity of our algorithm.
Looking at equation \eqref{eq:reduceddensity}, one sees that eight
sums are involved. However, the sum over $i_{1}$ will only yield
a multiplicity, which leaves seven sums. Furthermore, the matrices
created by the sum over $\alpha$ (which is enclosed in parentheses
above)

\[
\sum_{\alpha=1}^{2s+1}\,\mathrm{e}^{-\beta E_{\alpha}^{(s)}}{a_{\alpha;m}^{(s)}}^{\!\!\!*}a_{\alpha;m'}^{(s)},
\]
can be computed once (for all the different values of $s$), and stored
in memory. These matrices are nothing but the density matrices in
the spin-$s$ sectors (up to a factor of $1/Z$). The computational
cost of this is comparatively negligible, and one is left with a sum
over six variables, namely $s_{1}$, $m_{1}$, $m_{1}'$, $s_{2}$,
$s$ and $m$, which all take a number of values scaling linearly
with $N$ (at most). So we expect that our algorithm roughly scales
as $N^{6}$ in computation time. However, there we made an important
assumption, namely that we already know the Clebsch-Gordon coefficients.
Indeed, if we count them, we realize that we actually need in the
order of $N^{5}$ different ones of them. That means we need a very
efficient way of calculating them, otherwise this becomes the bottleneck
of the algorithm. In fact, even calculating all the coefficients for
a given system size in advance and storing them is not attractive:
we would require a huge amount of storage and reading from such storage
would not be very fast.

It is however indeed possible to calculate Clebsch-Gordon coefficients
efficiently, with just a small and constant number of operations per
coefficient. To do this, one needs to calculate them recursively.
However, the required recursion has nontrivial stability properties;
it is only stable in some of the possible directions, otherwise numerical
inaccuracies pile up very quickly. In the end, we arrived at the same
algorithm as is also described by Schulten and Gordon \cite{schulten1975}.

\section{\label{sec:lmg-analytical}Analytical calculations in the classical
limit}

In this section (which has been relegated to the end of the chapter
like the previous one, in order not to interrupt the discussion of
results), let us understand the analytical treatment of the classical
(quantum fluctuation free) case $h=0$. In this case, the LMG Hamiltonian
\eqref{eq:lmggeneral} simplifies to 
\[
H=-\frac{S_{x}^{2}}{N}\;.
\]
Its eigenstates are now those of $S_{x}$, and can be chosen as separable
states $|p,i\rangle$, with $p$ spins pointing in the $-x$ direction
and $(N-p)$ spins pointing in the $+x$ direction. The variable $i$
allows to distinguish between the $\left(\begin{array}{c}
N\\
p
\end{array}\right)$ such states which are degenerate and have eigenenergy $E_{p}=-N\left(\tfrac{p}{N}-\tfrac{1}{2}\right)^{2}$.

To obtain the mutual information, again one first needs to compute
the total entropy of the system at finite temperature. Let us again
start with the partition function 
\begin{equation}
Z(\beta)=\mathrm{Tr}\:{\rm e}^{-\beta H}=\sum_{p=0}^{N}\left(\begin{array}{c}
N\\
p
\end{array}\right){\rm e}^{-\beta E_{p}},\label{eq:Z_zero_field}
\end{equation}
where as before, $\beta=1/T$ denotes the inverse temperature $(k_{B}=1)$.
Despite the apparent simplicity of $H$, it is difficult to compute
$Z$ analytically for arbitrary finite $N$. That is why we will now
focus on the physically relevant large-$N$ limit to analyze the thermodynamical
limit and its neighbourhood. Computations get easier in this limit
because the discrete sum in \eqref{eq:Z_zero_field} can be approximated
by an integral with, as is well-known, an error of the order $1/N^{2}$.
This is fortunate since, in order to compute the mutual information
in the present problem, we will need to compute the first correction
to the infinite $N$ limit.

The very first step in the calculation is to perform a large-$N$
expansion of the binomials 
\begin{equation}
\left(\begin{array}{c}
N\\
p
\end{array}\right)=\frac{N!}{p!\:(N-p)!}=\frac{\Gamma(N+1)}{\Gamma(p+1)\Gamma(N-p+1)}.\label{eq:bin}
\end{equation}
 Using the series expansion of the Euler $\Gamma$ function at large
$N$ (or the Stirling formula) and skipping terms of relative order
$1/N^{2}$, one gets 
\begin{eqnarray}
\left(\begin{array}{c}
N\\
p
\end{array}\right) & = & \sqrt{\frac{2}{\pi N}}\frac{1}{\sqrt{1-4\varepsilon^{2}}}\times{\rm e}^{-N\left[\left(\frac{1}{2}-\varepsilon\right)\log\left((\frac{1}{2}-\varepsilon\right)+\left(\frac{1}{2}+\varepsilon\right)\log\left(\frac{1}{2}+\varepsilon\right)\right]}\nonumber \\
 &  & \times\left[1-\frac{1}{N}\frac{3+4\varepsilon^{2}}{12(1-4\varepsilon^{2})}+O\left(1/N^{2}\right)\right]\label{eq:bin_approx}
\end{eqnarray}
where we set $\varepsilon=\frac{p}{N}-\frac{1}{2}$. Then, replacing
the sum by an integral in equation \eqref{eq:Z_zero_field} gives

\begin{eqnarray*}
Z(\beta) & = & 2^{N}\sqrt{\frac{2N}{\pi}}\int_{-\frac{1}{2}}^{+\frac{1}{2}}\frac{{\rm d}\varepsilon}{\sqrt{1-4\varepsilon^{2}}}\times\left[1-\frac{1}{N}\frac{3+4\varepsilon^{2}}{12(1-4\varepsilon^{2})}+O\left(1/N^{2}\right)\right]\mathrm{e}^{-N\varphi(\varepsilon)},\quad\mathrm{\textrm{with}}\\
\varphi(\varepsilon) & = & \left(\frac{1}{2}-\varepsilon\right)\log\left(\frac{1}{2}-\varepsilon\right)+\left(\frac{1}{2}+\varepsilon\right)\log\left(\frac{1}{2}+\varepsilon\right)+\log2-\beta\varepsilon^{2}.
\end{eqnarray*}
Furthermore, the exponential term $\mathrm{e}^{-N\varphi(\varepsilon)}$
allows us to extend the integration range to $\mathbb{R}$. Indeed,
the effect of this extension consists in exponentially small terms
$(\propto\mathrm{e}^{-\alpha N}$) which are negligible compared to
the error we already made by approximating the sum by an integral.
The resulting integral can be evaluated using the standard Laplace's
method (also known as saddle-point or stationary-phase approximation)
though one has to take care of computing the subleading corrections
\cite{Bender99}.

At and above the critical temperature ($\beta\leqslant\beta_{\mathrm{c}}(h=0)=2$),
the minimum of $\varphi(\varepsilon)$ is found for $\varepsilon=0$
and the result of the saddle-point approximation reads

\begin{equation}
Z(\beta<2)=2^{N}\sqrt{\frac{2}{2-\beta}}\left[1-\frac{1}{N}\frac{\beta^{2}}{4(2-\beta)^{2}}+O(1/N^{2})\right],\label{eq:Z_highT}
\end{equation}
and

\begin{eqnarray}
Z(\beta=2) & = & 2^{N-1}\frac{3^{1/4}N^{1/4}\Gamma(1/4)}{\sqrt{\pi}}\bigg[1+\frac{2\sqrt{3}\:\Gamma(3/4)}{5\sqrt{N}\:\Gamma(1/4)}\nonumber \\
 &  & -\frac{1}{280N}-\frac{\Gamma(3/4)}{20\sqrt{3}N^{3/2}\:\Gamma(1/4)}+O(1/N^{2})\bigg].\nonumber \\
\end{eqnarray}
Note that, in this approach, $Z(\beta=2)$ does not coincide with
$\lim_{\beta\to2}Z(\beta<2)$ (which is divergent) since the integral
arising from the saddle-point approximation is not Gaussian anymore
at criticality. Furthermore, we see that $Z(\beta=2)/2^{N}$ diverges
with a non-trivial finite-size scaling exponent, as $N^{1/4}$. This
result, which is obtained by directly computing $Z(\beta=2)$, is
consistent with equation \eqref{eq:Z_highT}. Indeed, adapting the
finite-size scaling argument developed in references \cite{Dusuel04_3,Dusuel05_2}
for the zero-temperature problem, one can write (for $\beta$ smaller
than $2$ but close to 2) $Z(\beta\lesssim2)/2^{N}=z_{\infty}f[N(2-\beta)^{2}]$
where $z_{\infty}=\sqrt{2/(\beta-2)}$ is the thermodynamical limit
value of $Z(\beta<2)/2^{N}$, and $f$ is a scaling function. As this
quantity cannot be singular for finite $N$, one must have $f(x)\sim x^{1/4}$
so that the singularity at $\beta=2$ disappears~: $Z(\beta\lesssim2)/2^{N}\sim(\beta-2)^{-1/2}[N(2-\beta)^{2}]^{1/4}\sim N^{1/4}$.

In the low-temperature phase ($\beta>2$), $\varphi(\varepsilon)$
has two symmetric minima, the position of which can only be computed
numerically. Once determined, one can still perform the Gaussian integral
resulting from the second-order Taylor expansion around these minima,
which yields a {}``numerically exact\textquotedbl{} result in the
infinite-$N$ limit. As a consequence, we only give analytical expressions
for $\beta\leqslant2$, but we can still obtain exact numerical results
for $\beta>2$, as we saw e.g in figure \ref{fig:lmg_mi_T}(a).

To compute the entropy in the thermodynamical limit, let us start
by computing the internal energy of the system using the same approximation
\eqref{eq:bin_approx} of the binomials. The internal energy is

\[
U(\beta)=\mathrm{Tr}(\rho H)=-\frac{\partial\ln Z(\beta)}{\partial\beta}
\]
where $\rho=\tfrac{1}{Z}\mathrm{e}^{-\beta H}$ is the thermal density
matrix. For $\beta<2$, it can be obtained from equation \eqref{eq:Z_highT}
but, as before, special care must be taken to deal with the critical
case for which one can show that

\begin{eqnarray*}
U(\beta=2) & = & -\frac{\sqrt{3N}\:\Gamma(3/4)}{2\Gamma(1/4)}\bigg[1-\frac{2\sqrt{3}\:\Gamma(3/4)}{5\sqrt{N}\:\Gamma(1/4)}+12\frac{\Gamma(1/4)^{2}+7\:\Gamma(3/4)^{2}}{175N\:\Gamma(1/4)^{2}}\\
 &  & -\frac{10\:\Gamma(1/4)^{4}+32\:\Gamma(1/4)^{2}\:\Gamma(3/4)^{2}+504\:\Gamma(3/4)^{4}}{875\sqrt{3}N^{3/2}\:\Gamma(1/4)^{3}\Gamma(3/4)}+O(1/N^{2})\bigg]
\end{eqnarray*}
 so that the internal energy diverges as $N^{1/2}$.

The desired entropy can now be computed (analytically for $\beta\leqslant2$
and numerically for $\beta>2$) as

\begin{equation}
\mathcal{E}_{AB}(\beta)=-\mathrm{Tr}(\rho\log_{2}\rho)=\log_{2}Z(\beta)+\frac{\beta U(\beta)}{\log2}.\label{eq:entrop}
\end{equation}
Computing the mutual information is still a challenge, the most difficult
part of which lies in the derivation of the large-$N$ behavior of
the subsystem entropies like

\[
\mathcal{E}_{A}(\beta)=-\mathrm{Tr}(\rho_{A}\log_{2}\rho_{A})
\]
where $\rho_{A}=\mathrm{Tr}_{B}\rho$ is the reduced density matrix
(and correspondingly for subsystem $B$). It is therefore mandatory
to find an expression for $\rho_{A}$ (and $\rho_{B}$). With this
aim in mind, let us split the system into two parts $A$ and $B$
containing $N_{A}$ and $N_{B}=N-N_{A}$ spins respectively. Next,
let us decompose the eigenstates of $H$ as \mbox{$|p,i\rangle=|p_{A},i_{A}\rangle_{A}\otimes|p_{B},i_{B}\rangle_{B}$}
with $p_{A}+p_{B}=p$ and where $|p_{j},i_{j}\rangle_{j}$ denotes
a state of subsystem $j=A,B$ that has $p_{j}$ spins pointing in
the $-x$ direction. Variable $i_{j}$ can take $\left(\begin{array}{c}
N_{j}\\
p_{j}
\end{array}\right)$ values, describing which spins point in the $-x$ direction. This
decomposition allows one to write the reduced density matrix as

\begin{eqnarray*}
\rho_{A} & = & \frac{1}{Z}\mathrm{Tr}_{B}\sum_{p,i}\mathrm{e}^{N\beta\left(\frac{p}{N}-\frac{1}{2}\right)^{2}}|p,i\rangle\langle p,i|,\\
 & = & \frac{1}{Z}\mathrm{Tr}_{B}\sum_{p_{A},i_{A},p_{B},i_{B}}\mathrm{e}^{N\beta\left(\frac{p_{A}}{N}+\frac{p_{B}}{N}-\frac{1}{2}\right)^{2}}\times\\
 &  & \quad|p_{A},i_{A}\rangle_{A}\otimes|p_{B},i_{B}\rangle_{B}\,\mbox{}_{A}\langle p_{A},i_{A}|\otimes\mbox{}_{B}\langle p_{B},i_{B}|,\\
 & = & \sum_{p_{A},i_{A}}R(p_{A})|p_{A},i_{A}\rangle_{A}\,\mbox{}_{A}\langle p_{A},i_{A}|.
\end{eqnarray*}
Here, we introduced the quantity

\[
R(p_{A})=\frac{1}{Z}\sum_{p_{B}=0}^{N_{B}}\left(\begin{array}{c}
N_{B}\\
p_{B}
\end{array}\right)\mathrm{e}^{N\beta\left(\frac{p_{A}}{N}+\frac{p_{B}}{N}-\frac{1}{2}\right)^{2}}\;.
\]
The subsystem entropy is then

\[
\mathcal{E}_{A}(\beta)=-\sum_{p_{A}=0}^{N_{A}}\left(\begin{array}{c}
N_{A}\\
p_{A}
\end{array}\right)R(p_{A})\log_{2}R(p_{A})\;.
\]
Following the same line of reasoning as for the partition function
calculation and replacing the binomials by the same form as (\ref{eq:bin_approx}),
one can again use Laplace's method to obtain the large-$N$ behavior
of the partial entropy. Things are now rather more involved since
one has to deal with a double sum and therefore with a two-variable
integral. After some algebra, we get

\begin{eqnarray*}
\mathcal{E}_{A}(\beta<2) & = & \tau N-\frac{1}{2\log2}\bigg\{\frac{\beta\:\tau}{2-\beta}+\log\bigg[\frac{2-\beta}{2-\beta(1-\tau)}\bigg]\bigg\}+O(1/N)\\
\end{eqnarray*}
and

\[
\mathcal{E}_{A}(\beta=2)=\tau N-\tau\sqrt{N}\frac{\sqrt{3}}{\log2}\frac{\Gamma(3/4)}{\Gamma(1/4)}+\frac{1}{4}\log_{2}N+O(N^{0})
\]
where we have identified $N_{A}/N=\tau$, and $N_{B}/N=1-\tau$.

Finally, noting that $\mathcal{E}_{B}$ is obtained from $\mathcal{E}_{A}$
by exchanging $\tau\leftrightarrow(1-\tau)$ and using (\ref{eq:entrop}),
one obtains the following expressions of the mutual information

\begin{equation}
I(\beta<2)=\frac{1}{2}\log_{2}\bigg\{\frac{[2-\beta\:\tau][2-\beta(1-\tau)]}{2(2-\beta)}\bigg\}+O(1/N),\label{eq:mi_ana}
\end{equation}
and

\begin{equation}
I(\beta=2)=\frac{1}{4}\log_{2}N+O(N^{0}).\label{eq:Icrit}
\end{equation}

\chapter{\label{chap:m-n-models}$m,n$-models}

In this chapter, we will consider a class of generalizations of the
Lipkin-Meshkov-Glick model that was discussed in the previous chapter.
Again, the ground state entanglement properties of these models have
received some attention recently \cite{Filippone11}. We will now
examine what happens in the finite-temperature case.

\section{Hamiltonians}

Let us consider the Hamiltonian

\[
H=-N(\cos\omega(2S_{x}/N)^{m}+K_{m,n}\sin\omega(2S_{z}/N)^{n})
\]
where

\begin{eqnarray*}
K_{2,1} & = & 2\;,\\
K_{m>2,1} & = & m^{m/2}(m-2)^{m/2-1}(m-1)^{1-m}\;,\\
K_{m\geq2,n\geq2} & = & 1\;.
\end{eqnarray*}
These values have been chosen such that the relevant parameter range
for $\omega$ is always $[0,\pi/2]$ and the zero-temperature phase
transition always occurs exactly in the middle of it, at $\omega=\pi/4$. 

In particular, the $(m=2,n=1)$ case is again our LMG model in a slightly
re-parametrized form. We can consider the other cases generalizations
of the LMG model, in the sense that when we imagine the model derived
from spin-$1/2$ particles again, they describe interactions between
not just two spins, but $m$ or $n$ spins each instead. What is important
is that in every case the all-essential property $[H,\mathbf{S}^{2}]=0$
is preserved, and therefore the numerical methods of section \ref{sec:numerical_algorithm_lmg}
still remain applicable. In fact, even more symmetries are preserved:
the spin-flip symmetry in $z$-direction still exists in the cases
where $m$ is even, and there occurs a corresponding spin-flip symmetry
in $x$-direction for even $n$.

\section{Calculating the location of the phase transition}

We are of course again mainly interested in the behaviour of the mutual
information in the vicinity of the phase transition. For this, we
first have to know where exactly the phase transition is located.
It is in principle straightforward to generalize the procedure used
in the Lipkin model to higher spins; however, the equations become
significantly more complicated. In particular, there are in general
many more possible solutions for the magnetization and it will now
be much more important to check which possible solution for the magnetization
has the lowest free energy. It becomes very hard to write down analytical
solutions, but since we have no particular interest in the analytical
form anyway, we can just implement it numerically, which is possible
without problems. Let us recall all the steps from the LMG case and
see where things are different.

Again, we start by introducing order parameters $m_{x}=\left\langle \sigma_{x}\right\rangle $,
and now also $m_{z}=\left\langle \sigma_{z}\right\rangle $. The mean
field approach is just the same as in section \ref{sub:lmg-mean-field},
replacing $\sigma_{x}^{(i)}=m_{x}+(\sigma_{x}^{(i)}-m_{x})$ and $\sigma_{z}^{(i)}=m_{z}+(\sigma_{z}^{(i)}-m_{z})$
in the Hamiltonian, ignoring quadratic and higher terms in the {}``deviations''
$(\sigma_{x}^{(i)}-m_{x})$ and $(\sigma_{z}^{(i)}-m_{z})$. This
again yields a Hamiltonian of the form

\[
H=-\boldsymbol{h}_{\mathrm{eff}}\cdot\boldsymbol{S}+H_{0}
\]
which consists of a linearly spin-dependent part with 
\[
\boldsymbol{h}_{\mathrm{eff}}=(2mm_{x}^{m-1}\cos\omega,0,2K_{m,n}nm_{z}^{n-1}\sin\omega)
\]
and a part that involves only the mean-field magnetizations $m_{x}$,
$m_{z}$
\[
H_{0}=N[(m-1)m_{x}^{m}\cos\omega+K_{m,n}(n-1)m_{z}^{n}\sin\omega]\;.
\]
Again, we demand self-consistency as before, and find the mean-field
magnetizations (order parameters) as the self-consistent solution
that has the lowest free energy. The actual equations become rather
messy but can be handled efficiently numerically.

A phase transition occurs wherever an infinitesimal change of parameters
makes at least one of the order parameters change from zero to some
finite value, which can also easily be determined numerically.

\section{First order versus continuous (second order) phase transition}

In the LMG model, we were dealing with a \emph{continuous} phase transition.
This was manifesting in the fact that the order parameter went from
zero to a finite value in a continuous manner; this continuity does
not, however, extend to derivatives \cite{Yeomans92}.

\begin{figure}
\begin{centering}
\includegraphics[width=0.75\textwidth]{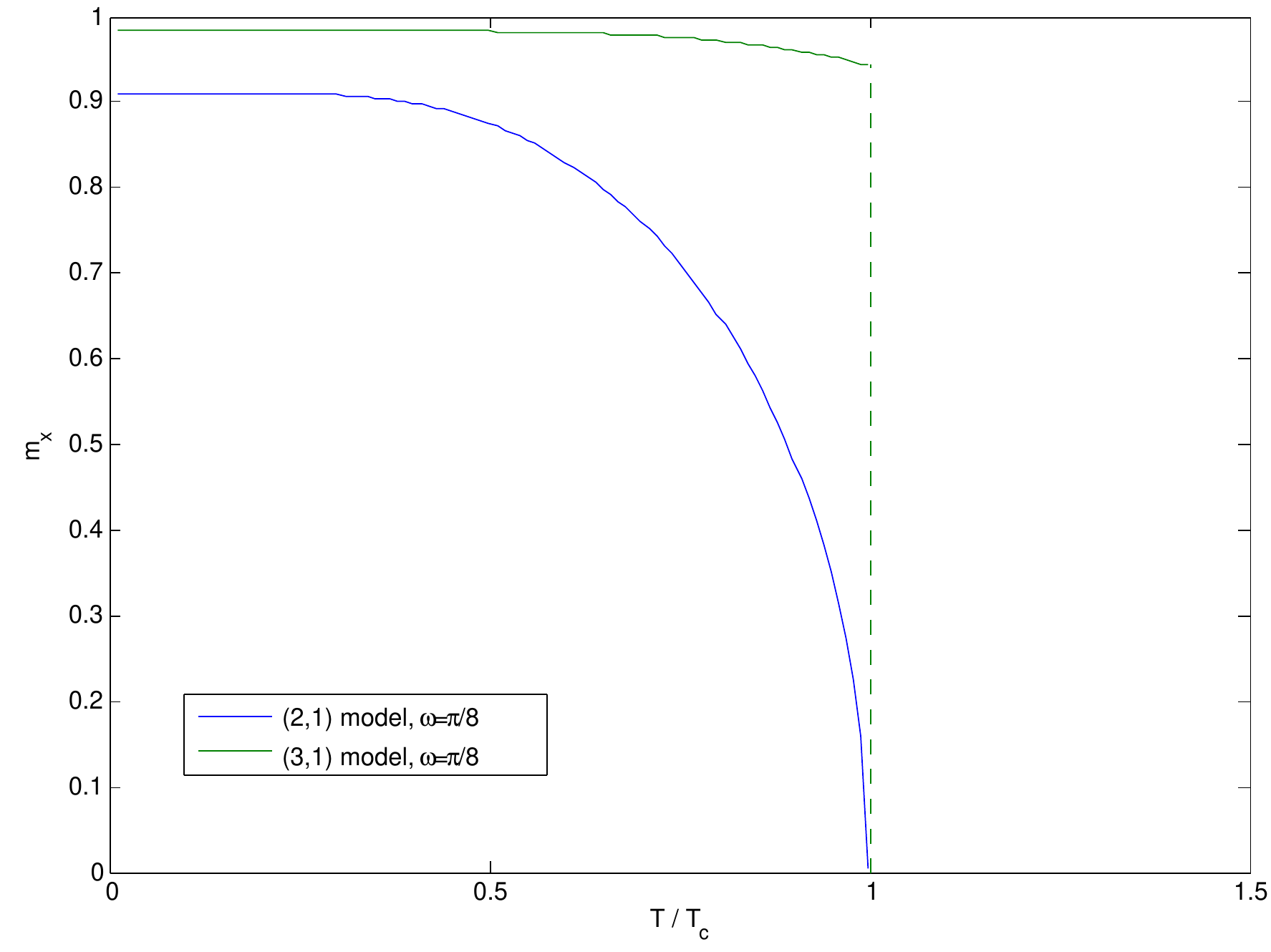}
\par\end{centering}

\caption{\label{fig:op-discont}Behavior of the order parameter in a first-order
vs. a second-order phase transition}

\end{figure}

However, many of the $m$-$n$-models we will now study do not exhibit
such a continuous (sometimes still called \emph{second-order}) transition.
Instead the order parameter is discontinuous at the phase transition,
as shown for the (3,1)-model in figure \ref{fig:op-discont}. It is
therefore also quite impossible to define anything like critical exponents
-- there is no universal behaviour for such \emph{first-order} transitions.

In the next section we will see that the behaviour of the mutual information
will also be very different in those models that show a first-order
transition rather than a continuous one.

\section{Results}

\begin{figure}
\includegraphics[width=1\textwidth]{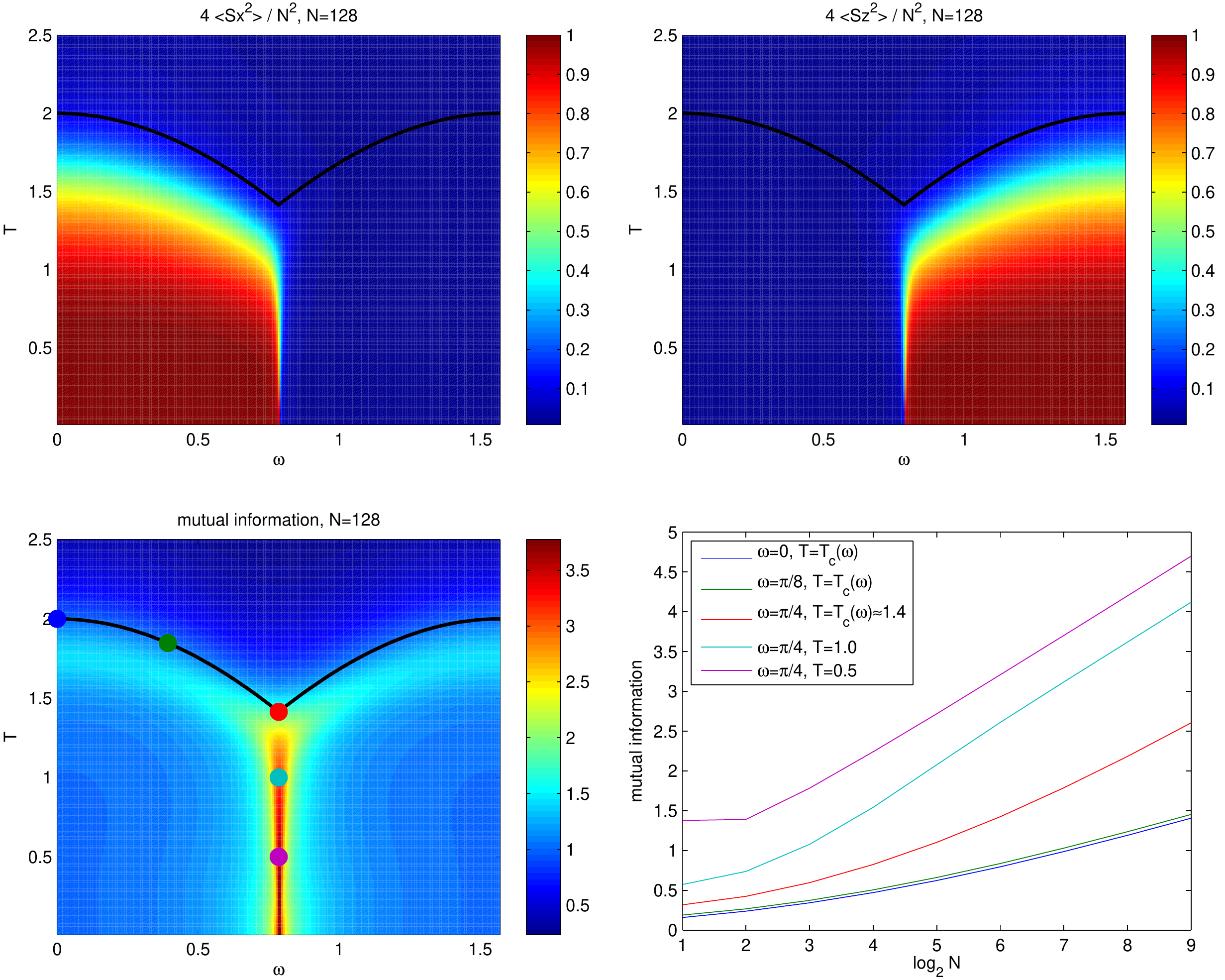}

\caption{$m=2,n=2$: Collective version of the quantum compass model \cite{kugel1982jahn}:
competing $x$- and $z$-interactions (of two spins each). There are
two essentially identical ordered phases and a high-temperature symmetric
phase. Again the mutual information works very well to point out the
different phases and their transititions (note that the ordered phases
are characterized by a mutual information of one, in contrast to the
unordered phase). At the phase transition, the mutual information
still seems to diverge logarithmically; interestingly, at the points
with $\omega=\pi/4$ the coefficient in front of the logarithm seems
to be $1/2$ rather than $1/4$ (just like the contributions from
both phases are added). }
\end{figure}

\begin{figure}
\includegraphics[width=1\textwidth]{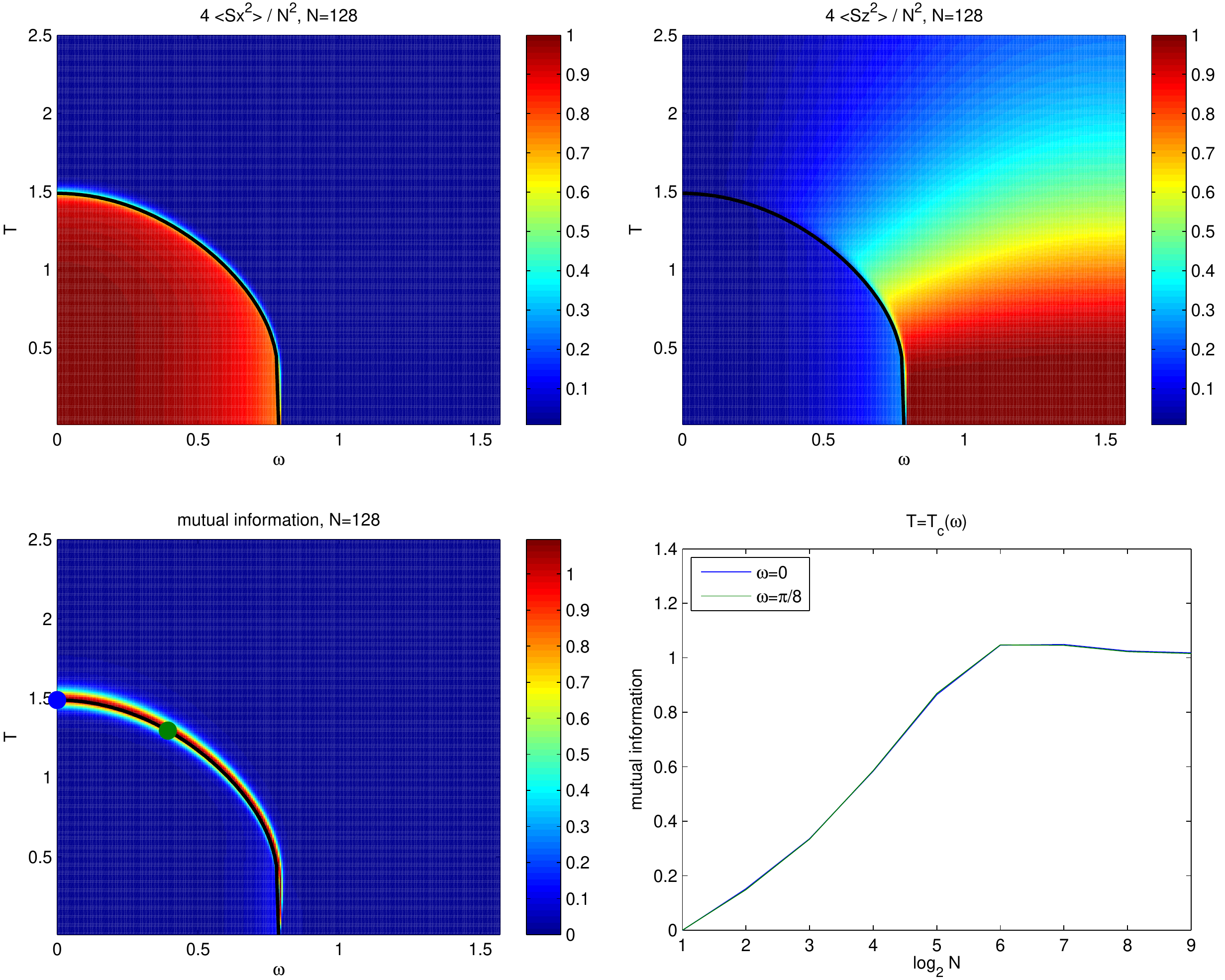}

\caption{$m=3,n=1$: Generalization of the LMG model in the sense that we have
a 3-spin interaction in an external field. Note that there are again
only two phases, and for an odd number of interacting Ising spins
even the one that was {}``ordered'' in the LMG model has zero mutual
information (a 3-spin interaction no longer orients all spins like
a 2-spin ferromagnetic interaction did). Note that the transition
has become much sharper. It is now a first-order transition rather
than a continuous/second-order one, and that also implies that there
are no longer any interesting exponents: while the mutual information
still pinpoints the phase transition very nicely, it no longer diverges
at the phase transition, but simply goes to a constant finite value
(of 1) in the thermodynamic limit. Everywhere else it goes to zero.}
\end{figure}

\begin{figure}
\includegraphics[width=1\textwidth]{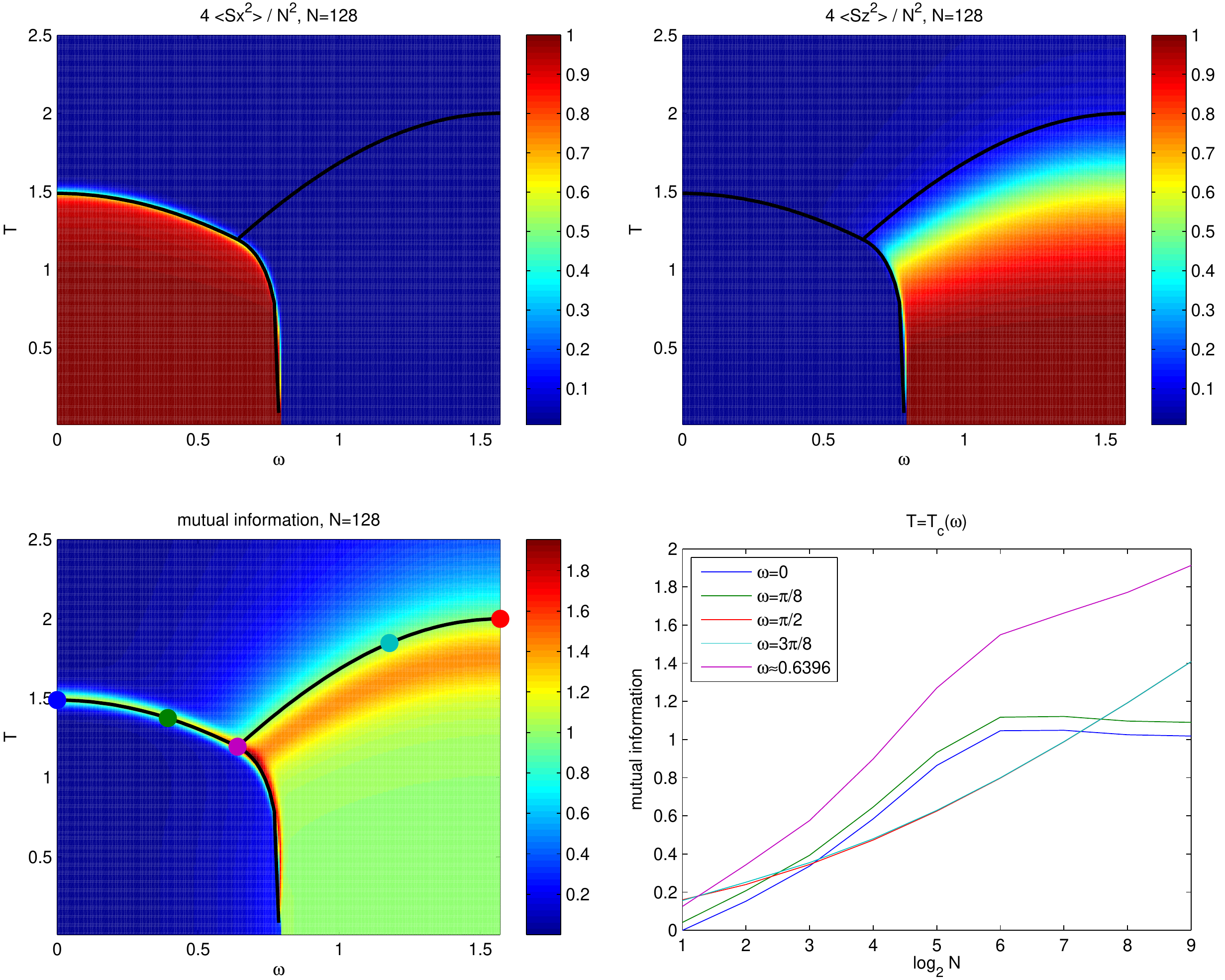}\caption{$m=3,n=2$: Competing 3-spin and 2-spin interactions: In the regime
where the two-spin interaction dominates, we get a phase with one
bit of mutual information again. Note how the mutual information very
clearly shows that there are three distinct phases again. This model
allows for a nice comparison between the sharp first-order phase transition,
where the mutual information does not diverge, and the far {}``wider''
second-order phase transition with logarithmically diverging mutual
information.}
\end{figure}

\begin{figure}
\includegraphics[width=1\textwidth]{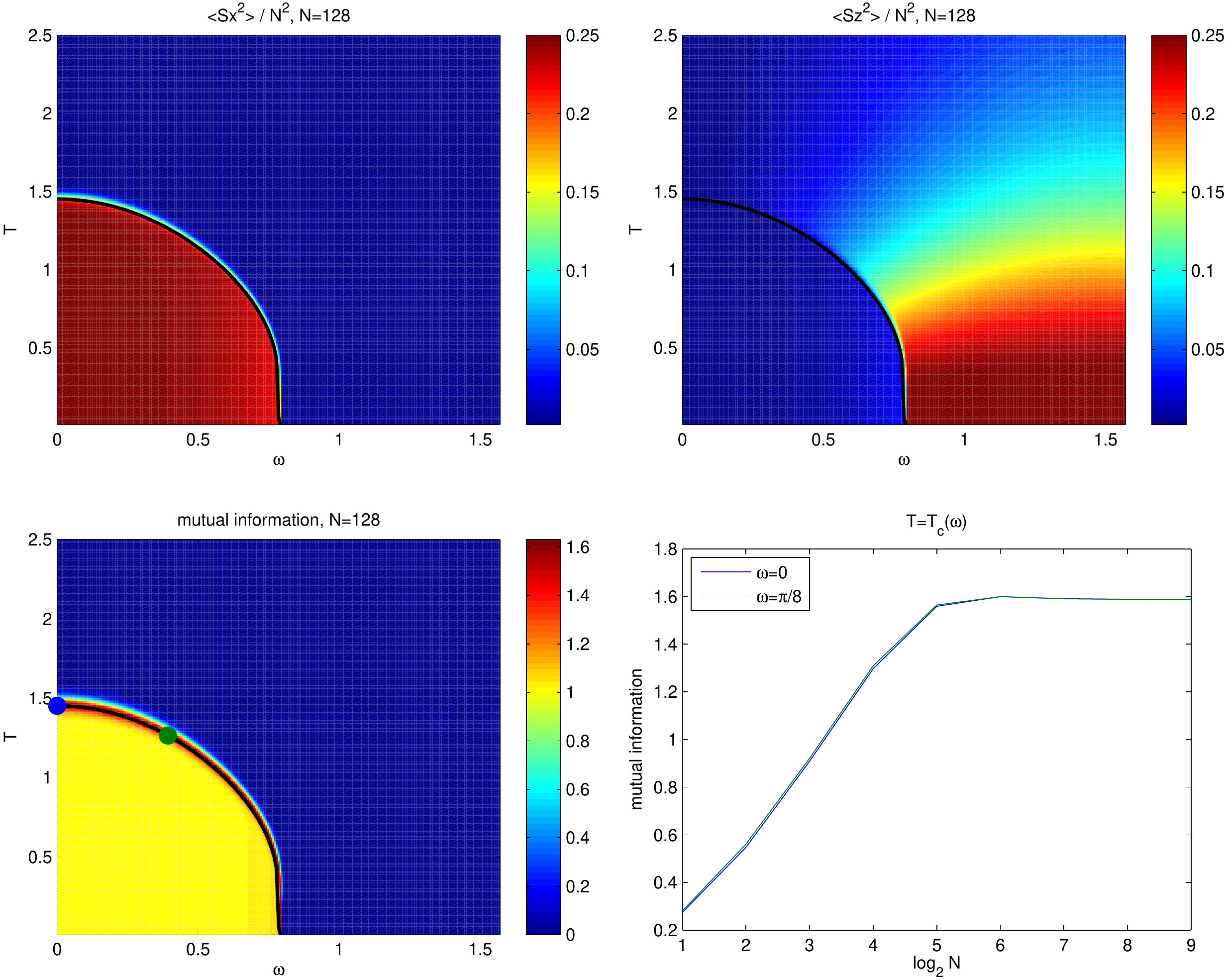}\caption{$m=4,n=1$: This figure has been included to show that also for even
exponents (where we have nonzero mutual information in the ordered
phase) one can again find a saturation of mutual information at the
first order phase transition, although now at a value larger than
1.}
\end{figure}

As mentioned, the LMG model is equivalent to the (2,1)-model. Since
it has been studied extensively in the previous chapter, we just remember
that we got a good idea about its phase diagram by looking at the
order parameter $m_{x}$ in figure \ref{fig:lmg_op_256}. Since we
can now also have an interaction in $S_{z}$-direction, we will for
the new models also plot the corresponding order parameter $m_{z}$.
As in section \ref{sub:LMG-Order-parameter}, we calculate the order
parameters using $\langle S_{x}^{2}\rangle$ and $\langle S_{z}^{2}\rangle$
to avoid any confusion by symmetry-breaking.

The essential behaviour that we found for the mutual information in
the LMG model was displayed in figure \ref{fig:lmg_mi_256}, where
you can see how it beautifully follows the phase transition, and in
the thermodynamic limit diverges logarithmically at the critical temperature,
as was examined in figure \ref{fig:lmg_mi_N}.

We will therefore now be studying the same sort of plots for the (2,2)-,
(3,1)-, (3,2)-, and (4,1)-models. The numerically determined locations
of the phase transitions will be marked by solid black lines.

The Hamiltonian is clearly symmetric in $m$ and $n$, so that e.g.
the (2,3)-model is exactly the same as the (3,2)-model. It also turns
out that models with even higher $m$ and $n$ will not show any interesting
new features, which is why we restrict ourselves to this selection.
Since some of the density plots will need finer resolution (i.e.,
more data points) than those in the last chapter, we reduced the system
sizes we studied to $N=128$, in order to speed up computations.

What do we find? The particular observations for each model are noted
in the caption below it, so let us just summarize the general behaviour
here: only a two-spin-interaction gives rise to a continuous phase
transition, where the mutual information diverges logarithmically.

Higher-order interactions still give rise to distinct phases, but
with a first-order transition between them, which shows up as much
sharper in the plots. In fact the mutual information captures also
this type of phase transition very well. However, one would not expect
any critical behaviour, and the mutual information shows none, it
just goes to a constant value.

Again, this sort of comparably trivial behaviour in the first-order
transitions corresponds with what has been found for the quantum phase
transitions in the same models in \cite{Filippone11}, where in particular
the entanglement entropy has been found to diverge only in continuous
transitions.

\chapter*{\label{chap:conclusion}Concluding remarks and outlook}

\addcontentsline{toc}{chapter}{Concluding remarks and outlook}

I hope that I have managed convince you that mutual information is
interesting! We have seen that is a well-founded correlation measure.
It is not easy to calculate though -- the biggest parts of this thesis
were dedicated to the problems of calculating it, in what are after
all rather simple model systems. I think it was worth the effort though:
We have clearly seen that mutual information has pronounced features
around a phase transition -- be it in the classical spin systems on
a lattice or in the fully-connected models. In the latter, we even
saw analytical results that support the argument that it is a natural
generalization of entanglement entropy.

For the systems studied, I certainly would not claim that it is very
useful for detecting the phase transitions, because these systems
are rather well understood and there are far easier ways to find their
phase transitions. We can however extract more information than just
the location of the phase transition -- we also learn about the amount
of correlations in the different phases and at the transition between
them: remember for example the Kurmann point in section \ref{sec:ising-to-xy}
and the markedly different behaviour at first- and second-order phase
transitions we saw in chapter \ref{chap:m-n-models}. Most crucially,
we get all this without having needed any intuition about order parameters.
That means such information-theory-based methods immediately suggest
themselves for the study of any sort of novel phases, which are not
yet as well-understood -- similarly as entanglement entropy has turned
out essential for ground-state studies in topological models \cite{kitaev2006topological,levin2006detecting}.
Of course that will in general be even more challenging, but I think
we have so far only seen a very small part of what is possible at
the point where (quantum) information theory and condensed matter
intersect.

\cleardoublepage{}

\clearpage{}

\bibliographystyle{unsrt}
\addcontentsline{toc}{chapter}{\bibname}\bibliography{ref}

\begin{thebibliography}{100}

\bibitem{Wilms11}
J.~Wilms, M.~Troyer, and F.~Verstraete.
\newblock Mutual information in classical spin models.
\newblock {\em Journal of Statistical Mechanics: Theory and Experiment},
  2011:P10011, 2011.

\bibitem{wilms2012finite}
J.~Wilms, J.~Vidal, F.~Verstraete, and S.~Dusuel.
\newblock Finite-temperature mutual information in a simple phase transition.
\newblock {\em Journal of Statistical Mechanics: Theory and Experiment},
  2012:P01023, 2012.

\bibitem{vedral1997quantifying}
V.~Vedral, M.~B. Plenio, M.~A. Rippin, and P.~L. Knight.
\newblock Quantifying entanglement.
\newblock {\em Phys. Rev. Lett.}, 78:2275--2279, 1997.

\bibitem{coverthomas}
T.~M. Cover and J.~A. Thomas.
\newblock {\em Elements of Information Theory}.
\newblock Wiley, 1991.

\bibitem{Shannon48}
C.~E. Shannon.
\newblock A mathematical theory of communication.
\newblock {\em The Bell System Technical Journal}, 27:379, 1948.

\bibitem{nielsenchuang2000}
M.~A. Nielsen and I.~L. Chuang.
\newblock {\em Quantum Computation and Quantum Information}.
\newblock Cambridge University Press, 2000.

\bibitem{cerf1997negative}
N.~J. Cerf and C.~Adami.
\newblock Negative entropy and information in quantum mechanics.
\newblock {\em Physical Review Letters}, 79(26):5194--5197, 1997.

\bibitem{renyi1960}
A.~R{\'e}nyi.
\newblock On measures of entropy and information.
\newblock In {\em Proceedings of the 4th Berkeley Symposium on Mathematics,
  Statistics and Probability}, page 547, 1961.

\bibitem{renyi1965}
A.~R{\'e}nyi.
\newblock On the foundations of information theory.
\newblock {\em Revue de l'Institut International de Statistique}, pages 1--14,
  1965.

\bibitem{viola1997alignment}
P.~Viola and W.~M. Wells~III.
\newblock Alignment by maximization of mutual information.
\newblock {\em International Journal of Computer Vision}, 24(2):137--154, 1997.

\bibitem{viola1995alignment}
P.~Viola.
\newblock {\em Alignment by maximization of mutual information}.
\newblock PhD thesis, Massachusetts Institute of Technology, 1995.

\bibitem{wells1996multi}
W.~M. Wells~III, P.~Viola, H.~Atsumi, S.~Nakajima, and R.~Kikinis.
\newblock Multi-modal volume registration by maximization of mutual
  information.
\newblock {\em Medical Image Analysis}, 1(1):35--51, 1996.

\bibitem{collignon1995automated}
A.~Collignon, F.~Maes, D.~Delaere, D.~Vandermeulen, P.~Suetens, and G.~Marchal.
\newblock Automated multi-modality image registration based on information
  theory.
\newblock In {\em Information Processing in Medical Imaging}, volume~3, pages
  264--274, 1995.

\bibitem{collignon1998multi}
A.~Collignon.
\newblock {\em Multi-modality medical image registration by maximization of
  mutual information}.
\newblock PhD thesis, Catholic University Leuven, Leuven, Belgium, 1998.

\bibitem{fraser1986independent}
A.~M. Fraser and H.~L. Swinney.
\newblock Independent coordinates for strange attractors from mutual
  information.
\newblock {\em Physical Review A}, 33(2):1134, 1986.

\bibitem{michel11}
J.~B. Michel, Y.~K. Shen, A.P. Aiden, A.~Veres, M.~K. Gray, J.~P. Pickett,
  D.~Hoiberg, D.~Clancy, P.~Norvig, J.~Orwant, et~al.
\newblock Quantitative analysis of culture using millions of digitized books.
\newblock {\em Science}, 331(6014):176, 2011.

\bibitem{matsuda1996}
H.~Matsuda, K.~Kudo, R.~Nakamura, O.~Yamakawa, and T.~Murata.
\newblock Mutual information of {Ising} systems.
\newblock {\em International Journal of Theoretical Physics}, 35(4):839--845,
  1996.

\bibitem{sole1996}
R.~V. Sol{\'e}, S.~C. Manrubia, B.~Luque, J.~Delgado, and J.~Bascompte.
\newblock Phase transitions and complex systems.
\newblock {\em Complexity}, 1(4):13, 1996.

\bibitem{arnold1996}
D.~V. Arnold.
\newblock Information-theoretic analysis of phase transitions.
\newblock {\em Complex Systems}, 10:143, 1996.

\bibitem{Hastings10}
M.~B. Hastings, I.~Gonzalez, A.~B. Kallin, and R.~G. Melko.
\newblock Measuring {R{\'e}nyi} entanglement entropy in {Quantum Monte Carlo}
  simulations.
\newblock {\em Physical Review Letters}, 104:157201, 2010.

\bibitem{Melko10}
R.~G. Melko, A.~B. Kallin, and M.~B. Hastings.
\newblock Finite-size scaling of mutual information in {Monte Carlo
  simulations}: {Application to the spin-1/2 XXZ model}.
\newblock {\em Physical Review B}, 82:100409, 2010.

\bibitem{Singh11}
R.~R.~P. Singh, M.~B. Hastings, A.~B. Kallin, and R.~G. Melko.
\newblock Finite-temperature critical behavior of mutual information.
\newblock {\em Physical Review Letters}, 106:135701, 2011.

\bibitem{smoothrenyi}
R.~Renner and S.~Wolf.
\newblock {Smooth R{\'e}nyi entropy and applications}.
\newblock In {\em International Symposium on Information Theory, 2004.
  Proceedings}, 2004.

\bibitem{Koenig09}
R.~K\"onig, R.~Renner, and C.~Schaffner.
\newblock The operational meaning of min- and max-entropy.
\newblock {\em IEEE Trans. Inf. Theory}, 55:4337, 2009.

\bibitem{hartley1928}
R.~V.~L. Hartley.
\newblock Transmission of information.
\newblock {\em The Bell System Technical Journal}, 7:535, 1928.

\bibitem{arealaws}
M.~M. Wolf, F.~Verstraete, M.~B. Hastings, and J.~I. Cirac.
\newblock {Area laws in quantum systems: mutual information and correlations}.
\newblock {\em Physical Review Letters}, 100(7):70502, 2008.

\bibitem{vidal2003entanglement}
G.~Vidal, J.~I. Latorre, E.~Rico, and A.~Kitaev.
\newblock {Entanglement in quantum critical phenomena}.
\newblock {\em Physical Review Letters}, 90(22):227902, 2003.

\bibitem{calabrese2004entanglement}
P.~Calabrese and J.~Cardy.
\newblock {Entanglement entropy and quantum field theory}.
\newblock {\em Journal of Statistical Mechanics: Theory and Experiment},
  2004:P06002, 2004.

\bibitem{plenio2005entropy}
M.~B. Plenio, J.~Eisert, J.~Dreissig, and M.~Cramer.
\newblock {Entropy, entanglement, and area: analytical results for harmonic
  lattice systems}.
\newblock {\em Physical Review Letters}, 94(6):60503, 2005.

\bibitem{its2005entanglement}
A.~R. Its, B.~Q. Jin, and V.~E. Korepin.
\newblock {Entanglement in the {XY} spin chain}.
\newblock {\em Journal of Physics A: Mathematical and General}, 38:2975, 2005.

\bibitem{verstraete2006matrix}
F.~Verstraete and J.~I. Cirac.
\newblock {Matrix product states represent ground states faithfully}.
\newblock {\em Physical Review B}, 73(9):94423, 2006.

\bibitem{verstraete2006criticality}
F.~Verstraete, M.~M. Wolf, D.~Perez-Garcia, and J.~I. Cirac.
\newblock {Criticality, the area law, and the computational power of projected
  entangled pair states}.
\newblock {\em Physical Review Letters}, 96(22):220601, 2006.

\bibitem{hastings2007area}
M.~B. Hastings.
\newblock {An area law for one-dimensional quantum systems}.
\newblock {\em Journal of Statistical Mechanics: Theory and Experiment},
  2007:P08024, 2007.

\bibitem{eisert2010colloquium}
J.~Eisert, M.~Cramer, and M.~B. Plenio.
\newblock {Colloquium: Area laws for the entanglement entropy}.
\newblock {\em Reviews of Modern Physics}, 82:277--306, 2010.

\bibitem{stephanshannon}
J.~M. St{\'e}phan, S.~Furukawa, G.~Misguich, and V.~Pasquier.
\newblock {Shannon and entanglement entropies of one-and two-dimensional
  critical wave functions}.
\newblock {\em Physical Review B}, 80(18):184421, 2009.

\bibitem{renyi2dising}
J.~M. St{\'e}phan, G.~Misguich, and V.~Pasquier.
\newblock R{\'e}nyi entropy of a line in two-dimensional ising models.
\newblock {\em Physical Review B}, 82(12):125455, 2010.

\bibitem{Latorre05_2}
J.~I. Latorre, R.~Or\'us, E.~Rico, and J.~Vidal.
\newblock Entanglement entropy in the {Lipkin-Meshkov-Glick model}.
\newblock {\em Physical Review A}, 71:064101, 2005.

\bibitem{Barthel06_2}
T.~Barthel, S.~Dusuel, and J.~Vidal.
\newblock Entanglement entropy beyond the free case.
\newblock {\em Physical Review Letters}, 97:220402, 2006.

\bibitem{Vidal07}
J.~Vidal, S.~Dusuel, and T.~Barthel.
\newblock Entanglement entropy in collective models.
\newblock {\em Journal of Statistical Mechanics: Theory and Experiment}, 2007.

\bibitem{Orus08_2}
R.~Or\'us, S.~Dusuel, and J.~Vidal.
\newblock Equivalence of critical scaling laws for many-body entanglement in
  the {Lipkin-Meshkov-Glick} model.
\newblock {\em Physical Review Letters}, 101:025701, 2008.

\bibitem{metropolis1953}
N.~Metropolis, A.~W. Rosenbluth, M.~N. Rosenbluth, A.~H. Teller, and E.~Teller.
\newblock Equation of state calculations by fast computing machines.
\newblock {\em The Journal of Chemical Physics}, 21:1087, 1953.

\bibitem{mcbook1}
D.~P. Landau and K.~Binder.
\newblock {\em A guide to {Monte Carlo} simulations in statistical physics}.
\newblock Cambridge University Press, 2005.

\bibitem{mpsreview}
F.~Verstraete, V.~Murg, and J.~I. Cirac.
\newblock {Matrix product states, projected entangled pair states, and
  variational renormalization group methods for quantum spin systems}.
\newblock {\em Advances in Physics}, 57(2):143--224, 2008.

\bibitem{pirvu2010matrix}
B.~Pirvu, V.~Murg, J.~I. Cirac, and F.~Verstraete.
\newblock Matrix product operator representations.
\newblock {\em New Journal of Physics}, 12:025012, 2010.

\bibitem{kramerswannier1941-1}
H.~A. Kramers and G.~H. Wannier.
\newblock Statistics of the two-dimensional ferromagnet. {Part I}.
\newblock {\em Physical Review}, 60:252--262, 1941.

\bibitem{kramerswannier1941-2}
H.~A. Kramers and G.~H. Wannier.
\newblock Statistics of the two-dimensional ferromagnet. {Part II}.
\newblock {\em Physical Review}, 60(3):263--276, 1941.

\bibitem{onsager1944}
L.~Onsager.
\newblock Crystal statistics. {I}. {A} two-dimensional model with an
  order-disorder transition.
\newblock {\em Physical Review}, 65(3-4):117, 1944.

\bibitem{exactlysolvedbaxter1982}
R.~J. Baxter.
\newblock {\em {Exactly solved models in statistical mechanics}}.
\newblock Academic Press London, 1982.

\bibitem{nishino1}
T.~Nishino.
\newblock {Density matrix renormalization group method for 2D classical
  models}.
\newblock {\em Journal of the Physical Society of Japan}, 64(10):3598--3601,
  1995.

\bibitem{nishino2}
T.~Nishino and K.~Okunishi.
\newblock {Product wave function renormalization group}.
\newblock {\em Journal of the Physical Society of Japan}, 64(11):4084--4087,
  1995.

\bibitem{pbcmps}
B.~Pirvu, F.~Verstraete, and G.~Vidal.
\newblock Exploiting translational invariance in {Matrix Product State}
  simulations of spin chains with periodic boundary conditions.
\newblock {\em Physical Review B}, 83(12):125104, 2011.

\bibitem{sandvik2007}
A.~W. Sandvik and G.~Vidal.
\newblock Variational quantum {Monte Carlo} simulations with tensor-network
  states.
\newblock {\em Physical Review Letters}, 99(22):220602, 2007.

\bibitem{schuch2008}
N.~Schuch, M.~M. Wolf, F.~Verstraete, and J.~I. Cirac.
\newblock Simulation of quantum many-body systems with strings of operators and
  {Monte Carlo} tensor contractions.
\newblock {\em Physical Review Letters}, 100(4):40501, 2008.

\bibitem{ising1925}
E.~Ising.
\newblock {Beitrag zur Theorie des Ferromagnetismus}.
\newblock {\em Zeitschrift f{\"u}r Physik}, 31(1):253--258, 1925.

\bibitem{valiant2002quantum}
L.~G. Valiant.
\newblock Quantum circuits that can be simulated classically in polynomial
  time.
\newblock {\em SIAM Journal on Computing}, 31:1229, 2002.

\bibitem{knill2001fermionic}
E.~Knill.
\newblock Fermionic linear optics and matchgates.
\newblock {\em Arxiv preprint quant-ph/0108033}, 2001.

\bibitem{terhal2002classical}
B.~M. Terhal and D.~P. DiVincenzo.
\newblock Classical simulation of noninteracting-fermion quantum circuits.
\newblock {\em Physical Review A}, 65(3):032325, 2002.

\bibitem{bravyi2002fermionic}
S.~Bravyi and A.~Kitaev.
\newblock Fermionic quantum computation.
\newblock {\em Annals of Physics}, 298(1):210--226, 2002.

\bibitem{bravyi2005lagrangian}
S.~Bravyi.
\newblock Lagrangian representation for fermionic linear optics.
\newblock {\em Quantum Information \& Computation}, 5(3):216--238, 2005.

\bibitem{jozsa2008matchgates}
R.~Jozsa and A.~Miyake.
\newblock Matchgates and classical simulation of quantum circuits.
\newblock {\em Proceedings of the Royal Society A: Mathematical, Physical and
  Engineering Science}, 464(2100):3089--3106, 2008.

\bibitem{van2009quantum}
M.~van~den Nest, W.~D{\"u}r, R.~Raussendorf, and H.~J. Briegel.
\newblock Quantum algorithms for spin models and simulable gate sets for
  quantum computation.
\newblock {\em Physical Review A}, 80(5):52334, 2009.

\bibitem{jozsa2010matchgate}
R.~Jozsa, B.~Kraus, A.~Miyake, and J.~Watrous.
\newblock Matchgate and space-bounded quantum computations are equivalent.
\newblock {\em Proceedings of the Royal Society A: Mathematical, Physical and
  Engineering Science}, 466(2115):809--830, 2010.

\bibitem{wick1950evaluation}
G.~C. Wick.
\newblock The evaluation of the collision matrix.
\newblock {\em Physical Review}, 80(2):268--272, 1950.

\bibitem{majorana1937teoria}
E.~Majorana.
\newblock Teoria simmetrica dell elettrone e del positrone.
\newblock {\em Nuovo Cimento}, 14(4):171--184, 1937.

\bibitem{mccoy1973tdim}
B.~M. McCoy and T.~T. Wu.
\newblock {\em {The two-dimensional Ising model}}.
\newblock Harvard University Press, 1973.

\bibitem{fisher1966dspim}
M.~E. Fisher.
\newblock {On the dimer solution of planar Ising models}.
\newblock {\em Journal of Mathematical Physics}, 7:1776, 1966.

\bibitem{fisher1961statistical}
M.~E. Fisher.
\newblock Statistical mechanics of dimers on a plane lattice.
\newblock {\em Physical Review}, 124(6):1664, 1961.

\bibitem{kasteleyn1961statistics}
P.~W. Kasteleyn.
\newblock The statistics of dimers on a lattice:: I. the number of dimer
  arrangements on a quadratic lattice.
\newblock {\em Physica}, 27(12):1209--1225, 1961.

\bibitem{temperley1961dimer}
H.~N.~V. Temperley and M.~E. Fisher.
\newblock Dimer problem in statistical mechanics-an exact result.
\newblock {\em Philosophical Magazine}, 6(68):1061--1063, 1961.

\bibitem{ieee754}
{IEEE Standard for Floating-Point Arithmetic}.
\newblock Technical report, Microprocessor Standards Committee of the IEEE
  Computer Society, 2008.

\bibitem{fortuinkasteleyn}
C.~M. Fortuin and P.~W. Kasteleyn.
\newblock {On the random-cluster model: {I. Introduction and relation to other
  models}}.
\newblock {\em Physica}, 57(4):536--564, 1972.

\bibitem{swendsenwang}
R.~H. Swendsen and J.~S. Wang.
\newblock {Nonuniversal critical dynamics in {Monte Carlo} simulations}.
\newblock {\em Physical Review Letters}, 58(2):86--88, 1987.

\bibitem{ALPS}
F.~Albuquerque et~al. (ALPS~collaboration).
\newblock {The ALPS project release 1.3: open source software for strongly
  correlated systems}.
\newblock {\em Journal of Magnetism and Magnetic Materials}, 310:1187, 2007.

\bibitem{ALPSscheduler}
M.~Troyer, B.~Ammon, and E.~Heeb.
\newblock {Parallel object oriented Monte Carlo Simulations}.
\newblock {\em Lecture Notes in Computer Science}, 1505:191, 1998.

\bibitem{Potts51}
R.~B. Potts.
\newblock {\em The Mathematical Investigation of Some Cooperative Phenomena}.
\newblock PhD thesis, University of Oxford, 1951.

\bibitem{potts1952some}
R.~B. Potts.
\newblock Some generalized order-disorder transformations.
\newblock {\em Mathematical Proceedings of the Cambridge Philosophical
  Society}, 48(01):106--109, 1952.

\bibitem{pottsrmpwu}
F.~Y. Wu.
\newblock {The Potts model}.
\newblock {\em Reviews of Modern Physics}, 54(1):235--268, 1982.

\bibitem{clock456}
J.~Tobochnik.
\newblock {Properties of the q-state clock model for q=4, 5, and 6}.
\newblock {\em Physical Review B}, 26(11):6201--6207, 1982.

\bibitem{clock456erratum}
J.~Tobochnik.
\newblock {Erratum: Properties of the q-state clock model for q=4, 5, and 6}.
\newblock {\em Physical Review B}, 27(11):6972, 1983.

\bibitem{clock6}
M.~S.~S. Challa and D.~P. Landau.
\newblock {Critical behavior of the six-state clock model in two dimensions}.
\newblock {\em Physical Review B}, 33(1):437--443, 1986.

\bibitem{betts1964exact}
D.~D. Betts.
\newblock The exact solution of some lattice statistics models with four states
  per site.
\newblock {\em Canadian Journal of Physics}, 42(8):1564--1572, 1964.

\bibitem{clock4}
M.~Suzuki.
\newblock {Solution of Potts Model for Phase Transition}.
\newblock {\em Progress of Theoretical Physics}, 37:770--772, 1967.

\bibitem{Lipkin65}
H.~J. Lipkin, N.~Meshkov, and A.~J. Glick.
\newblock Validity of many-body approximation methods for a solvable model:
  {(I). Exact solutions and perturbation theory}.
\newblock {\em Nuclear Physics}, 62:188, 1965.

\bibitem{Meshkov65}
N.~Meshkov, A.~J. Glick, and H.~J. Lipkin.
\newblock Validity of many-body approximation methods for a solvable model:
  {(II). Linearization procedures}.
\newblock {\em Nuclear Physics}, 62:199, 1965.

\bibitem{Glick65}
A.~J. Glick, H.~J. Lipkin, and N.~Meshkov.
\newblock Validity of many-body approximation methods for a solvable model:
  {(III). Diagram summations}.
\newblock {\em Nuclear Physics}, 62:211, 1965.

\bibitem{Stigler80}
S.~Stigler.
\newblock Stigler's law of eponymy.
\newblock In T.~F. Gieryn, editor, {\em Science and social structure: a
  festschrift for Robert K. Merton}. New York Academy of Sciences, 1980.

\bibitem{Fallieros59}
S.~Fallieros.
\newblock {\em Collective oscillations in O-16}.
\newblock PhD thesis, University of Maryland, 1959.

\bibitem{Maruhn10}
J.~A. Maruhn, P.~Reinhard, and E.~Suraud.
\newblock {\em Simple Models of Many-Fermion Systems}.
\newblock Springer, 2010.

\bibitem{cirac1998quantum}
J.~I. Cirac, M.~Lewenstein, K.~M\o{}lmer, and P.~Zoller.
\newblock Quantum superposition states of bose-einstein condensates.
\newblock {\em Physical Review A}, 57:1208--1218, 1998.

\bibitem{larson2010circuit}
J.~Larson.
\newblock Circuit {QED} scheme for the realization of the {Lipkin-Meshkov-Glick
  model}.
\newblock {\em Europhysics Letters}, 90:54001, 2010.

\bibitem{Vidal04_1}
J.~Vidal, G.~Palacios, and R.~Mosseri.
\newblock Entanglement in a second order quantum phase transition.
\newblock {\em Physical Review A}, 69:022107, 2004.

\bibitem{Dusuel04_3}
S.~Dusuel and J.~Vidal.
\newblock Finite-size scaling exponents of the {Lipkin-Meshkov-Glick} model.
\newblock {\em Physical Review Letters}, 93:237204, 2004.

\bibitem{Dusuel05_2}
S.~Dusuel and J.~Vidal.
\newblock Continuous unitary transformations and finite-size scaling exponents
  in the {Lipkin-Meshkov-Glick model}.
\newblock {\em Physical Review B}, 71:224420, 2005.

\bibitem{Wichterich10}
H.~Wichterich, J.~Vidal, and S.~Bose.
\newblock Universality of the negativity in the {Lipkin-Meshkov-Glick model}.
\newblock {\em Physical Review A}, 81:032311, 2010.

\bibitem{Botet82}
R.~Botet, R.~Jullien, and P.~Pfeuty.
\newblock Size scaling for infinitely coordinated systems.
\newblock {\em Physical Review Letters}, 49(7):478, 1982.

\bibitem{Botet83}
R.~Botet and R.~Jullien.
\newblock Large-size critical behavior of infinitely coordinated systems.
\newblock {\em Physical Review B}, 28:3955, 1983.

\bibitem{Dutta11}
A.~Dutta, U.~Divakaran, D.~Sen, B.~K. Chakrabarti, T.~F. Rosenbaum, and
  G.~Aeppli.
\newblock Transverse field spin models: From statistical physics to quantum
  information.
\newblock 2010.
\newblock arXiv:1012.0653.

\bibitem{Quan09}
H.~T. Quan and F.~M. Cucchietti.
\newblock Quantum fidelity and thermal phase transitions.
\newblock {\em Physical Review E}, 79:031101, 2009.

\bibitem{moler1978nineteen}
C.~Moler and C.~Van~Loan.
\newblock Nineteen dubious ways to compute the exponential of a matrix.
\newblock {\em SIAM Review}, pages 801--836, 1978.

\bibitem{moler2003nineteen}
C.~Moler and C.~Van~Loan.
\newblock Nineteen dubious ways to compute the exponential of a matrix,
  twenty-five years later.
\newblock {\em SIAM Review}, pages 3--49, 2003.

\bibitem{Wootters98}
W.~K. Wootters.
\newblock Entanglement of formation of an arbitrary state of two qubits.
\newblock {\em Physical Review Letters}, 80:2245, 1998.

\bibitem{Matera08}
J.~M. Matera, R.~Rossignoli, and N.~Canosa.
\newblock Thermal entanglement in fully connected spin systems and its
  random-phase-approximation description.
\newblock {\em Physical Review A}, 78:012316, 2008.

\bibitem{Kwok08}
H.-M. Kwok, W.-Q. Ning, S.-J. Gu, and H.-Q. Lin.
\newblock Quantum criticality of the {Lipkin-Meshkov-Glick model} in terms of
  fidelity susceptibility.
\newblock {\em Physical Review E}, 78:032103, 2008.

\bibitem{Ma08}
J.~Ma, L.~Xu, H.-N. Xiong, and X.~Wang.
\newblock Reduced fidelity susceptibility and its finite-size scaling
  behaviors.
\newblock {\em Physical Review E}, 78:051126, 2008.

\bibitem{kurmann1982}
J.~Kurmann, H.~Thomas, and G.~M\"uller.
\newblock Antiferromagnetic long-range order in the anisotropic quantum spin
  chain.
\newblock {\em Physica A: Statistical and Theoretical Physics}, 112(1-2):235,
  1982.

\bibitem{li2008entanglement}
H.~Li and F.~D.~M. Haldane.
\newblock Entanglement spectrum as a generalization of entanglement entropy:
  Identification of topological order in non-abelian fractional quantum hall
  effect states.
\newblock {\em Physical Review Letters}, 101(1):10504, 2008.

\bibitem{calabrese2008entanglement}
P.~Calabrese and A.~Lefevre.
\newblock Entanglement spectrum in one-dimensional systems.
\newblock {\em Physical Review A}, 78(3):032329, 2008.

\bibitem{fidkowski2010entanglement}
L.~Fidkowski.
\newblock Entanglement spectrum of topological insulators and superconductors.
\newblock {\em Physical Review Letters}, 104:130502, 2010.

\bibitem{turner2010entanglement}
A.~M. Turner, Y.~Zhang, and A.~Vishwanath.
\newblock Entanglement and inversion symmetry in topological insulators.
\newblock {\em Physical Review B}, 82:241102, 2010.

\bibitem{schulten1975}
K.~Schulten and R.~G. Gordon.
\newblock Exact recursive evaluation of 3j-and 6j-coefficients for
  quantum-mechanical coupling of angular momenta.
\newblock {\em Journal of Mathematical Physics}, 16:1961, 1975.

\bibitem{Bender99}
C.~M. Bender and S.~A. Orszag.
\newblock {\em Advanced mathematical methods for scientists and engineers:
  Asymptotic methods and perturbation theory}.
\newblock Springer, 1978.

\bibitem{Filippone11}
M.~Filippone, S.~Dusuel, and J.~Vidal.
\newblock Quantum phase transitions in fully connected spin models: An
  entanglement perspective.
\newblock {\em Physical Review A}, 83:022327, 2011.

\bibitem{Yeomans92}
J.~M. Yeomans.
\newblock {\em Statistical mechanics of phase transitions}.
\newblock Oxford University Press, 1992.

\bibitem{kugel1982jahn}
K.~I. Kugel' and D.~I. Khomski{\u\i}.
\newblock The {Jahn-Teller} effect and magnetism: transition metal compounds.
\newblock {\em Soviet Physics Uspekhi}, 25(4):231--256, 1982.

\bibitem{kitaev2006topological}
A.~Kitaev and J.~Preskill.
\newblock {Topological entanglement entropy}.
\newblock {\em Physical Review Letters}, 96(11):110404, 2006.

\bibitem{levin2006detecting}
M.~Levin and X.~G. Wen.
\newblock {Detecting topological order in a ground state wave function}.
\newblock {\em Physical Review Letters}, 96(11):110405, 2006.

\end{thebibliography}

\chapter*{Acknowledgements}

\addcontentsline{toc}{chapter}{Acknowledgements}

Most of all I would like to thank Frank Verstraete, who was an extremely
knowledgeable and at the same time approachable advisor, and who created
a great open environment, scientifically and otherwise, in which it
was possible to freely pursue ideas.

I very much enjoyed working on various projects with Sébastien Dusuel,
Matthias Troyer, and Julien Vidal. I gained a lot of insight about
Rényi entropies from talking to Normand Beaudry and Renato Renner.
I am also grateful to Caslav Brukner, Hans Gerd Evertz, Barbara Kraus,
Maarten van den Nest, and Gerardo Ortiz for various other interesting
and helpful discussions.

Special thanks go to Markus Arndt, who created and managed the doctoral
program Complex Quantum Systems, which brought me to Vienna in the
first place, and in which I had the pleasure to participate.
\end{document}